\documentstyle[pre,aps,amsmath,amsfonts]{revtex} 
\begin{document}
\newcommand{\bea}{\begin{eqnarray}}
\newcommand{\eea}{\end{eqnarray}}

\title{An Integrable $U_q\bigl(\widehat{gl}(2\vert2)\bigr)_1$-Model:\\
Corner Transfer Matrices and Young Skew Diagrams}
\author{R.M. Gade\footnote{e-mail:renate.gade@t-online.de}
\\Lerchenfeldstra\ss e 12, 80538 Munich, Germany}
\maketitle

\begin{abstract}
The path space of an inhomogeneous vertex model constructed from the vector
representation
of $U_q\bigl(gl(2\vert2)\bigr)$ and its dual is studied for various choices of
composite vertices and assignments of $gl(2\vert2)$-weights.
At $q=0$, the corner transfer matrix Hamiltonian acts trigonally on the space
of half-infinite configurations subject to a particular boundary condition.
A weight-preserving one-to-one correspondence between the half-infinite
configurations and the weight states of a level-one module of $U_q\bigl(
\widehat{sl}(2\vert2)\bigr)/{\cal H}$ with grade $-n$ is found for $n\geq-3$ if
the grade $-n$ is identified with the diagonal element of the CTM Hamiltonian.
In each case, the module
can be decomposed into two irreducible level-one modules, one of them
including infinitely many weight states at fixed grade.
Based on a mapping of the path space onto pairs of border stripes, the
character of the reducible module is decomposed in terms of skew Schur functions.
Relying on an explicit verification for simple border stripes,
a correspondence between the paths and level-zero modules of
$U_q\bigl(\widehat{sl}(2\vert2)\bigr)$ constructed from an infinite-dimensional
$U_q\bigl(gl(2\vert2)\bigr)$-module is conjectured.
\end{abstract}


\section{Introduction}

A link between integrable lattice models and irreducible modules of quantum affine
algebras $U_q\bigl(\hat g\bigr)$
is provided by the spectra of the corner transfer matrix originally introduced in
\cite{bax}. The trace of the CTM of a homogeneous integrable vertex model based on $U_q
\bigl(\hat g\bigr)$  relates to an affine Lie algebra character
\cite{djkmo,fm}. Incorporating the concept of vertex operators,
a mathematical characterization of
physical objects including transfer matrices and N-point correlators
has been developed \cite{dav,jm}.  Descriptions of the same type have been given
for $U_q\bigl(\widehat{sl}(2)\bigr)$-vertex models built from two different spin
representations \cite{mix1,mix2} and for other
integrable models (see \cite{abf,ctm1,ctm2,jmo} and references in \cite{jm}).
More recently, the CTM spectrum of an inhomogeneous vertex model based on
the quantum affine superalgebra $U_q\bigl(\widehat{sl}(2\vert1)\bigr)$ has been
investigated \cite{gade3}. An analysis of all CTM-eigenvectors with
eigenvalue $\geq-4$ suggests a one-to-one correspondence between the space of
half-infinite configurations subject to a suitable boundary condition
and the level-one module $V(\Lambda_2)$ of
$U_q\bigl(\widehat{sl}(2\vert1)\bigr)$.

This study deals with an integrable vertex model characterized by alternating
sequences of the vector representation $W$ of $U_q\bigl(gl(2\vert2)\bigr)$ and
its dual $W^*$ in horizontal as well as vertical direction. The R-matrices
attributed to composite vertices are required to provide
well-defined invertible maps of the tensor products of evaluation
modules in the limit $q\to0$. For two choices of composite vertices, this is
achieved by a suitable adjustment of the inhomogeneity in the spectral parameters.
In the limit $q\to0$, the action of the corner transfer matrix Hamiltonian in
the space of the half-infinite configurations becomes triangular with respect
to the canonical basis. Moreover, the diagonal elements decouple into a
contribution depending only on the part governed by the modules $W$
and into a second contribution due to the modules $W^*$ only. Thus the present
model clearly differs from the $U_q\bigl(\widehat{sl}(2\vert1)\bigr)$-model
in \cite{gade3} whose CTM Hamiltonian
does not show a triangular action in the canonical basis even for $q\to0$.
For either choice of the composite vertices,
two assignments of $U_q\bigl(gl(2\vert2)\bigr)$-weights are considered.
With respect to two assignments,
the $U_q\bigl(gl(2\vert2)\bigr)$-weights of all configurations with the diagonal
element of the CTM Hamiltonian given by $-n$ are collected for $n=0,1,2,3$.
They are found in one-to-one correspondence
with the weight states of a reducible level-one module $\tilde V(\Lambda_0)$
at grade $-n$. Two reducible modules
$\tilde V(\Lambda_1+\Lambda_4)$ and $\tilde V(2\Lambda_0-\Lambda_{3}+\Lambda_4)$
account for the remaining assignments of weights.
All are modules of
$U_q\bigl(\widehat{sl}(2\vert2)\bigr)/{\cal H}$ with ${\cal H}$ denoting the
center of the algebra.
Relying on these results, the correspondence may be assumed to hold at any grade.
Each of the reducible modules can be decomposed into two irreducible level-one
modules, one of them containing a finite-dimensional, the other an infinite-dimensional
$U_q\bigl(sl(2\vert2)\bigr)$-module as their grade-zero subspace. Hence the
reducible modules are nonintegrable. The occurrence
of nonintegrable modules appears to be a typical feature of inhomogeneous vertex
models associated to $U_q\bigl(\widehat{gl}(m\vert m')\bigr)$ with $m,m'>1$.
Relations between $\tilde V(\Lambda_0)$, $\tilde V(\Lambda_1
+\Lambda_4)$ and $\tilde V(2\Lambda_0-\Lambda_3+\Lambda_4)$ and their
decompositions are readily inferred from boson realizations of the
reducible modules. 
All subsequent investigations specialize to $\tilde V(\Lambda_0)$.

Skew Young diagrams have been employed in various related contexts. They label
particular modules of the Yangian $Y\bigl(sl(N)\bigr)$ called the tame
modules \cite{cher,nata1}. The $\widehat{sl}(N)$-character of an irreducible
tame module can be expressed as a product of skew Schur functions
\cite{nata1,nata2,naka1}.
The latter are involved in the
spectral decomposition of the path space of $U_q\bigl(\widehat{sl}(N)\bigr)$-vertex
models with respect to an infinite family of mutually commuting Hamiltonians
defined by the local energy function \cite{naka1,ara,naka2}. Taking advantage of
a one-to-one correspondence between the paths and a suitable class of skew
Young diagrams, the characters specifying the degeneracy of the spectrum of
level-one models can be written in terms of skew Schur functions. Identification
of these characters with the characters of irreducible tame modules yields
a link to the $Y\bigl(sl(N)\bigr)$-module structure of the level-one
integrable modules of $\widehat{sl}(N)$ underlying the path space of the
model. Skew Young diagrams also label the decomposition of integrable modules
at higher levels \cite{naka2,ugl}. For integrable level-one 
$\widehat{sl}(2)$-modules, the action of the Yangian algebra and the resulting
$Y\bigl(sl(2)\bigr)$-structure have first been described in \cite{pas1} and \cite{pas2}
(further references on Yangian module structures are found in \cite{ugl}).
Integrable level-one modules of the quantum affine algebras $U_q\bigl(\widehat{sl}
(N)\bigr)$ and $U_q\bigl(\widehat{gl}(N)\bigr)$ show an analogous structure
emerging under a level-zero action defined in terms of
Cherednik operators \cite{pet1,pet2,sai1,sai2,tak,vv,tug1,tug2}. The analysis in
\cite{tak} exploits correspondences
between skew Young diagrams of border strip type 
and irreducible level zero modules. These modules are given as subspaces of
tensor products of evaluation modules obtained from the vector representation
of $U_q\bigl(sl(N)\bigr)$.

The paths of the $U_q\bigl(\widehat{gl}(2\vert2)\bigr)$-model studied here correspond
to pairs of border stripes, one of them related to the component of the path
contributed by the $W$-modules and the other to the $W^*$-component. 
Each semi-standard super tableau of a finite border strip subject to a particular
condition on the labeling of the lowest leftmost box is mapped onto the $W$-part
of a configuration and vice versa. The full set of semi-standard super tableaux
for a given border strip with $M$ boxes corresponds to a specific subspace
of $W_{x_1}\otimes W_{x_2}\otimes\ldots\otimes W_{x_M}$. Here $W_{x_i}$ denote
evaluation modules associated to $W$ with the ratios of the spectral parameters
$x_i$ determined by the shape of the border strip.

Similarly, $W^*$-path components may be attributed to semi-standard super
tableaux of a finite border strip. Two conditions are imposed on the length and
height of the two border stripes combined for the labeling of paths. Such
pairs of semi-standard super tableaux map onto a subspace of the paths
provided that the tableaux related to the $W$-component fulfill the condition
mentioned above.
This subspace in turn contains a subspace of paths
corresponding to the weight states of the irreducible level-one module
$V(\Lambda_0)$. To each of the pairs of border stripes, a tensor product
$W_{x_1}\otimes W_{x_2}\otimes\ldots\otimes W_{x_M}\otimes W^*_{\bar x_1}\otimes
W^*_{\bar x_2}\otimes\ldots\otimes W^*_{\bar x_M}$ is assigned. Here the ratios of the
spectral parameters $\bar x_i$ of the evaluation modules $W^*_{\bar x_i}$ follow
from the shape of the second border strip. The ratio $x_1/\bar x_1$
is fixed by the requirement that $\Delta^{(2M-1)}(H^1_s+H^3_s)$ annihilates
the tensor product for any $s\neq0$. An examination of pairs of border stripes
with simple shapes suggests that the weight states of $V(\Lambda_0)$ are in
one-to-one correspondence with certain subspaces of the tensor products.

To extend the map between semi-standard super tableaux and half-infinite configurations
to the complete path space of the model, a particular type of infinite border
stripes is introduced. The set of all semi-standard super tableaux
of these border stripes exactly corresponds to the entire space of $W^*$-path
components. Based on this correspondence, a character decomposition of $\tilde V(\Lambda_0)$
similar to the one presented in \cite{naka1} for the $U_q\bigl(\widehat{sl}(N)
\bigr)$-models is given. This expression can be expected to reflect a decomposition
of the level-one module $\tilde V(\Lambda_0)$ into modules of a level-zero
action. As a preparation of such an analysis, level-zero modules of $U_q\bigl(\widehat{sl}
(2\vert2)\bigr)$ matching the entire configuration space are constructed.
Particular  $U_q\bigl(\widehat{sl}(2\vert2)\bigr)$-modules are introduced
as tensor products of evaluation modules $V_{y_i}$ obtained from an infinite-dimensional
$U_q\bigl(gl(2\vert2)\bigr)$-module $V$.
This class of tensor products is supplemented by a one-dimensional level-zero
module of $U_q\bigl(\widehat{sl}(2\vert2)\bigr)$.
The $U_q\bigl(gl(2\vert2)\bigr)$-weights
present in these $U_q\bigl(\widehat{sl}(2\vert2)\bigr)$-modules correspond
to the $W^*$-path components combined with the unique $W$-path component
with vanishing diagonal element of the corner transfer matrix Hamiltonian. 
Suitable tensor products for general diagonal elements involve evaluation
modules $W^*_{\bar x_i}$ in addition. Then tensor products of evaluation modules
$W_{x_i}$, $W^*_{\bar x_i}$ and $V_{y_i}$
with a specific prescription for the ratios of the spectral parameters
are associated to pairs of border stripes.
Certain classes of $U_q\bigl(\widehat{sl}(2\vert2)\bigr)$-modules are
introduced as parts of these tensor products closed under the evaluation
action or as subquotients. Based on an explicit verification for 
border stripes with a simple form, it is conjectured that the $U_q\bigl(gl(2\vert2)
\bigr)$-weights found in the level-zero modules related to a pair of border
stripes exactly correspond to the weights
of the configurations attributed to the same pair.

The new findings presented here can be summarized as follows:
\begin{enumerate}
\item
Reducible nonintegrable $U_q\bigl(\widehat{gl}(2\vert2)\bigr)$-modules with level
one are related to the corner transfer
matrices of an inhomogeneous vertex model based on this algebra. The CTM acts on the space
$\Omega_A$ of half-infinite configurations subject to a particular boundary
condition. Each of the level-one modules
can be decomposed into two irreducible level-one modules, one of them
weakly integrable. Two generalized
energy functions attributed to the $R$-matrices give rise to a quasiparticle
character formula for the reducible modules.
\item
The entire space $\Omega_A$ of half-infinite configurations can be mapped onto
a particular subset of all semi-standard super tableaux of a pair of border
stripes and vice versa. Each pair consists of a finite and an infinite border
strip. 
The map allows to express the character of the reducible
level-one module in terms of skew Schur functions.
Both character formulae for the reducible module split into a factor exclusively
due to the $W$-components and another factor arising from the $W^*$-components.
One or two level-zero module(s) are assigned to each pair of border stripes.
Their $U_q\bigl(gl(2\vert2)\bigr)$-weights coincide with the weights of
all configurations related to the same pair of border stripes.
Partial results are obtained for the irreducible weakly integrable submodule.
\end{enumerate}
All results are obtained by examination of weight structures at grades with
low absolute values. Assuming that the results are valid at all grades, the above statements
are formulated in terms of four conjectures
found in Sections \ref{sec:module} and at the end of Section
\ref{sec:bs2},\ref{sec:bos3a},\ref{sec:bos3b}.

Section \ref{sec:mod} collects definitions of the quantum affine superalgebra and the
vector representation. Then the construction of the vertex model is described. Section
\ref{sec:ctm} proceeds with an analysis of the spectra of the CTM Hamiltonian. 
Section \ref{sec:module} deals with the mapping onto the level-one modules
and Section \ref{sec:bos} presents boson realizations for all level-one modules
involved.
In Section \ref{sec:bs1}, the border stripes relevant to the $U_q\bigl(\widehat{gl}
(2\vert2)\bigr)$-model and their semi-standard super tableaux are specified.
$U_q\bigl(\widehat{sl}(2\vert2)\bigr)$-modules related to them are constructed
by means of the evaluation modules $W_{x_i}$ or, alternatively, of $W^*_{\bar x_i}$.
Both types of modules are combined in Section \ref{sec:bs2}, where their relation to the
irreducible level-one module $V(\Lambda_0)$ is discussed for grades $\geq-3$.
Section \ref{sec:bs3} provides a character formula for $\tilde V(\Lambda_0)$
in terms of skew Schur functions.
Introducing the infinite-dimensional $U_q\bigl(gl(2\vert2)\bigr)$-module $V$, 
the $U_q\bigl(\widehat{sl}(2\vert2)\bigr)$-modules
involving $W_{x_i}$, $W^*_{\bar x_i}$ and $V_{y_i}$ are characterized.

\section{The model}
\label{sec:mod}

The basic construction of the vertex model is covered by the quantum affine
superalgebra $U'_q\bigl(\widehat{sl}(2\vert2)\bigr)$.
Defining  relations for affine Lie superalgebras \cite{kac3} and quantum affine
superalgebras with each choice 
of simple roots can be found in \cite{yamane}.
In terms of its
Chevalley basis, $U'_q\bigl(\widehat{sl}(2\vert2)\bigr)$ is
introduced as the associative ${\mathbb{Z}}_2$-graded algebra over ${\mathbb{C}}[[q-1]]$
generated by $e_i,f_i,q^{\pm h_i},\;i=0,1,2,3$, through the defining relations
\bea
q^{h_i}\,q^{h_j}&=&q^{h_j}\,q^{h_i}\qquad q^{h_i}e_j\,q^{-h_i}=q^{a_{ij}}\,e_j\qquad
q^{h_i}f_j\,q^{-h_i}=q^{-a_{ij}}\,f_j\cr
\noalign{\bigskip}
[e_i,f_j]&=&\delta_{i,j}{q^{h_i}-q^{-h_i}\over q-q^{-1}}\cr
\noalign{\bigskip}
[e_1,e_3]&=&[e_0,e_2]=[f_1,f_3]=[f_0,f_2]=0
\label{eq:def1}
\eea
\bea
\bigl[[e_0,e_1]_q,[e_0,e_3]_q\bigr]=0&&\qquad\qquad\bigl[[f_0,f_1]_{q^{-1}},[f_0,f_3]_{
q^{-1}}\bigr]=0\cr
\noalign{\smallskip}
\bigl[[e_1,e_2]_q,[e_1,e_0]_q\bigr]=0&&\qquad\qquad\bigl[[f_1,f_2]_{q^{-1}},[f_1,f_0]_{
q^{-1}}\bigr]=0\cr
\noalign{\smallskip}
\bigl[[e_2,e_1]_q,[e_2,e_3]_q\bigr]=0&&\qquad\qquad\bigl[[f_2,f_1]_{q^{-1}},[f_2,f_3]_{
q^{-1}}\bigr]=0\cr
\noalign{\smallskip}
\bigl[[e_3,e_2]_q,[e_3,e_0]_q\bigr]=0&&\qquad\qquad\bigl[[
f_3,f_2]_{q^{-1}},[f_3,f_0]_{
q^{-1}}\bigr]=0
\label{eq:def2}
\eea
The matrix elements $a_{ij}$ in (\ref{eq:def1}) depend on the choice of simple roots.
With all simple roots odd, the matrix $a$ is given by
\bea
a=\left(\begin{array}{rrrr}
0&-1&0&1\cr
\noalign{\smallskip}
-1&0&1&0\cr
\noalign{\smallskip}
0&1&0&-1\cr
\noalign{\smallskip}
1&0&-1&0\end{array}\right)
\label{eq:a}
\eea
and the ${\mathbb{Z}}_2$-grading assigns the value $1$ to all $e_i,f_i$ and the value $0$
to $q^{\pm h_i}$. In equation
(\ref{eq:def1}), $[\,,]$ denotes the Lie superbracket $[x,y]=
xy-(-1]^{\vert x\vert\cdot\vert y\vert}yx$. The $q$-deformed superbrackets in the
Serre relations (\ref{eq:def2}) are defined by
\bea
[e_i,e_j]_q&=&e_ie_j+q^{a_{ij}}\,e_je_1\cr
\noalign{\medskip}
[f_i,f_j]_{q^{-1}}&=&f_if_j+q^{-a_{ij}}\,f_jf_i
\eea
$U'_q\bigl(\widehat{sl}(2\vert2)\bigr)$ can be equipped with a graded Hopf algebra
structure introducing the coproduct
\bea
\Delta(e_i)=q^{h_i}\otimes e_i+e_i\otimes1\qquad\Delta(f_i)=f_i\otimes q^{-h_i}+
1\otimes f_i\qquad\Delta(q^{\pm h_i})=q^{\pm h_i}\otimes q^{\pm h_i}
\label{eq:codef}
\eea
the antipode
\bea
S(e_i)=-q^{-h_i}\,e_i\qquad S(f_i)=-f_i\,q^{h_i}\qquad S(q^{\pm h_i})=q^{\mp h_i}
\label{eq:antipode}
\eea
with the property $S(xy)=(-1)^{\vert x\vert\cdot\vert y\vert}\,S(y)S(x)$ and the
counit
\bea
\epsilon(e_i)=\epsilon(f_i)=\epsilon(h_i)=0\qquad\epsilon(1)=1
\eea
The coproduct satisfies $\Delta(xy)=\Delta(x)\Delta(y)$ with the product operation
defined by $(x_1\otimes x_2)(y_1\otimes y_2)=(-1)^{\vert x_2\vert\cdot\vert y_1
\vert}\,x_1y_1\otimes x_2y_2$. Due to the special choice of simple roots, the antipode
fulfills $S^2=id$.
Including the grading operator $d$ with the
properties
\bea
[d,e_i]&=&\delta_{i,0}\,e_i\qquad[d,f_i]=-\delta_{i,0}\,f_i\qquad [d,h_i]=[d,d]=0
\label{eq:d}
\eea
yields the quantum affine superalgebra $U_q\bigl(\widehat{sl}(2\vert2)\bigr)$.
Coproduct and antipode of the grading operator are defined by
$\Delta(d)=d\otimes1+1\otimes d$ and $S(d)=-d$. 

$U'_q\bigl(\widehat{sl}(2\vert2)\bigr)$ is a subalgebra of the quantum affine
superalgebra $U'_q\bigl(\widehat{gl}(2\vert2)\bigr)$ which may be introduced
by means of L-operators or by Drinfeld generators. The latter are preferred here
since the Drinfeld basis \cite{drin1,yamane} underlies the free boson realizations of a
quantum affine (super)algebra. For $U'_q\bigl(\widehat{sl}(m\vert
m')\bigr)$, the Drinfeld basis has been provided in \cite{yamane}.

$U'_q\bigl(\widehat{gl}(2\vert2)\bigr)$ is defined as the associative
${\mathbb{Z}}_2$-graded algebra over ${\mathbb{C}}[[q-1]]$ with generators $E^{i,\pm}_n,\,
i=1,2,3;\,n\in{\mathbb{Z}}$ and $\Psi^{j,\pm}_{\pm n},\,j=1,2,3,4;\,n\in{\mathbb{Z}}_+$
and the central element $q^c$.
In terms of the generating functions
\bea
E^{i,\pm}(z)=\sum_{n\in{\bf Z}}E^{i,\pm}_n\,z^{-n-1}\qquad\qquad
\Psi^{j,\pm}(z)=\sum_{n\geq0}\Psi^{j,\pm}_{\pm n}\,z^{\mp n}
\eea
the defining relations read
\bea
\Psi^{j,+}_0\,\Psi^{j,-}_0&=&\Psi^{j,-}_0\,\Psi^{j,+}_0=1\cr
\noalign{\bigskip}
\Psi^{j_1,\pm}(z)\,\Psi^{j_2,\pm}(w)&=&\Psi^{j_2,\pm}(w)\,\Psi^{j_1,\pm}(z)\cr
\noalign{\bigskip}
\Psi^{j_1,+}(z)\,\Psi^{j_2,-}(w)&=&{\bigl(z-wq^{c+a_{j_1,j_2}}\bigr)\bigl(z-w
q^{-c-a_{j_1,j_2}}\bigr)\over\bigl(z-wq^{c-a_{j_1,j_2}}\bigr)\bigl(z-wq^{-c+a_{
j_1,j_2}}\bigr)}\,\Psi^{j_2,-}(w)\,\Psi^{j_1,+}(z)
\label{eq:def3}
\eea
\bea
\Psi^{j,+}(z)\,E^{l,+}(w)&=&q^{a_{jl}}{z-wq^{-{c\over2}-a_{jl}}\over z-wq^{-{c\over
2}+a_{jl}}}\,E^{l,+}(w)\,\Psi^{j,+}(z)\cr
\noalign{\bigskip}
\Psi^{j,+}(z)\,E^{l,-}(w)&=&q^{-a_{jl}}{z-wq^{{c\over2}+a_{jl}}\over z-wq^{{c\over
2}-a_{jl}}}\,E^{l,-}(w)\,\Psi^{j,+}(z)\cr
\noalign{\bigskip}
\Psi^{j,-}(z)\,E^{l,+}(w)&=&q^{a_{jl}}{z-wq^{{c\over2}-a_{jl}}\over z-wq^{{c\over
2}+a_{jl}}}\,E^{l,+}(w)\,\Psi^{j,-}(z)\cr
\noalign{\bigskip}
\Psi^{j,-}(z)\,E^{l,-}(w)&=&q^{-a_{jl}}{z-wq^{-{c\over2}+a_{jl}}\over z-wq^{-{c\over
2}-a_{jl}}}\,E^{l,-}(w)\,\Psi^{j,-}(z)
\label{eq:def4}
\eea
\bea
\bigl[E^{l_1,+}(z),E^{l_2,-}(w)\bigr]&=&{1\over wz}{\delta_{l_1,l_2}\over q-q^{-1}}
\biggl(\sum_{n\in{\mathbb{Z}}}\Bigl(q^c{w\over z}\Bigr)^n\,\Psi^{l_1,+}\bigl(q^{c\over2}
w\bigr)-\sum_{n\in{\mathbb{Z}}}\Bigl(q^{-c}{w\over z}\Bigr)^n\,\Psi^{l_1,-}\bigl(q^{-
{c\over2}}w\bigr)\biggr)
\label{eq:def5}
\eea
and
\bea
E^{l_1,\pm}(z)\,E^{l_2,\pm}(w)&=&-E^{l_2,\pm}(w)\,E^{l_1,\pm}(z)\qquad\mbox{for}\;\;
a_{l_1,l_2}=0\cr
\noalign{\bigskip}
\bigl(wq^{\pm a_{l_1,l_2}}-z\bigr)\,E^{l_1,\pm}(z)\,E^{l_2,\pm}(w)&=&\bigl(
zq^{\pm a_{l_1,l_2}}-w\bigr)\,E^{l_2,\pm}(w)\,E^{l_1,\pm}(z)\cr
\noalign{\bigskip}
\Bigl[\bigl[E^{2,\pm}(z),E^{1,\pm}(z')\bigr]_{q^{\pm1}},\bigl[E^{2,\pm}(w),
E^{3,\pm}(w')\bigr]_{q^{\pm1}}\Bigr]&+&\Bigl[\bigl[E^{2,\pm}(w),E^{1,\pm}(z')
\bigr]_{q^{\pm1}},\bigl[E^{2,\pm}(z),E^{3,\pm}(w')\bigr]_{q^{\pm1}}\Bigr]=0
\label{eq:def6}
\eea
In (\ref{eq:def3})-(\ref{eq:def6}), $j,j_1,j_2=1,2,3,4$; $l,l_1,l_2=1,2,3$ and
$a_{14}=a_{41}=a_{34}=a_{43}=1$, $a_{24}=a_{42}=a_{44}=0$.
$U'_q\bigl(\widehat{gl}(2\vert2)\bigr)$ supplemented by the grading operator
$d$ will be referred to as $U_q\bigl(\widehat{gl}(2\vert2)\bigr)$. The operator
$d$ satisfies
\bea
w^{-d}\,E^{l,\pm}(z)\,w^{d}=w\,E^{l,\pm}(wz)\qquad w^{-d}\,\Psi^{j,\pm}(z)\,w^d=
\Psi^{j,\pm}(wz)
\label{eq:d2}
\eea
In terms of the basis $\{\tau_1,
\tau_2,\tau_3,\tau_4\}$ with the bilinear form $(\tau_i,\tau_j)=-(-1)^i\delta_{i,j}$, 
the classical roots $\bar{\alpha}_i$ are expressed by
$\bar{\alpha}_i=-(-1)^i(\tau_i+\tau_{i+1})$ for $i=1,2,3$ and $\bar{\alpha}_4=
 \tau_1-\tau_4$. The classical weights $\bar{\Lambda}_i$ read
$\bar{\Lambda}_i=\sum_{j=1}^i\tau_j-{1\over2}\sum_{j=1}^4\tau_j$ with $i=1,2,3,4$.
The affine root
$\delta$ and the affine weight $\Lambda_0$ satisfy $(\Lambda_0,\Lambda_0)=
(\delta,\delta)=(\tau_i,\Lambda_0)=(\tau_i,\delta)=0$ and $(\Lambda_0,\delta)=1$.
Then the set of simple roots is given by $\alpha_0=\delta-\bar{\alpha}_1-\bar{\alpha}_2
-\bar{\alpha}_3$ and $\alpha_i=\bar{\alpha}_i$ for $i=1,2,3,4$. The free abelian group
$P=\oplus_{i=0}^4{\mathbb{Z}}\Lambda_i+{\mathbb{Z}}\delta$ with $\Lambda_i=
\bar{\Lambda}_i+\Lambda_0$ for $i=1,2,3$ and $\Lambda_4=\bar{\Lambda}_4$ is called
the weight lattice. Via $(\,,)$, its dual lattice $P^*=\oplus_{i=0}^4{\mathbb{Z}}h_i+
{\mathbb{Z}}d$ can be identified with $P$ by setting $\alpha_i=h_i$ and $d=\Lambda_0$.

Discarding the subset of generators $\Psi^{4,\pm}(z)$ from (\ref{eq:def3})-(\ref{eq:def6})
yields
a definition of $U'_q\bigl(\widehat{sl}(2\vert2)\bigr)$. Chevalley and Drinfeld basis
of $U'_q\bigl(\widehat{sl}(2\vert2)\bigr)$ are related by 
\bea
q^{\pm h_l}&=&\Psi^{l,\pm}_0\qquad\;\; q^{\pm h_0}=q^{\pm(c-h_1-h_2-h_3)}\cr
\noalign{\bigskip}
e_l&=&E^{l,+}_0\qquad\;\;f_l=E^{l,-}_0\qquad\;\;l=1,2,3\cr
\noalign{\bigskip}
e_0&=&\Bigl[E^{3,-}_0,\bigl[E^{2,-}_0,E^{1,-}_1\bigr]_q\Bigr]_q\,q^{-h_1-h_2-h_3}\cr
\noalign{\bigskip}
f_0&=&q^{h_1+h_2+h_3}\,\Bigl[\bigl[E^{1,+}_{-1},E^{2,+}_0\bigr]_{q^{-1}},E^{3,+}_0
\Bigr]_{q^{-1}}
\label{eq:basrel}
\eea
with the super brackets defined as
\bea
&&\Bigl[E^{l_3,\pm}_{n''},\bigl[E^{l_2,\pm}_{n'},E^{l_1,\pm}_{n}\bigr]_p\Bigr]_p=
-p^{a_{l_1,l_2}+a_{l_1,l_3}+a_{l_2,l_3}}\,\Bigl[\bigl[E^{l_1,\pm}_{n},E^{l_2,\pm}_{
n'}\bigr]_{p^{-1}},E^{l_3,\pm}_{n''}\Bigr]_{p^{-1}}=\cr
\noalign{\bigskip}      
&&\qquad=E^{l_3,\pm}_{n''}\,\Bigl(E^{l_2,\pm}_{n'}E^{l_1,\pm}_{n}+p^{a_{l_1,l_2}}
E^{l_1,\pm}_{n}E^{l_2,\pm}_{n'}\Bigr)-p^{a_{l_1,l_3}+a_{l_2,l_3}}\Bigl(E^{l_2,
\pm}_{n'}E^{l_1,\pm}_{n}+p^{a_{l_1,l_2}}E^{l_1,\pm}_{n}E^{l_2,\pm}_{n'}\Bigr)\,
E^{l_3,\pm}_{n''}
\eea
for $p=q$ or $p=q^{-1}$. $\Psi^{4,\pm}_0$ may be 
reintroduced as a Chevalley generator:
\bea
q^{\pm h_4}\equiv\Psi^{4,\pm}_0
\label{eq:h4def}
\eea
The quantum superalgebras generated by $\bigl\{e_l,f_l,q^{\pm h_l},
\,l=1,2,3\bigr\}$ and $\bigl\{e_l,f_l,q^{\pm h_j},\,l=1,2,3;j=1,2,3,4\bigr\}$
are denoted by $U_q\bigl(sl(2\vert2)\bigr)$ and $U_q\bigl(gl(2\vert2)\bigr)$,
respectively. For several purposes, a reformulation of
the generating functions $\Psi^{j,\pm}(z)$ in terms of $q^{\pm h_j}$
and generators $H^j_n$ with $j=1,2,3,4;\,n\in{\mathbb{Z}}\backslash\{0\}$ proves
useful:
\bea
\Psi^{j,+}(z)&=&q^{h_j}\,\exp\Bigl((q-q^{-1})\sum_{n>0}H^j_n\,z^{-n}\Bigr)\cr
\noalign{\bigskip}
\Psi^{j,-}(z)&=&q^{-h_j}\,\exp\Bigl(-(q-q^{-1})\sum_{n>0}H^j_{-n}\,z^n\Bigr)
\label{eq:Hdef}
\eea
A graded Hopf algebra structure for $U_q\bigl(\widehat{gl}(2\vert2)\bigr)$
with the choice of simple roots adopted here is found
in \cite{gade2}. There the universal R-matrix is expressed in terms of the 
generators $\{c,E^{i,\pm}_n,\,H^j_{\pm n'},\,h_j,\;i=1,2,3,\,j=1,2,3,4,\,n\in {\mathbb{Z}},
n'\in{\mathbb{Z}}_+\}$ and the grading operator $d$.

As is readily verified from (\ref{eq:def3}) and (\ref{eq:def4}), for each $n$ the
sum $H^1_n+H^3_n$ commutes with any generator of $U'_q\bigl(\widehat{sl}
(2\vert2)\bigr)$. All combinations $H^1_n+H^3_n,\;n\in{\mathbb{Z}}\backslash\{0\}$, commute
among themselves and generate the commutative algebra ${\cal H}$.
Therefore, a representation
of $U'_q\bigl(\widehat{sl}(2\vert2)\bigr)$ can be realized as the tensor product
of a one-dimensional representation of ${\cal H}$ and a representation of
$U'_q\bigl(\widehat{sl}(2\vert2)\bigr)/{\cal H}$. In the present study, explicite
reference will be made mainly to $U'_q\bigl(\widehat{sl}(2\vert2)\bigr)$. For
the discussion of weight spaces, the quantum super algebra $U_q\bigl(gl(2\vert2)
\bigr)$ is required as well. In this context it is useful to supplement the
Chevalley basis (\ref{eq:def1}), (\ref{eq:def2})  of $U'_q\bigl(\widehat{sl}(2\vert2)\bigr)$
by $q^{\pm h_4}$. Then the first line of (\ref{eq:def1}) applies to $i=0,1,2,3,4$ and
$j=0,1,2,3$ with $a_{04}=a_{40}=-2$.

If $c$ acts as a scalar on a module of the quantum affine superalgebra, then
this scalar is referred to as the level of the module.
For the affine Lie superalgebra $\widehat{gl}(m\vert m')$, 
the irreducible, nonzero level modules with finite-dimensional weight spaces
are classified in \cite{eswfut}. Since the notion of integrability formulated
for affine algebras \cite{kang}
proves too restrictive for affine Lie superalgebras \cite{kacwak1},
the concept of weak integrability has been introduced \cite{kacwak2}. 
The affine Lie superalgebra $\hat g$ is associated to the Lie superalgebra
$g=g_{\bar 0}+g_{\bar 1}$ with an even symmetric bilinear form $(\cdot
\vert\cdot)$. Even means that $(g_{\bar 0}\vert g_{\bar 1})=0$ and symmetric
indicates that $(\cdot\vert\cdot)$ is symmetric on $g_{\bar 0}$ and skewsymmetric
on $g_{\bar 1}$ \cite{kac}. Furthermore, $g_{\bar 0}$ is taken reductive, i.e. $g_{\bar 0}=
\oplus_{j=0}^Jg_{\bar 0,j}$, where $g_{\bar 0,0}$ is abelian and 
$g_{\bar 0,j}$ with $j>0$ are simple Lie algebras. 
Given a subset $J'\subset
\{1,2,\ldots,J\}$, a $\hat g$-module $M$ is called weakly integrable
(or $J'$-integrable), if
\begin{enumerate}
\item the Cartan subalgebra $\hat h$ of $\hat g$ is diagonalizable on $M$
\item $G$ is locally
finite on $M$
\item $M$ is integrable as a $\hat g_{\bar 0,j}$-module for
all $j\in J'$
\end{enumerate}
Here $\hat g_{\bar 0,j}$ denotes the affine Lie algebra associated
to $g_{\bar 0,j}$.
As shown in \cite{eswfut}, any irreducible weakly integrable module is a highest
weight module. In \cite{kacwak2},
all irreducible weakly integrable modules with level one have been
classified for $\widehat{gl}(m\vert m')$. To each $s\in\mathbb{Z}$ there
corresponds exactly one such module. 
Character formulae of various types are available
for these modules \cite{kacwak2}, \cite{cheng}.

For the quantum affine superalgebra $U_q\bigl(\widehat{sl}(2\vert1)\bigr)$, a free
boson realization at level one has been employed in \cite{shiretal} to classify
infinitely many highest weight vectors. Based on an investigation of the Fock spaces
by means of the technique outlined in \cite{bouw}, character formulae have been
conjectured for the corresponding level-one modules of 
$U_q\bigl(\widehat{sl}(2\vert1)\bigr)$. This type of analysis has also been applied
to $U_q\bigl(\widehat{gl}(2\vert2)\bigr)$ in \cite{zha}. 
The classifications and character formulae in \cite{eswfut}, \cite{kacwak2} and \cite{shiretal}
refer to the standard set of
simple roots for $sl(m\vert m')$ containing only one unique odd simple root.
In contrast, \cite{zha} employs the simple root set characterized by (\ref{eq:a}).
It turns out that all
but one irreducible level-one modules relevant to the model studied below are found 
among the (infinitely many) modules listed in \cite{zha}. 
A convenient tool to transform between the different
simple root systems and the related modules is provided by the application of odd
reflections \cite{penser,kacwak2}. 

Starting with \cite{kac2}, the representation theory of $sl(m\vert m')$ or
$gl(m\vert m')$ has been widely studied. References can be found in \cite{kacwak1}
or \cite{suz}, \cite{cheng2}, for example.

A four-dimensional module $W$ of $U'_q\bigl(\widehat{sl}(2\vert2)\bigr)$ with basis
$\{w_j\}_{0\leq j\leq3}$ is provided by
\bea
\begin{array}{rlrl}
h_k\,w_j&=(-1)^j\,\bigl(\delta_{j,k-1}-\delta_{j,k}\bigr)\,w_j
\qquad\qquad\;\;&h_0\,w_j&=-(h_1+h_2+h_3)\,w_j
\cr
\noalign{\bigskip}
f_{k}\,w_{k-1}&=w_{k}\qquad\qquad\;\;&f_0\,w_3&=-q^{-1}\,w_0\cr
\noalign{\bigskip}
e_{k}\,w_{k}&=(-1)^{k-1}\,w_{k-1}\qquad\qquad\;\;&e_0\,w_0&=q\,w_3
\label{eq:wdef1}
\end{array}
\eea
with $k=1,2,3;\,j=0,1,2,3$.
Furthermore, $W^*$ is introduced as the dual space of $W$ endowed with the
$U'_q\bigl(\widehat{sl}(2\vert2)\bigr)$-$\Bigl(\mbox{or}\;U'_q\bigl(\widehat{gl}(2\vert2)
\bigr)$-$\Bigr)$structure
\bea
\langle aw^*\vert w\rangle=(-1)^{\vert a\vert\cdot\vert w^*\vert}\,\langle
w^*\vert S(a)w\rangle \qquad\forall a\in U'_q\bigl(\widehat{sl}(2\vert2)\bigr)\;
\Bigl(\mbox{or}\;a
\in U'_q\bigl(\widehat{gl}(2\vert2)\bigr)\Bigr)
\eea
Its basis $\{w^*_j\}_{0\leq j\leq 3}$ is chosen such that the action of
$U'_q\bigl(\widehat{sl}(2\vert2)\bigr)$ reads
\bea
\begin{array}{rlrl}
h_k\,w^*_j&=(-1)^j\,\bigl(\delta_{j,k}-\delta_{j,k-1}\bigr)\,w^*_j
\qquad\qquad\;\;&h_0\,w^*_j&=-(h_1+h_2+h_3)\,w^*_j\cr
\noalign{\bigskip}
f_{k}\,w^*_{k}&=(-1)^{k-1}q^{(-1)^{k-1}}\,w^*_{k-1}\qquad\qquad\;\;&
f_0\,w^*_0&=q^{-2}\,w^*_3\cr
\noalign{\bigskip}
e_{k}\,w^*_{k-1}&=-q^{(-1)^k}\,w^*_{k}\qquad\qquad\;\;&e_0\,w^*_3&=q^2\,w^*_0
\end{array}
\label{eq:wdef2}
\eea
where $k=1,2,3;\,j=0,1,2,3$.
The ${\mathbb{Z}}_2$-grading on $W$ and $W^*$ is fixed by $\vert w_j\vert=
\vert w^*_j\vert={1\over2}\bigl(1-(-1)^j\bigr)$.
A $U_q\bigl(gl(2\vert2)\bigr)$-structure on $W$ and $W^*$ is obtained from the
left columns of (\ref{eq:wdef1}) and (\ref{eq:wdef2}) and
\bea
h_4\,w_j=(\delta_{j,0}-\delta_{j,3})\,w_j\qquad\;\;
h_4\,w^*_j=(\delta_{j,3}-\delta_{j,0})\,w^*_j
\label{eq:glstr}
\eea
On the evaluation modules $W_z=W\otimes {\mathbb{C}}[z,z^{-1}]$ and
$W^*_z=W^*\otimes {\mathbb{C}}[z,z^{-1}]$
a $U_q\bigl(\widehat{sl}(2\vert2)\bigr)$-structure can be introduced via
\bea
\begin{array}{rlrl}
e_k\bigl(v_j\otimes z^n\bigr)&=e_k\,v_j\otimes z^{n+\delta_{k,0}}\qquad&f_k
\bigl(v_j\otimes z^n\bigr)&=f_k\,v_j\otimes z^{n-\delta_{k,0}}\cr
\noalign{\bigskip}
h_k\bigl(v_j\otimes z^n\bigr)&=h_k\,v_j\otimes z^n\qquad&d\bigl(v_j\otimes z^n
\bigr)&=n\,v_j\otimes z^n
\end{array}
\label{eq:wevmo1}
\eea 
for $j,k=0,1,2,3$ and $v_j=w_j$ or $v_j=w^*_j$. Throughout the paper, the structure
(\ref{eq:wevmo1}) will be called the level-zero action of $U_q\bigl(\widehat{sl}
(2\vert2)\bigr)$. In contrast, the $q$-analogies of Yangian actions defined on
irreducible level-one modules of $U_q\bigl(\widehat{sl}(N)\bigr)$ are referred
to as level-zero action in references \cite{pet1,sai1,sai2,tak,vv,tug1,tug2}.

For the tensor product of two evaluation modules $V^{(1)}_{z_1}$ and $V^{(2)}_{z_2}$, 
the R-matrix $R(z_1/z_2)\in End\bigl(V^{(1)}_{z_1}\otimes V^{(2)}_{z_2}\bigr)$
intertwines the action of $U_q\bigl(\widehat{sl}(2\vert2)\bigr)$ $\Bigl(\mbox{or}\;
U_q\bigl(\widehat{gl}(2\vert2)\bigr)\Bigr)$:
\bea
R(z_1/z_2)\,\Delta(a)=\Delta'(a)\,R(z_1/z_2)\qquad\qquad\forall a\in
U_q\bigl(\widehat{sl}(2\vert2)\bigr) \Bigl(\mbox{or}\;
U_q\bigl(\widehat{gl}(2\vert2)\bigr)\Bigr)
\label{eq:rdef}
\eea
where $\Delta'=\sigma\circ\Delta$ and $\sigma(a\otimes b)=(-1)^{\vert a\vert\cdot
\vert b\vert}\,b\otimes a$. In the following, the choice of evaluation modules will
be indicated by subscripts. The construction of the vertex model involves the
matrix elements of  $R_{WW}(z)$, $R_{WW^*}(z)$, $R_{W^*W}(z)$ and $R_{W^*W^*}(z)$
introduced by
\bea
R_{WW}(z_1/z_2)\bigl(w_{j_1}\otimes w_{k_1}\bigr)&=&\sum_{j_2,k_2=0}^3
R^{j_2,k_2}_{j_1,k_1}(z_1/z_2)\,w_{j_2}\otimes w_{k_2}\cr
\noalign{\bigskip}
R_{WW^*}(z_1/z_2)\bigl(w_{j_1}\otimes w^*_{k_1}\bigr)&=&\sum_{j_2,k_2=0}^3
R^{j_2,k_2^*}_{j_1,k_1^*}(z_1/z_2)\,w_{j_2}\otimes w^*_{k_2}\cr
\noalign{\bigskip}
R_{W^*W}(z_1/z_2)\bigl(w^*_{j_1}\otimes w_{j_1}\bigr)&=&\sum_{j_2,k_2=0}^3
R^{j_2^*,k_2}_{j_1^*,k_1}(z_1/z_2)\,w^*_{j_2}\otimes w_{k_2}\cr
\noalign{\bigskip}
R_{W^*W^*}(z_1/z_2)\bigl(w^*_{j_1}\otimes w^*_{k_1}\bigr)&=&\sum_{j_2,k_2=0}^3
R^{j_2^*,k^*_2}_{j_1^*,k^*_1}(z_1/z_2)\,w^*_{j_2}\otimes w^*_{k_2}
\eea
Up to a scalar multiple, the solutions of the intertwining condition (\ref{eq:rdef})
on $W_{z_1}\otimes W_{z_2}$ and on $W_{z_1}\otimes W^*_{z_2}$ with $z=z_1/z_2$
are given by
\bea
R_{0,0}^{0,0}(z)&=&R_{2,2}^{2,2}(z)=1\qquad\;\;R_{1,1}^{1,1}(z)=R_{3,3}^{3,3}(z)=
{q^2-z\over 1-q^2z}\cr
\noalign{\bigskip}
R_{j,k}^{j,k}(z)&=&{q(1-z)\over1-q^2z}\qquad j\neq k\cr
\noalign{\bigskip}
R_{j,k}^{k,j}(z)&=&-(-1)^{\vert j\vert\cdot\vert k\vert}\,
{z(q^2-1)\over1-q^2z}\qquad j<k\cr
\noalign{\bigskip}
R_{j,k}^{k,j}(z)&=&-(-1)^{\vert j\vert\cdot\vert k\vert}\,{q^2-1\over1-q^2z}\qquad j>k
\label{eq:r1}
\eea
and
\bea
R_{0,0^*}^{0,0^*}(z)&=&R_{2,2^*}^{2,2^*}(z)=1\qquad\;\;R_{1,1^*}^{1,1^*}(z)=R_{3,
3^*}^{3
,3^*}(z)={1-q^2z\over q^2-z}\cr
\noalign{\bigskip}
R_{j,k^*}^{j,k^*}(z)&=&{q(1-z)\over q^2-z}\qquad j\neq k\cr
\noalign{\bigskip}
R_{j,j^*}^{k,k^*}(z)&=&(-1)^{\vert j\vert}\,{q^2-1\over q^2-z}\qquad j<k\cr
\noalign{\bigskip}
R_{j,j^*}^{k,k^*}(z)&=&(-1)^{\vert j\vert}\,{z(q^2-1)\over q^2-z}\qquad j>k
\label{eq:r2}
\eea
where $j,k=0,1,2,3$ and $\vert j\vert\equiv{1\over2}\bigl(1-(-1)^j\bigr)$. 
All other matrix elements of $R_{WW}(z)$ and $R_{WW^*}(z)$ vanish. The matrix
elements of $R_{W^*W}(z)$ and $R_{W^*W^*}(z)$ are obtained from (\ref{eq:r1})
and (\ref{eq:r2}) via
\bea
R_{j_1^*,j_2}^{k_1^*,k_2}(z)&=&(-1)^{\vert j_1\vert+\vert k_1\vert}\,R_{k_1,k^*_2}^{
j_1,j^*_2}(z)\cr
\noalign{\bigskip}
R_{j^*_1,j^*_2}^{k^*_1,k^*_2}(z)&=&R_{k_1,k_2}^{j_1,j_2}(z)
\label{eq:rrel}
\eea
The R-matrix elements (\ref{eq:r1})-(\ref{eq:rrel}) yield Boltzmann weights
for an integrable vertex model with the modules $W$ and $W^*$ attributed alternately
to both its horizontal and vertical lines (see Fig. 1).
An analogous model has been investigated
for the quantum affine superalgebra $U_q\bigl(\widehat{sl}(2\vert1)\bigr)$ in
\cite{gade3}. For further analysis, four neighboring
vertices of different type are collected in composite vertices of type $A$ or $B$
as illustrated in
Fig. 2. A Boltzmann weight depending on the configuration of basis elements
$\{w_j\}_{0\leq j\leq3}$ or $\{w^*_j\}_{0\leq j\leq3}$ on the outer links
as well as on the spectral parameters assigned to the elementary vertices is
provided by
\bea
R_{\;j_1,j^*_2;\,j_3,j^*_4}^{k_1,k^*_2;\,k_3,k^*_4}(w,z)&=&\sum_{l_1,l_2,l_3,l_4=0}^3
\bar R_{\;l_1,l^*_4}^{k^*_2,k_3}(w^{-1}z)\,\bar R_{l^*_3,j^*_4}^{l^*_4,k^*_4}(z)\,
\bar R_{j_1,l_2}^{k_1,l_1}(z)\,\bar R_{j^*_2,j_3}^{l_2,l^*_3}(wz)\qquad\;\;\mbox{type}\;A
\cr
\noalign{\bigskip}
R_{\;j^*_1,j_2;\,j^*_3,j_4}^{k^*_1,k_2;\,k^*_3,k_4}(w,z)&=&\sum_{l_1,l_2,l_3,l_4=0}^3
\bar R_{\;l_1^*,l_4}^{k_2,k^*_3}(wz)\,\bar R_{j^*_1,l^*_2}^{k^*_1,l^*_1}(z)\,\bar R_{l_3,
j_4}^{l_4,k_4}(z)\,\bar R_{j_2,j^*_3}^{l^*_2,l_3}(w^{-1}z)\qquad\;\;\mbox{type}\;B
\label{eq:compvert}
\eea
In (\ref{eq:compvert}), $\bar R(z)=\sigma\cdot R(z)$.
The assignment of indices and spectral parameters is shown in Fig. 2. As a
consequence of the initial condition
$\bar R_{j_1,j_2}^{k_1,k_2}(1)=\bar R_{j_1^*,j_2^*}^{k_1^*,k_2^*}(1)=
\delta_{j_1,k_1}\delta_{j_2,k_2}$ and the unitarity relation
\bea
\sum_{l_1,l_2=0}^3\bar R_{l_1,l^*_2}^{k^*_1,k_2}(w^{-1})\,
\bar R_{j^*_1,j_2}^{l_1,l^*_2}(w)=\delta_{j_1,k_1}\delta_{j_2,k_2}
\cdot{q^2(1-w)^2\over(q^2-w)(1-q^2w)}
\label{eq:unit}
\eea
the composite R-matrices at $z=1$ act as a scalar multiples on $W\otimes W^*\otimes W
\otimes W^*$ or $W^*\otimes W\otimes W^*\otimes W$:
\bea
R_{\;j_1,j^*_2;\,j_3,j^*_4}^{k_1,k^*_2;\,k_3,k^*_4}(w,1)&=&\delta_{j_1,k_1}
\delta_{j_2,k_2}\delta_{j_3,k_3}\delta_{j_4,k_4}\cdot{q^2(1-w)^2\over(q^2-w)(1-
q^2w)}\cr
\noalign{\bigskip}
R_{\;j_1^*,j_2;\,j^*_3,j_4}^{k^*_1,k_2;\,k^*_3,k_4}(w,1)&=&\delta_{j_1,k_1}
\delta_{j_2,k_2}\delta_{j_3,k_3}\delta_{j_4,k_4}\cdot{q^2(1-w)^2\over(q^2-w)(1-
q^2w)}
\label{eq:init}
\eea
Equation (\ref{eq:init}) states an initial condition for the composite
R-matrices.

\vskip 1cm
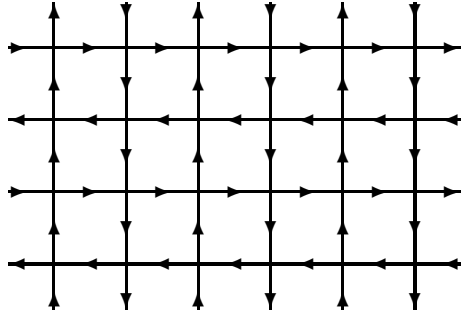
\begin{center}
\setlength{\unitlength}{0.8cm}
\begin{picture}(7.4,5)
\thicklines
\put(0,1.95){\vector(1,0){0.3}}
\put(0,4.35){\vector(1,0){0.3}}
\multiput(0.3,1.95)(1.2,0){6}{\vector(1,0){1.2}}
\multiput(0.3,4.35)(1.2,0){6}{\vector(1,0){1.2}}
\put(7.5,0.75){\vector(-1,0){0.3}}
\put(7.5,3.15){\vector(-1,0){0.3}}
\multiput(7.2,0.75)(-1.2,0){6}{\vector(-1,0){1.2}}
\multiput(7.2,3.15)(-1.2,0){6}{\vector(-1,0){1.2}}
\multiput(0.75,0)(2.4,0){3}{\vector(0,1){0.3}}
\multiput(0.75,0.3)(2.4,0){3}{\vector(0,1){1.2}}
\multiput(0.75,1.5)(2.4,0){3}{\vector(0,1){1.2}}
\multiput(0.75,2.7)(2.4,0){3}{\vector(0,1){1.2}}
\multiput(0.75,3.9)(2.4,0){3}{\vector(0,1){1.2}}
\multiput(1.95,5.1)(2.4,0){3}{\vector(0,-1){0.3}}
\multiput(1.95,4.8)(2.4,0){3}{\vector(0,-1){1.2}}
\multiput(1.95,3.6)(2.4,0){3}{\vector(0,-1){1.2}}
\multiput(1.95,2.4)(2.4,0){3}{\vector(0,-1){1.2}}
\multiput(1.95,1.2)(2.4,0){3}{\vector(0,-1){1.2}}
\end{picture}\par
\vskip 0.4cm
Fig. 1: The lattice model
\end{center}

\begin{center}
\setlength{\unitlength}{1cm}
\begin{picture}(10,6)
\thicklines
\put(0.8,1.4){\line(1,0){0.1}}
\put(2.1,1.4){\vector(-1,0){1.2}}
\put(3.3,1.4){\vector(-1,0){1.2}}
\put(3.7,1.4){\vector(-1,0){0.4}}
\put(0.8,3.2){\vector(1,0){0.4}}
\put(1.2,3.2){\vector(1,0){1.2}}
\put(2.4,3.2){\vector(1,0){1.2}}
\put(3.6,3.2){\line(1,0){0.1}}
\put(1.35,0.9){\vector(0,1){0.4}}
\put(1.35,1.3){\vector(0,1){1.2}}
\put(1.35,2.5){\vector(0,1){1.2}}
\put(1.35,3.7){\line(0,1){0.1}}
\put(3.15,3.8){\vector(0,-1){0.45}}
\put(3.15,3.4){\vector(0,-1){1.2}}
\put(3.15,2.2){\vector(0,-1){1.25}}
\put(3.15,1){\line(0,-1){0.1}}
\put(1.5,1.6){$wz$}
\put(2.22,2.8){$w^{-1}z$}
\put(1.5,2.8){$z$}
\put(2.85,1.6){$z$}
\put(1.7,4.6){Type A}

\put(6,1.4){\vector(1,0){0.4}}
\put(6.4,1.4){\vector(1,0){1.2}}
\put(7.6,1.4){\vector(1,0){1.25}}
\put(8.8,1.4){\line(1,0){0.1}}
\put(8.9,3.2){\vector(-1,0){0.4}}
\put(8.5,3.2){\vector(-1,0){1.2}}
\put(7.3,3.2){\vector(-1,0){1.2}}
\put(6.1,3.2){\line(-1,0){0.1}}
\put(6.55,3.8){\vector(0,-1){0.45}}
\put(6.55,3.4){\vector(0,-1){1.2}}
\put(6.55,2.2){\vector(0,-1){1.25}}
\put(6.55,1){\line(0,-1){0.1}}
\put(8.33,0.9){\vector(0,1){0.4}}
\put(8.33,1.3){\vector(0,1){1.2}}
\put(8.33,2.5){\vector(0,1){1.2}}
\put(8.33,3.7){\line(0,1){0.1}}
\put(6.7,1.6){$w^{-1}z$}
\put(7.8,2.8){$wz$}
\put(6.7,2.8){$z$}
\put(8.05,1.6){$z$}
\put(6.9,4.6){Type B}
\end{picture}\par
Spectral parameters
\end{center}

\begin{center}
\setlength{\unitlength}{1cm}
\begin{picture}(10,6)
\thicklines
\put(0.8,1.4){\line(1,0){0.1}}
\put(2.1,1.4){\vector(-1,0){1.2}}
\put(3.3,1.4){\vector(-1,0){1.2}}
\put(3.7,1.4){\vector(-1,0){0.4}}
\put(0.8,3.2){\vector(1,0){0.4}}
\put(1.2,3.2){\vector(1,0){1.2}}
\put(2.4,3.2){\vector(1,0){1.2}}
\put(3.6,3.2){\line(1,0){0.1}}
\put(1.35,0.9){\vector(0,1){0.4}}
\put(1.35,1.3){\vector(0,1){1.2}}
\put(1.35,2.5){\vector(0,1){1.2}}
\put(1.35,3.7){\line(0,1){0.1}}
\put(3.15,3.8){\vector(0,-1){0.45}}
\put(3.15,3.4){\vector(0,-1){1.2}}
\put(3.15,2.2){\vector(0,-1){1.25}}
\put(3.15,1){\line(0,-1){0.1}}
\put(0.45,3.15){$j_1$}
\put(0.4,1.35){$j_2^*$}
\put(1.25,0.5){$j_3$}
\put(3.05,0.5){$j_4^*$}
\put(1.27,4){$k_1$}
\put(3,4){$k_2^*$}
\put(3.95,3.15){$k_3$}
\put(3.9,1.35){$k_4^*$}
\put(1.7,4.6){Type A}

\put(6,1.4){\vector(1,0){0.4}}
\put(6.4,1.4){\vector(1,0){1.2}}
\put(7.6,1.4){\vector(1,0){1.25}}
\put(8.8,1.4){\line(1,0){0.1}}
\put(8.9,3.2){\vector(-1,0){0.4}}
\put(8.5,3.2){\vector(-1,0){1.2}}
\put(7.3,3.2){\vector(-1,0){1.2}}
\put(6.1,3.2){\line(-1,0){0.1}}
\put(6.55,3.8){\vector(0,-1){0.45}}
\put(6.55,3.4){\vector(0,-1){1.2}}
\put(6.55,2.2){\vector(0,-1){1.25}}
\put(6.55,1){\line(0,-1){0.1}}
\put(8.33,0.9){\vector(0,1){0.4}}
\put(8.33,1.3){\vector(0,1){1.2}}
\put(8.33,2.5){\vector(0,1){1.2}}
\put(8.33,3.7){\line(0,1){0.1}}
\put(5.6,3.15){$j_1^*$}
\put(5.65,1.35){$j_2$}
\put(6.45,0.5){$j_3^*$}
\put(8.28,0.5){$j_4$}
\put(6.45,4){$k_1^*$}
\put(8.25,4){$k_2$}
\put(9.1,3.15){$k_3^*$}
\put(9.13,1.35){$k_4$}
\put(6.9,4.6){Type B}
\end{picture}\par
Assignment of indices
\vskip 0.6cm
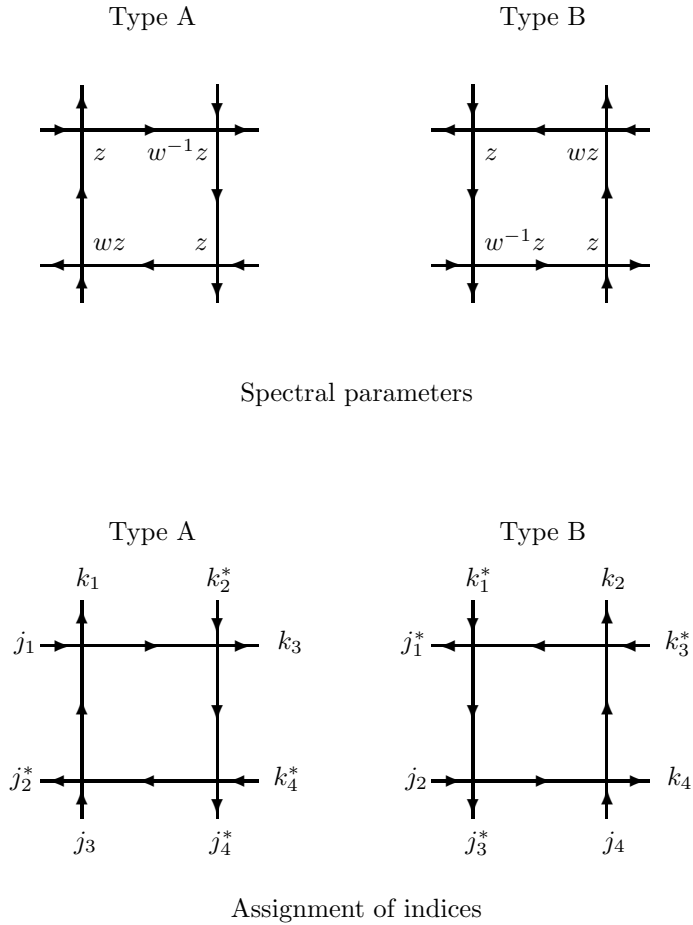
Fig.2: The composite vertices
\end{center}

\section{Corner transfer matrix in the limit $q\to0$}
\label{sec:ctm}

For vertex models based on quantum affine algebras, the space of states has been
interpreted in the framework of the corresponding representation theory
\cite{jm,djkmo}.
The analysis relies on the spectrum of the corner transfer matrix Hamiltonian
\cite{bax}
acting on half-infinite tensor products of evaluation modules subject to a
boundary condition. Quite
generally, the key to its evaluation is the limit of vanishing deformation
parameter $q$.
This section considers the same limit for the inhomogeneous
vertex model specified in the previous section. First a lattice decomposed into
composite vertices of type A with the corresponding R-matrix denoted by $R(w,z)$
will be examined.

The limits of each R-matrix element (\ref{eq:r1})-(\ref{eq:rrel}) are well
defined. However, in the limit $q\to 0$, the composite R-matrix $R(w,z)$ does
not give rise to an invertible map $(W\otimes W^*\bigr)^{\otimes2}\longrightarrow
(W\otimes W^*\bigr)^{\otimes2}$. An invertible map emerges taking the limit
$q\to0$ of $R(q^2w,z)$. While some single matrix elements on $W\otimes W^*$ or
$W^*\otimes W$ diverge, the composite R-matrix $R(q^2w,z)$ always remains
well-defined. This observation applies to the analogous construction for the
quantum affine superalgebra $U_q\bigl(\widehat{sl}(2\vert1)\bigr)$ as well
\cite{gade3}.
To define the corner transfer matrix Hamiltonian,
consider the triangular section $A_N$ of the lattice model introduced in
Sect. \ref{sec:mod} with a vertical (horizontal)
boundary formed by $2N+1$ horizontal (vertical) links on the boundaries of the
upper left quadrant. A Hamiltonian $h^{(N,q)}_{CTM}(w):\,W^*\otimes \bigl(W\otimes W^*
\bigr)^{\otimes(N+1)}\longrightarrow W^*\otimes \bigl(W\otimes W^*\bigr)^{\otimes(N+1)}$
associated to the section $A_N$ is introduced by
\bea
h^{(N,q)}_{CTM}(w)=(2N+3)\tilde h_{2N+3,2N+2}+\sum_{\hat N=1}^N\hat N\,h_{2\hat N+2,2\hat N+1,
2\hat N,2\hat N-1}(w)
\label{eq:hctmdef}
\eea
In (\ref{eq:hctmdef}),
$\tilde h_{2N+3,2N+2}$ denotes the operator $\tilde h$ obtained from $\bar R_{W^*W^*}(z)=
1+(z-1)\tilde h+O\bigl((z-1)^2\bigr)$ acting on the two leftmost components
of $W^*\otimes \bigl(W\otimes W^*\bigr)^{\otimes(N+1)}$ and
$h_{2\hat N+2,2\hat N+1,2\hat N,2\hat N-1}(w)=
1^{\otimes(2N-2\hat N+1)}
\otimes h^{(q)}(w)\otimes 1^{\otimes(2\hat N-2)}$ is defined by the
expansion
\bea
{(1-wz)(z-q^4w)\over(1-q^2wz)(z-q^2w)}\,R(q^2w,z)=1+(z-1)\,h^{(q)}(w)+O\bigl(
(z-1)^2\bigr)
\label{eq:hctmdef2}
\eea
The prefactor on the lhs takes into account the scalar function in the unitarity
relation (\ref{eq:unit}). From (\ref{eq:hctmdef}), the corner transfer matrix
Hamiltonian is obtained by performing the limit $N\to\infty$ with respect to a
suitable boundary condition. To facilitate the limit $q\to 0$,
$R(q^2w,z)$ may be decomposed according to

\bea
R_{\;j_1,j^*_2;\,j_3,j^*_4}^{k_1,k^*_2;\,k_3,k^*_4}(q^{2}w,z)=
{(1-w)(1-q^4w)\over(1-q^2w)^2}\,
\sum_{l_1,l_2=0}^3Y_{\;j_1,l^*_2,l_1}^{k_1,k^*_2,k_3}(w,w^{-1}z)\,
X_{j^*_2,j_3,j^*_4}^{l^*_2,l_1,k^*_4}(wz,w^{-1})
\label{eq:rdec1}
\eea
with
\bea
X_{\;l^*_1,l_2,j^*_4}^{k^*_2,k_3,k^*_4}(w_1,w_2)&\equiv&\sum_{l_1,l_2,l_3=0}^3
\bar R_{\;l_3,l^*_5}^{k^*_2,k_3}(q^{-2}w_2)\,\bar R_{l^*_4,j^*_4}^{l^*_5,k^*_4}(w_1w_2)
\,\bar R_{l^*_1,l_2}^{l_3,l^*_4}(q^{2}w_1)\cr
\noalign{\bigskip}
Y_{\;j_1,j^*_2,j_3}^{k_1,l^*_1,l_2}(w_1,w_2)&\equiv&\sum_{l_1,l_2,l_3=0}^3
\bar R_{l_5,l^*_4}^{l^*_1,l_2}(q^{-2}w_2)\,\bar R_{j_1,l_3}^{k_1,l_5}(w_1w_2)\,\bar R_{j^*_2,
j_3}^{l_3,l^*_4}(q^{2}w_1)
\label{eq:rdec2}
\eea
Equation (\ref{eq:rdec1}) implies a decomposition of $h^{(q)}(w)$ introduced in
(\ref{eq:hctmdef2}):
\bea
h^{(q)}(w)=h_X(w)+h_Y(w)
\label{eq:hdec}
\eea
where
\bea
{(1-q^4w)(1-wz)\over(1-q^2w)(1-q^2wz)}\,X(wz,w^{-1})&=&1+(z-1)\,h_X(w)+O\bigl(
(z-1)^2\bigr)\cr
\noalign{\bigskip}
{(1-w)(q^4w-z)\over(1-q^2w)(q^2w-z)}\,Y(w,w^{-1}z)&=&1+(z-1)\,h_Y(w)+O\bigl(
(z-1)^2\bigr)
\label{eq:xyhdef}
\eea
For convenience, the limits of the matrix elements of $X(w_1,w_2)$
and $Y(w_1,w_2)$  may be denoted by
\bea
x_{j_1^*,j_2,j^*_3}^{k_1^*,k_2,k^*_3}(w_1,w_2)&\equiv& \lim_{q\to0}
X_{\,j_1^*,j_2,j^*_3}^{k_1^*,k_2,k^*_3}(w_1,w_2)\cr
\noalign{\bigskip}
y_{\,j_1,j^*_2,j_3}^{k_1,k^*_2,k_3}(w_1,w_2)&\equiv&\lim_{q\to0}
Y_{\;j_1,j^*_2,j_3}^{k_1,k^*_2,k_3}(w_1,w_2)
\label{eq:limdef1}
\eea
Insertion of equations (\ref{eq:r1})-(\ref{eq:rrel}) in (\ref{eq:rdec2}) yields
\bea
x_{j^*_1,j_2,j^*_3}^{j^*_1,j_2,j^*_3}(wz,w^{-1})=y_{j^*_3,j_2,j^*_1}^{j^*_3,j_2,j^*_1}
(w,w^{-1}z)&=&{z\over 1-w}\qquad\mbox{if}\;\;j_1>j_3\;\;\mbox{or}\;\;j_1=j_3=1,3
\cr
\noalign{\bigskip}
x_{j^*_1,j_2,j^*_3}^{j^*_1,j_2,j^*_3}(wz,w^{-1})=y_{j^*_3,j_2,j^*_1}^{j^*_3,j_2,j^*_1}
(w,w^{-1}z)&=&{1\over 1-w}\qquad\mbox{if}\;\;j_1<j_3\;\;\mbox{or}\;\;j_1=j_3=0,2
\label{eq:lim1}
\eea
and
\bea
x_{j^*_1,j_1,j^*_3}^{j^*_3,j_2,j^*_2}(wz,w^{-1})=y_{j_3,j^*_1,j_1}^{j_2,j^*_2,j_3}
(w,w^{-1}z)
&=&0\qquad\qquad\;\mbox{if}\;\;j_1>j_2\;\;\mbox{or}\;\;0,2=j_2=j_1\neq j_3\cr
\noalign{\bigskip}
x_{j^*_1,j_1,j^*_3}^{j^*_3,j_2,j^*_2}(wz,w^{-1})=y_{j_3,j^*_1,j_1}^{j_2,j^*_2,j_3}
(w,w^{-1}z)
&=&{1-z\over1-w}\qquad\mbox{if}\;\;j_1<j_2\;\;\mbox{or}\;\;1,3=j_2=j_1\neq j_3
\label{eq:lim2}
\eea
and
\bea
x_{j^*_3,j_1,j^*_1}^{j^*_2,j_2,j^*_3}(wz,w^{-1})=y_{j_1,j^*_1,j_3}^{j_3,j^*_2,j_2}
(w,w^{-1}z)&=&
-(-1)^{\vert j_1\vert+\vert j_2\vert}\,{1-z\over1-w}\qquad\mbox{if}\;\;j_1<j_2
\;\;\mbox{or}\;\;0,2=j_2=j_1\neq j_3\cr
\noalign{\bigskip}
x_{j^*_3,j_1,j^*_1}^{j^*_2,j_2,j^*_3}(wz,w^{-1})=y_{j_1,j^*_1,j_3}^{j_3,j^*_2,j_2}
(w,w^{-1}z)&=&0\qquad\qquad\qquad\qquad\qquad\,\mbox{if}\;\;j_1>j_2
\;\;\mbox{or}\;\;1,3=j_2=j_1\neq j_3
\label{eq:lim3}
\eea

Clearly, in the limit $q\to0$, $h^{(q)}(w)$ and $h^{(N,q)}_{CTM}(w)$ become
independent
of $w$. Hence they may be denoted by
\bea
h\equiv\lim_{q\to0}h^{(q)}(w)
\eea
and
\bea
h^{(N)}_{CTM}\equiv\lim_{q\to0}h^{(N,q)}_{CTM}(w)=\lim_{q\to0}\biggl((2N+3)\tilde h_{
2N+3,2N+2}+\sum_{\hat N=1}^N\hat Nh_{2\hat N+2,2\hat N+1,2\hat N,2\hat N-1}(w)\biggr)
\label{eq:h0def}
\eea
With (\ref{eq:lim1})-(\ref{eq:lim3}), it is easily verified that
the matrix elements of $h^{(N)}_{CTM}$ form a
triangular matrix. A particular configuration $\bigl(w_{j_{2N+3}}^*\otimes
\ldots\otimes w_{j_4}\otimes
w^*_{j_3}\otimes w_{j_2}\otimes w^*_{j_1}\bigr)$ may be abbreviated by
$\bigl(j^*_{2N+3},\ldots,j_4,j^*_3,j_2,j^*_1\bigr)$. The matrix element
$h_{(j^*_{2N+3},\ldots,j_4,j^*_3,j_2,j^*_1);(k^*_{2N+3},\ldots,k_4,k^*_3,k_2,
k^*_1)}$ of $h^{(N)}_{CTM}$ is defined by
\bea
h^{(N)}_{CTM}\,(j^*_{2N+3},\ldots,j_4,j^*_3,j_2,j^*_1)=\sum_{k_r=0}^3
h_{(j^*_{2N+3},\ldots,j_4,j^*_3,j_2,j^*_1);(k^*_{2N+3},\ldots,k_4,k^*_3,k_2,
k^*_1)}\,(k^*_{2N+3},\ldots,k_4,k^*_3,k_2,k^*_1)
\label{eq:heldef1}
\eea
Suppose $h_{(j^*_{2N+3},\ldots,j_4,j^*_3,j_2,j^*_1);(k^*_{2N+3}\ldots,k_4,k^*_3,
k_2,k^*_1)}\neq 0$. Then either
\bea
\bigl(j^*_{2N+3},\ldots,j_4,j^*_3,j_2,j^*_1\bigr)=
\bigl(k^*_{2N+3},\ldots,k_4,k^*_3,k_2,k^*_1\bigr)\qquad\mbox{or}\qquad \sum_{r=1}^{2N+3}j_r<
\sum_{r=1}^{2N+3}k_r
\eea
or
$\bigl(k^*_{2N+3},\ldots,k_4,k^*_3,k_2,k^*_1\bigr)$ results from
$\bigl(j^*_{2N+3},\ldots,j_4,j^*_3,j_2,j^*_1\bigr)$ by one of the following
eight replacements:
\bea
\begin{array}{lllll}
(j^*,0,0^*)&\longrightarrow(0^*,0,j^*)\qquad\qquad&(0,0^*,j)&\longrightarrow(j,0^*,0)
\qquad&j\neq0\cr
\noalign{\medskip}
(j^*,2,2^*)&\longrightarrow(2^*,2,j^*)\qquad\qquad&(2,2^*,j)&\longrightarrow(j,2^*,2)
\qquad&j\neq2\cr
\noalign{\medskip}
(j,1^*,1)&\longrightarrow(1,1^*,j)\qquad\qquad&(1^*,1,j^*)&\longrightarrow(j^*,1,1^*)
\qquad&j\neq1\cr
\noalign{\medskip}
(j,3^*,3)&\longrightarrow(3,3^*,j)\qquad\qquad&(3^*,3,j^*)&\longrightarrow(j^*,3,3^*)
\qquad& j\neq3
\end{array}
\eea
Thus the matrix $h_{(j_{2N+3},\ldots,j_4,j^*_3,j_2,j_1^*);(k_{2N+3},\ldots,k_4,k^*_3,
k_2,k^*_1)}$ is triangular. In particular, $(3^*,\ldots,3,3^*,3,3^*)$ is an
eigenvector of $h^{(N)}_{CTM}$. According to (\ref{eq:lim1}),
the diagonal elements $h_{(j^*_{2N+3},\ldots,j_4,j^*_3,j_2,
j^*_1);(j^*_{2N+3},\ldots,j_4,j^*_3,j_2,j^*_1)}$ decouple into a contribution
depending only on the entries $(j^*_{2N+3},j^*_{2N+1},\ldots,j^*_3,j^*_1)$ and into a
second one depending only on the remaining entries $(j_{2N+2},j_{2N},\ldots,j_4,
j_2)$. The diagonal elements are restricted by
\bea
h_{(j^*_{2N+3},\ldots,j_4,j^*_3,j_2,j^*_1);(j^*_{2N+3},\ldots,j_4,j^*_3,j_2,j^*_1)}
\leq h_{(3^*,3,3^*,3,\ldots,3^*,3,3^*);(3^*,3,3^*,3,\ldots,3^*,3,3^*)}=2
\label{eq:cond1}
\eea
for any $(j^*_{2N+3},\ldots,j_4,j^*_3,j_2,j^*_1)$. 
Regarding $\bigl(\ldots\otimes w_3\otimes w^*_3\otimes w_3\otimes w^*_3\bigr)$ as a
ground state configuration, a $CTM$-Hamiltonian at $q=0$ may be defined on the set
$\Omega_A$ of all configurations $\bigl(\ldots\otimes w_{j_4}\otimes w^*_{j_3}
\otimes w_{j_2}\otimes w^*_{j_1}\bigr)$
in the half-infinite tensor product
$\ldots\otimes W\otimes W^*\otimes W\otimes W^*$ with $j_{2r}=3$ and $j^*_{2r-1} 
=3^*$ for almost all $r$.
With respect to this restricted set, the limit $N\to\infty$ can be performed for
(\ref{eq:h0def}) after replacing ${1\over2}h^{(q)}(w)$ by ${1\over2}h^{(q)}(w)-1$
in the definition of $h_{2\hat N+2,2\hat N+1,2\hat N,2\hat N-1}(w)$
and discarding the part containing $\lim_{q\to0}
\tilde h$. This amounts to rescaling the matrix $R(q^2w,z)$ by a factor $z^{-2}$
which does not affect integrability. Renormalizations of that type are familiar
from the vertex models related to quantum affine algebras \cite{fm}.
The diagonal elements  $h^{\mathit{ren}}_{(\ldots,j_4,j^*_3,j_2,j^*_1);
(\ldots,j_4,j^*_3,j_2,j^*_1)}$ of the resulting CTM-Hamiltonian $h_{CTM}$ are
obtained from (\ref{eq:hdec}), (\ref{eq:xyhdef}) and (\ref{eq:lim1}):
\bea
h^{\mathit{ren}}_{(\ldots,j_4,j^*_3,j_2,j^*_1);(\ldots,j_4,j^*_3,j_2,j^*_1)}=-\sum_
{r=1}^{\infty}r\bigl(x_{j_{2r+1},j_{2r-1}}+y_{j_{2r+2},j_{2r}}\bigr)
\label{eq:hdiag1}
\eea
with
\bea
x_{j_1,j_2}=y_{j_2,j_1}=\left\{\begin{array}{lll}
0\qquad&j_1>j_2\qquad\mbox{or}\;\;&j_1=j_2=1,3\cr
\noalign{\medskip}
1\qquad&j_1<j_2\qquad\mbox{or}\;\;&j_1=j_2=0,2
\end{array} \right.
\label{eq:hdiag2}
\eea
The decomposition of the diagonal elements (\ref{eq:hdiag1})
into two parts depending either only on the even or on the odd components
of the tensor product $(\ldots\otimes W\otimes W^*\otimes W\otimes W^*)$ suggests
to classify the configurations 
$(\ldots\otimes w^*_{j_5}\otimes w^*_{j_3}\otimes w^*_{j_1})$ and
$(\ldots\otimes w_{j_6}\otimes w_{j_4}\otimes w_{j_2})$
according to the action of $x_{j_{2r+1},j_{2r-1}}$ and $y_{j_{2r+2},j_{2r}}$
on $(\ldots\otimes w_{j_4}\otimes w^*_{j_3}\otimes w_{j_2}\otimes
w^*_{j_1})\in\Omega_A$,
respectively. Writing $(\ldots,j_5^*,j^*_3,j^*_1)\equiv
(\ldots\otimes w^*_{j_5}\otimes w^*_{j_3}\otimes w^*_{j_1})$ and $(\ldots,j_6,j_4,
j_2)\equiv(\ldots\otimes w_{j_6}\otimes w_{j_4}\otimes w_{j_2})$, the sets
corresponding to the contribution $-n$ may be denoted by
\bea
\bigl\{\tau^*\bigr\}_{-n}&=&\Bigl\{(\ldots,j^*_5,j^*_3,j^*_1)\in(\ldots\otimes W^*\otimes
W^*\otimes W^*)\,\big\vert j^*_r=3^*\;\forall r>R
\in{\mathbb{N}}\;\;\mbox{and}\;\;\sum_{r=1}^{\infty}r\,x_{j_{2r+1},j_{2r-1}}=n\Bigr\}\cr
\noalign{\bigskip}
\bigl\{\tau\bigr\}_{-n}&=&\Bigl\{(\ldots,j_6,j_4,j_2)\in(\ldots\otimes W\otimes W\otimes W)\,
\big\vert j_r=3\;\forall r>R
\in{\mathbb{N}}\;\;\mbox{and}\;\;\sum_{r=1}^{\infty}r\,y_{j_{2r+2},j_{2r}}=n\Bigr\}
\label{eq:taudef}
\eea
with $n=0,1,2,\ldots$. Due to (\ref{eq:hdiag2}),
\bea
\bigl\{\tau\bigr\}_0&=&\Bigl\{(\ldots,3,3,3)\Bigr\}\cr
\noalign{\bigskip}
\bigl\{\tau\bigr\}_{-1}&=&\Bigl\{(\ldots,3,3,0),\,(\ldots,3,3,1),\,(\ldots,3,3,2)
\Bigr\}\cr
\noalign{\bigskip}
\bigl\{\tau\bigr\}_{-2}&=&\Bigl\{(\ldots,3,3,0,1),\,(\ldots,3,3,0,2),\,(\ldots,3,3,0,3),
\,(\ldots,3,3,1,1),\,(\ldots,3,3,1,2),\,(\ldots,3,3,1,3),\cr
\noalign{\bigskip}
&&\;\;(\ldots,3,3,2,3)\Bigr\}\cr
\noalign{\bigskip}
\bigl\{\tau\bigr\}_{-3}&=&\Bigl\{(\ldots,3,3,0,1,1),\,(\ldots,3,3,0,1,2),\,
(\ldots,3,3,0,1,3),\;(\ldots,3,3,0,2,3),\;(\ldots,3,3,0,3,3),\cr
\noalign{\bigskip}
&&\;\;(\ldots,3,3,1,1,1),\,(\ldots,3,3,1,1,2),\,(\ldots,3,3,1,1,3),\,
(\ldots,3,3,1,2,3),\,(\ldots,3,3,1,3,3),\cr
\noalign{\bigskip}
&&\;\;(\ldots,3,3,2,3,3),\cr
\noalign{\bigskip}
&&\;\;(\ldots,3,3,0,0),\,(\ldots,3,3,1,0),\,(\ldots,3,3,2,0),\,
(\ldots,3,3,2,1),\,(\ldots,3,3,2,2)\Bigr\}
\label{eq:taun}
\eea
In contrast to the sets $\bigl\{\tau\bigr\}_{-n}$,
each set $\bigl\{\tau^*\bigr\}_{-n}$ contains infinitely many configurations.
For example,
\bea
&&\bigl\{\tau^*\bigr\}_0=\Bigl\{(\ldots,3^*,3^*,3^*),\,(\ldots,3^*,3^*,2^*)\,
(\ldots,3^*,3^*,0^*),\,(\ldots,3^*,3^*,2^*,0^*),\cr
\noalign{\bigskip}
&&\qquad(\ldots,3^*,3^*,(1^*)^m),\,(\ldots,3^*,3^*,2^*,(1^*)^m),\,(\ldots,3^*,3^*,
(1^*)^m,0^*)\,(\ldots,3^*,3^*,2^*,(1^*)^m,0^*)\Bigr\}_{m\in\mathbb{N}}
\label{eq:taustar0}
\eea
In general, the set
$\{\tau^*\}_{-n}$ is more conveniently specified in terms of finitely many
subsets of configurations. For this purpose,
two types of subsets $\{\Theta_{1^*},j^*_{2R'-1},\ldots,j^*_3,j^*_1\}$
and $\{\Theta_{3^*},j^*_{2R'-1},\ldots,j^*_3,j^*_1\}$ with finite $R'$ are introduced:
\bea
&&\{\Theta_{1^*},j^*_{2R'-1},\ldots,j^*_3,j^*_1\}
\equiv\Bigl\{(\ldots,3^*,3^*,2^*,j^*_{2R'-1},\ldots,j^*_3,j^*_1),
(\ldots,3^*,3^*,(1^*)^m,j^*_{2R'-1},\ldots,j^*_3,j^*_1),\cr
\noalign{\bigskip}
&&\qquad\qquad\qquad\qquad\qquad\qquad\;\,(\ldots,3^*,3^*,
2^*,(1^*)^m,j^*_{2R'-1},\ldots,j^*_3,j^*_1)\Bigr\}_{m\in\mathbb{N}}\cr
\noalign{\bigskip}
&&\{\Theta_{3^*},j^*_{2R'-1},\ldots,j^*_3,j^*_1\}
\equiv\Bigl\{(\ldots,3^*,3^*,0^*,j^*_{2R'-1},\ldots,j^*_3,j^*_1),(\ldots,3^*,3^*,2^*,0^*,
j^*_{2R'-1},\ldots,j^*_3,j^*_1),\cr
\noalign{\bigskip}
&&\qquad\qquad\qquad\qquad
(\ldots,3^*,3^*,(1^*)^m,0^*,j^*_{2R'-1},\ldots,j^*_3,j^*_1),
(\ldots,3^*,3^*,2^*,(1^*)^m,0^*,j^*_{2R'-1},\ldots,j^*_3,j^*_1)\Bigr\}_{m\in\mathbb{N}}
\label{eq:thetadef}
\eea
Then the three sets $\bigl\{\tau^*\bigr\}_{-n}$ with $n=1,2,3$ read 
\bea
\bigl\{\tau^*\bigr\}_{-1}&=&\Bigl\{\{\Theta_{1^*},2^*\},\{\Theta_{1^*},3^*\},
\{\Theta_{3^*},0^*\},\{\Theta_{3^*},1^*\},\{\Theta_{3^*},2^*\},\{\Theta_{3^*},3^*\}
\Bigr\}\cr
\noalign{\bigskip}
\bigl\{\tau^*\bigr\}_{-2}&=&\Bigl\{\{\Theta_{1^*},3^*,0^*\},\{\Theta_{1^*},3^*,1^*\},
\{\Theta_{1^*},3^*,2^*\},\{\Theta_{1^*},3^*,3^*\},\{\Theta_{1^*},2^*,0^*\},
\{\Theta_{1^*},2^*,1^*\},\cr
\noalign{\bigskip}
&&\;\;\{\Theta_{3^*},3^*,0^*\},\{\Theta_{3^*},3^*,1^*\},\{\Theta_{3^*},3^*,2^*\},
\{\Theta_{3^*},3^*,3^*\},\cr
\noalign{\bigskip}
&&\;\;\{\Theta_{3^*},2^*,0^*\},\{\Theta_{3^*},2^*,1^*\},
\{\Theta_{3^*},1^*,0^*\},\{\Theta_{3^*},1^*,1^*\}\Bigr\}\cr
\noalign{\bigskip}
\bigl\{\tau^*\bigr\}_{-3}&=&\Bigl\{\{\Theta_{1^*},2^*,3^*\},\{\Theta_{1^*},2^*,2^*\},
\{\Theta_{1^*},3^*,3^*,k^*\},\{\Theta_{1^*},3^*,2^*,0^*\},\{\Theta_{1^*},3^*,2^*,
1^*\},\cr
\noalign{\bigskip}
&&\;\;\{\Theta_{1^*},3^*,1^*,0^*\},\{\Theta_{1^*},3^*,1^*,1^*\},\{\Theta_{1^*},
2^*,1^*,0^*\},\{\Theta_{1^*},2^*,1^*,1^*\},\cr
\noalign{\bigskip}
&&\;\;\{\Theta_{3^*},2^*,2^*\},\{\Theta_{3^*},2^*,3^*\},\{\Theta_{3^*},1^*,2^*\},
\{\Theta_{3^*},1^*,3^*\},\{\Theta_{3^*},0^*,k^*\},\{\Theta_{3^*},3^*,3^*,k^*\},\cr
\noalign{\bigskip}
&&\;\;
\{\Theta_{3^*},3^*,2^*,0^*\},\{\Theta_{3^*},3^*,2^*,1^*\},\{\Theta_{3^*},3^*,1^*,0^*\},
\{\Theta_{3^*},3^*,1^*,1^*\},\{\Theta_{3^*},2^*,1^*,0^*\},\{\Theta_{3^*},2^*,1^*,1^*\},
\cr
\noalign{\bigskip}
&&\;\;
\,\{\Theta_{3^*},1^*,1^*,0^*\},\{\Theta_{3^*},1^*,1^*,1^*\}\Bigr\}
\label{eq:taustarn}
\eea
Omitting the part $j^*_{2R'-1},\ldots,j^*_3,j^*_1$ in (\ref{eq:taudef}), the set
$\{\tau^*\}_0$ may be rewritten as
$\{\tau^*\}_0=\Bigl\{(\ldots,3^*,3^*,3^*),\{\Theta_1\},\{\Theta_3\}\Bigr\}$.

A $U_q\bigl(gl(2\vert2)\bigr)$-weight $\bar h^{A}=(\bar h_1^{A},\bar h_2^{A},
\bar h_3^{A},\bar h_4^{A})$ for any configuration $(\ldots\otimes w_{j_4}
\otimes w^*_{j_3}\otimes w_{j_2}\otimes w^*_{j_1})\in\Omega_A$ follows
unambiguously from
the $U_q\bigl(gl(2\vert2)\bigr)$-weight defined for a reference configuration.
For the
reference configuration $\bigl(\ldots\otimes w_3\otimes w^*_3
\otimes w_3\otimes w^*_3\bigr)$, two obvious choices $\bar h^A$ 
and $\bar h'^{A}$ are
\bea
\bar h^{A}_j&=&0\qquad\mbox{for}\;\;j=1,2,3,4\cr
\noalign{\bigskip}
\bar h'^{A}_j&=&\delta_{j,4}-\delta_{j,3}
\label{eq:weightch}
\eea
Due to (\ref{eq:wdef1}), (\ref{eq:wdef2}) and (\ref{eq:taun}), (\ref{eq:taustarn}), the
weights of the configurations with vanishing diagonal element of $h_{CTM}$
evaluated with respect to $\bar h^{A}$ are
given by
\bea
(-t,t,t,-t),\;(-t,t+1,t,-t-1),\;(-t-1,t,t+1,-t-2),\;
(-t-1,t+1,t+1,-t-3),\qquad t=0,1,2,\ldots
\label{eq:weizero}
\eea

The corner transfer matrix built from composite vertices of type B is analyzed
in a completely analogous manner. A half-infinite configuration $\bigl(\ldots
\otimes w^*_{j_4}\otimes w_{j_3}\otimes w^*_{j_2}\otimes w_{j_1}\bigr)$ may be
denoted shortly by $\bigl(\ldots,j^*_4,j_3,j^*_2,j_1\bigr)$. Again,
configurations with
entries $j_{2r}^*=3^*$ and $j_{2r-1}=3$ for almost all $r$ are selected. The
set of all these configurations is denoted by $\Omega_B$.
In the limit $q\to0$, the diagonal elements $h^{\mathit{ren}}_{(\ldots,j^*_4,j_3,j^*_2,j_1);\,
(\ldots,j^*_4,j_3,j^*_2,j_1)}$ of the corresponding corner transfer matrix
Hamiltonian are given by
\bea
h^{\mathit{ren}}_{(\ldots,j^*_4,j_3,j^*_2,j_1);\,
(\ldots,j^*_4,j_3,j^*_2,j_1)}=-\sum_{r=1}^{\infty}r(y_{j_{2r+1},j_{2r-1}}+x_{j_{
2r+2},j_{2r}})
\label{eq:hdiagb}
\eea
with $x_{j_1,j_2}$ and $y_{j_1,j_2}$ specified by (\ref{eq:hdiag2}). The
configuration $(\ldots,3^*,3,3^*,3)$ is an eigenvector 
if the limit is performed for a lattice composed from elementary R-matrices
$R_{WW}(z)$, $R_{W^*W^*}(z)$, $R_{WW^*}(q^{-2}w^{-1}z)$ and $R_{W^*W}(q^2wz)$.
Clearly, the classification in terms of the sets $\bigl\{\tau^*\bigr\}_{-n}$
and $\bigl\{\tau \bigr\}_{-n}$ expressed in (\ref{eq:taun})  and (\ref{eq:taustarn})
simultaneously
applies to the configurations $\bigl(\ldots,j^*_5,j_6,j^*_3,j_4,j^*_1,j_2\bigr)$.
Two reference weights $\bar h^{B}$ and $\bar h'^{B}$  for the 
configuration $\bigl(\ldots\otimes w_{3}^*\otimes
w_{3}\otimes w^*_{3}\otimes w_{3}\bigr)$  are fixed by
\bea
\bar h^{B}_j&=&0\qquad\mbox{for}\;\;j=1,2,3,4 \cr
\noalign{\bigskip}
\bar h'^{B}_j&=&\delta_{j,3}-\delta_{j,4}
\label{eq:weightch2}
\eea
With respect to $\bar h'^{B}$, the weights of the configurations in $\Omega_B$ with
vanishing diagonal element of $h_{CTM}$ are listed by
\bea
(-t,t,t+1,-t-1),\;(-t,t+1,t+1,-t-2),\;(-t-1,t,t+2,-t-3),\;(-t-1,t+1,t+2,-t-4)
\label{eq:weizero2}
\eea
with $t=0,1,2,\ldots$.

For any module of $U_q\bigl(gl(2\vert2)\bigr)$ or $U_q\bigl(\widehat{gl}(2\vert2)
\bigr)$, the eigenvalues of $h_4$ may be shifted by an arbitrary fixed value
without modifying the action of the remaining generators. Not being of any
relevance for the following, this freedom will not be mentioned further.

In context with quantum affine algebra models, the notation paths refers to
eigenvectors of the corner transfer matrix Hamiltonian. For convenience,
the configurations encountered in the present model may also be addressed to
as paths. Similarly, $x_{j_1,j_2}$ and $y_{j_1,j_2}$ in (\ref{eq:hdiag1}),
(\ref{eq:hdiag2}) may be called generalized energy functions.

\section{Level-one modules}
\label{sec:module}

The four-tuples (\ref{eq:weizero}) can be viewed as the weights of the vectors contained
in the one-dimensional $U_q\bigl(gl(2\vert2)\bigr)$-module $V_{\bar{\Lambda}_0}$ with highest
weight $(0,0,0,0)$ and the infinite-dimensional, irreducible $U_q\bigl(gl(2\vert2)
\bigr)$-module  $V_{\bar{\Lambda}_2-\bar{\Lambda}_4}$
with highest weight $(0,1,0,-1)$. All vectors contained in
$V_{\bar{\Lambda}_0}$ and $V_{\bar{\Lambda}_2-\bar{\Lambda}_4}$
together with the corresponding eigenvalues of $h_1,h_2,h_3,h_4$
consistent with $\bar h^{A}$ are listed by
\bea
\lambda_0\qquad\qquad&&(0,0,0,0)\cr
\noalign{\bigskip}
\lambda_2\qquad\qquad&&(0,1,0,-1)\cr
\noalign{\bigskip}
f_2\,\lambda_2\qquad\qquad&&(-1,1,1,-1)\cr
\noalign{\bigskip}
(f_3f_2)^m\,\lambda_2\qquad\qquad&&(-m,m+1,m,-m-1)\cr
\noalign{\bigskip}
f_2(f_3f_2)^m\,\lambda_2\qquad\qquad&&(-m-1,m+1,m+1,-m-1)\cr
\noalign{\bigskip}
f_1f_2\,\lambda_2\qquad\qquad&&(-1,0,1,-2)\cr
\noalign{\bigskip}
f_1(f_3f_2)^m\,\lambda_2\qquad\qquad&&(-m,m,m,-m-2)\cr
\noalign{\bigskip}
f_1f_2(f_3f_2)^m\,\lambda_2\qquad\qquad&&(-m-1,m,m+1,-m-2)\qquad\qquad\qquad m=1,2,3,\ldots
\label{eq:lamlist}
\eea
Here $\lambda_0$ and $\lambda_2$ denote the highest weight vectors of
$V_{\bar{\Lambda}_0}$ and $V_{\bar{\Lambda}_2-\bar{\Lambda}_4}$ with the properties
\bea
e_j\,\lambda_0=f_j\,\lambda_0=0\qquad j=1,2,3
\label{eq:muhw}
\eea
and
\bea
e_j\,\lambda_2=0\;\;\mbox{for}\;\;j=1,2,3\qquad\qquad f_j\,\lambda_2=0\;\;\mbox{for}
\;\;j=1,3
\label{eq;lamhw}
\eea
The weights collected in (\ref{eq:lamlist}) may be attributed to the vectors of
one reducible $U_q\bigl(gl(2\vert2)\bigr)$-module $\tilde V_{\bar{\Lambda}_0}$ with highest
weight $(0,0,0,0)$ rather than to two irreducible modules. $\tilde V_{\bar{\Lambda}_0}$ is
characterized by the highest weight vector ${\kappa}$ with the properties
\bea
h_j\,\kappa&=&0\;\;\mbox{for}\;\; j=1,2,3,4\qquad\qquad
e_j\,{\kappa}=0\;\;\mbox{for}\;\;j=1,2,3\qquad\qquad f_3\,{\kappa}\neq0\;\;
\qquad\qquad
f_1\,{\kappa}=f_2\,\kappa=0
\label{eq:tilkap}
\eea
Similarly, the level-one highest weight $U_q\bigl(\widehat{sl}(2\vert2)\bigr)
/{\cal H}$-modules $V(\Lambda_0)$ and $V(\Lambda_2-\Lambda_4)$ may be regarded
as the two irreducible submodules of a level-one module $\tilde V(\Lambda_0)$. 
The latter has a highest weight vector $\hat{\kappa}$ satisfying
\bea
h_j\,\hat{\kappa}=\delta_{j,0}\,\hat{\kappa}\qquad\mbox{for}\qquad j=0,1,2,3,4
\qquad\qquad d\,\hat{\kappa}=0
\label{eq:kaph}
\eea
and
\bea
e_j\,\hat{\kappa}=0\;\;\mbox{for}\;\;j=0,1,2,3\qquad\qquad f_j\,\hat{\kappa}
\neq0\;\;\mbox{for}\;\;j=0,3\qquad\qquad f_1\,\hat{\kappa}=f_2\,\hat{\kappa}=0
\label{eq:kapdef}
\eea
Thus the action of $f_j$ on $\hat{\kappa}$ provides all vectors in
$\tilde V(\Lambda_0)$. 
The module $\tilde V(\Lambda_0)$ contains $\tilde V_{\bar{\Lambda}_0}$ as its
zero-grade subspace. It turns out that the configuration space of the present
model is more conveniently characterized in terms of a reducible module. 
For this reason, the following analysis refers
to $\tilde V(\Lambda_0)$. In Sect. \ref{sec:bos}, the decomposition into
irreducible constituents is taken up again.
The eigenvalues of $h_j$ acting on a vector in a $U_q\bigl(\widehat{sl}(2\vert2)\bigr)/
{\cal H}$-module may be
denoted by $\bar h_j,\;j=0,1,2,3,4$. At level one, they are related by
$\bar h_0=1-\bar h_1-\bar h_2-\bar h_3$. For brevity, in the following
only the $U_q\bigl(gl(2\vert2)\bigr)$-weight $(\bar h_1,\bar h_2,\bar h_3,\bar h_4)$
of a vector in a level-one module will be indicated. The eigenvalue of the
grading operator $d$ on any weight vector of the module is referred to as its grade.
According to (\ref{eq:kaph}), (\ref{eq:kapdef}) and the defining relations
(\ref{eq:def1}), the module includes the two vectors
\bea
f_0\,\hat{\kappa}\qquad\mbox{and}\qquad f_0\,f_1f_3f_2f_3\,\hat{\kappa}
\label{eq:gradeone}
\eea
with grade $-1$ and weights $(1,0,-1,2)$ and $(0,1,0,-1)$, respectively.
Restricted by the Serre relations (\ref{eq:def2}), the action of $f_1,f_2,f_3$ on 
$f_0\,\hat{\kappa}$ generates the vectors
\bea
(f_2f_3)^m\,f_0\,\hat{\kappa}&&\qquad\qquad(-m+1,m,m-1,-m+2)\cr
\noalign{\bigskip}
f_3\,(f_2f_3)^m\,f_0\,\hat{\kappa}&&\qquad\qquad(-m+1,m+1,m-1,-m+1)\cr
\noalign{\bigskip}
f_1\,(f_2f_3)^m\,f_0\,\hat{\kappa}&&\qquad\qquad(-m+1,m-1,m-1,-m+1)\cr
\noalign{\bigskip}
f_1f_3\,(f_2f_3)^m\,f_0\,\hat{\kappa}&&\qquad\qquad(-m+1,m,m-1,-m)\cr
\noalign{\bigskip}
f_2\,(f_3f_2)^m\,f_1f_0\,\hat{\kappa}&&\qquad\qquad(-m,m-1,m,-m+1)\cr
\noalign{\bigskip}
(f_3f_2)^{m+1}\,f_1f_0\,\hat{\kappa}&&\qquad\qquad(-m,m,m,-m)\cr
\noalign{\bigskip}
f_1f_2\,(f_3f_2)^{m+1}\,f_1f_0\,\hat{\kappa}&&\qquad\qquad(-m-1,m-1,m+1,-m-1)\cr
\noalign{\bigskip}
f_1\,(f_3f_2)^{m+2}\,f_1f_0\,\hat{\kappa}&&\qquad\qquad(-m-1,m,m+1,-m-2)\cr
\noalign{\bigskip}
(f_3f_2)^{m+1}\,f_3f_1f_0\,\hat{\kappa}&&\qquad\qquad(-m,m+1,m,-m-1)\cr
\noalign{\bigskip}
f_2\,(f_3f_2)^m\,f_3f_1f_0\,\hat{\kappa}&&\qquad\qquad(-m,m,m,-m)\cr
\noalign{\bigskip}
f_1\,(f_3f_2)^{m+1}\,f_3f_1f_0\,\hat{\kappa}&&\qquad\qquad(-m,m,m,-m-2)\cr
\noalign{\bigskip}
f_1f_2\,(f_3f_2)^m\,f_3f_1f_0\,\hat{\kappa}&&\qquad\qquad(-m,m-1,m,-m-1)
\label{eq:list1}
\eea
with $m=0,1,2,3,\ldots$. The corresponding $U_q\bigl(gl(2\vert2)\bigr)$-weights are
listed in the right column. Taking into account (\ref{eq:taun}), (\ref{eq:weightch})
and (\ref{eq:weizero}), these weights are found in one-to-one correspondence with
the weights of all configurations $(\ldots\otimes w_{j_4}\otimes w^*_{j_3}\otimes w_{j_2}
\otimes w^*_{j_1})$ with $(\ldots,j_6,j_4,j_2) \in
\bigl\{\tau\bigr\}_{-1}$ and $(\ldots,j^*_5,j^*_3,j^*_1)\in\bigl\{\tau^*\bigr\}_0$.
The level-one module $\tilde V(\Lambda_0)$ does not contain $f_0f_3\,\hat{\kappa}$
or $f_0f_2f_3\,\hat{\kappa}=-f_2f_0f_3\,\hat{\kappa}$. Including these would yield
a module with more than two irreducible $U_q\bigl(\widehat{sl}(2\vert2)\bigr)/
{\cal H}$-submodules.
Use of (\ref{eq:def4}), (\ref{eq:basrel}) and (\ref{eq:Hdef}) yields
\bea
\bigl(H^1_{-1}+H^3_{-1}\bigr)\,\hat{\kappa}=q^{-{1\over 2}}\bigl(f_0f_1f_2f_3\,\hat{\kappa}+
f_3f_2f_1f_0\,\hat{\kappa}-f_1f_2f_3f_0\,\hat{\kappa}\bigr)
\label{eq:hcond1}
\eea
Regarding $\tilde V(\Lambda_0)$ as a $U_q\bigl(\widehat{sl}(2\vert2)\bigr)/
{\cal H}$-module, the rhs of (\ref{eq:hcond1}) is set to zero.
On $f_0f_1f_3f_2f_3\,\hat{\kappa}$, the action of $f_1,f_2,f_3$ produces some of the
vectors in (\ref{eq:list1}) and the further vectors
\bea
(f_2f_3)^m\,f_0f_1f_3f_2f_3\,\hat{\kappa}&&\qquad\qquad (-m,m+1,m,-m-1)\cr
\noalign{\bigskip}
f_3\,(f_2f_3)^m\,f_0f_1f_3f_2f_3\,\hat{\kappa}&&\qquad\qquad (-m,m+2,m,-m-2)\cr
\noalign{\bigskip}
f_1\,(f_2f_3)^{m+1}\,f_0f_1f_3f_2f_3\,\hat{\kappa}&&\qquad\qquad (-m-1,m+1,m+1,-m-3)\cr
\noalign{\bigskip}
f_1f_3\,(f_2f_3)^{m+1}\,f_0f_1f_3f_2f_3\,\hat{\kappa}&&\qquad\qquad (-m-1,m+2,m+1,-m-4)\cr
\noalign{\bigskip}
(f_3f_2)^{m+1}\,f_0f_1f_3f_2f_3\,\hat{\kappa}&&\qquad\qquad(-m-1,m+2,m+1,-m-2)\cr
\noalign{\bigskip}
f_2\,(f_3f_2)^{m}\,f_0f_1f_3f_2f_3\,\hat{\kappa}&&\qquad\qquad(-m-1,m+1,m+1,-m-1)\cr
\noalign{\bigskip}
f_1\,(f_3f_2)^{m+2}\,f_0f_1f_3f_2f_3\,\hat{\kappa}&&\qquad\qquad(-m-2,m+2,m+2,-m-4)\cr
\noalign{\bigskip}
f_1f_2\,(f_3f_2)^{m+1}\,f_0f_1f_3f_2f_3\,\hat{\kappa}&&\qquad\qquad(-m-2,m+1,m+2,-m-3)\cr
\noalign{\bigskip}
(f_2f_3)^m\,f_1f_2\,f_0f_1f_3f_2f_3\,\hat{\kappa}&&\qquad\qquad(-m-1,m,m+1,-m-2)\cr
\noalign{\bigskip}
f_3\,(f_2f_3)^m\,f_1f_2\,f_0f_1f_3f_2f_3\,\hat{\kappa}&&\qquad\qquad(-m-1,m+1,m+1,-m-3)\cr
\noalign{\bigskip}
f_1\,(f_2f_3)^{m+1}\,f_1f_2\,f_0f_1f_3f_2f_3\,\hat{\kappa}&&\qquad\qquad(-m-2,m,m+2,-m-4)\cr
\noalign{\bigskip}
f_1f_3\,(f_2f_3)^{m+1}\,f_1f_2\,f_0f_1f_3f_2f_3\,\hat{\kappa}&&\qquad\qquad(-m-2,m+1,m+2,-m-5)
\label{eq:list2}
\eea
where $m=0,1,2,3,\ldots$.
Their weights given in the right column coincide with the
weights of all configurations $(\ldots\otimes w_{j_4}
\otimes w^*_{j_3}\otimes w_{j_2}\otimes w^*_{j_1})$ with $(\ldots,j_6,j_4,j_2)\in
\bigl\{\tau\bigr\}_0$ and $(\ldots,j^*_5,j^*_3,j^*_1)\in\bigl\{\tau^*\bigr\}_{-1}$.
Due to the Serre relations (\ref{eq:def2}), the action of $f_0$ does not give rise
to other vectors than those collected in (\ref{eq:list1}) and (\ref{eq:list2}).
Thus the weights of the configurations $(\ldots\otimes w_{j_4}\otimes w^*_{j_3}
\otimes w_{j_2}\otimes w^*_{j_1})$ with the diagonal element of $h_{CTM}$ given
by $-1$ and the weights of the vectors contained in the level-one module $\tilde V
(\Lambda_0)$ at grade $-1$ are found in one-to-one correspondence.
These weights may be listed separately for the irreducible submodules $V(\Lambda_0)$
and $V(\Lambda_2-\Lambda_4)$. The irreducible submodule
 $V(\Lambda_0)$ contains fourteen weights at grade $-1$. They
are given by $(1,0,-1,2)$, $(1,\pm1,-1,1)$,
$(1,0,-1,0)$, $(0,\pm1,-1,0)$, $(0,0,0,0)$,
$(0,\pm1,0,-1)$, $(-1,\pm1,1,-1)$, $(-1,0,1,0)$ and
$(-1,0,1,2)$, where $(0,0,0,0)$ occurs with multiplicity two. Excluding these fourteen
weights from the set of all weights listed in the right columns of (\ref{eq:list1})
and (\ref{eq:list2}) yields the set of weights found for the $U_q\bigl(\widehat{sl}(2\vert2
)\bigr)/{\cal H}$-module $V(\Lambda_2-\Lambda_4)$ at grade $-1$.

To specify the weights present in $\tilde V(\Lambda_0)$ at lower grades, it is
useful to introduce sets $\sigma(\bar h_1,\bar h_2,\bar h_3,\bar h_4)$ of $U_q\bigl(
gl(2\vert2)\bigr)$-weights:
\bea
\sigma(\bar h_1,\bar h_2,\bar h_3,\bar h_4)&=&\Bigl\{(\bar h_1-m,\bar h_2+m,\bar h_3+m,
\bar h_4-m),\cr
\noalign{\bigskip}
&&\;\;(\bar h_1-m,\bar h_2+m+1,\bar h_3+m,\bar h_4-m-1),\cr
\noalign{\bigskip}
&&\;\;(\bar h_1-m-1,\bar h_2+m,\bar h_3+m+1,\bar h_4-m-2),\cr
\noalign{\bigskip}
&&\;\;(\bar h_1-m-1,\bar h_2+m+1,\bar h_3+m+1,\bar h_4-m-3)\Bigr\}_{m\in{\mathbb{N}}_0}
\label{eq:setdef}
\eea
Then the weights at a given grade can be arranged in a finite number of sets
$\sigma(\bar h_1,\bar h_2,\bar h_3,\bar h_4)$ with $\bar h_i\in{\mathbb{Z}}$. For example,
the weights of the
zero-grade subspace $\tilde V_{\bar{\Lambda}_0}$ yield the set $\sigma(0,0,0,0)$. The weights
listed in (\ref{eq:list1}) and (\ref{eq:list2}) form the sets $\sigma(1,0,-1,2),
\,\sigma(1,-1,-1,1),\,\sigma(0,-1,0,1)$ and $\sigma(0,1,0,-1),\,\sigma(-1,1,1,-1),
\,\sigma(-1,0,1,-2)$, respectively.
All weight vectors at grade $-n$ can be expressed as polynomials
in $f_1,f_2,f_3$ acting on  finitely many vectors $\alpha_t^{(-n)}$, $t=1,2,3,\ldots$,
 obtained by the action of $f_0$ on vectors of grade $-(n-1)$.
 The assignment of $t$ is such that
the closer the positions of the generators $f_0$ are to $\kappa$,
the smaller the value of $t$ is.
Due to the Serre relations (\ref{eq:def2}), in some cases
polynomials acting on different $\alpha^{(-n)}_{t_i}$ with fixed $n$, $1\leq i\leq i'$
and $t_1<t_2<\ldots <t_{i'}$ are linear dependent.
The $\alpha^{(-n)}_{t_i}$ and the polynomials can be chosen such that
the polynomials on $\alpha^{(-n)}_{t_i}$ are linear independent for
$1\leq i\leq i'-1$. Then only the $i'-1$ weights corresponding to the latter
are listed below. 
An example is provided by $f_1\,f_0f_1f_3f_2f_3f_2f_1\,f_0\,\kappa$.
Applying (\ref{eq:def2}) repeatedly and taking
into account the weight structure (\ref{eq:list1}), (\ref{eq:list2})
 of the module at grade
$-1$ yields $f_1\,f_0f_1f_3f_2f_3f_2f_1\,f_0\,\kappa=
f_1f_2f_1\bigl(f_3f_2f_3f_0+[2]f_2f_3f_0f_3+f_2f_3f_0f_3-f_3f_0f_3f_2\bigr)f_1f_0
\,\kappa$. The second and third term can be rewritten using $[2]f_0f_3f_1f_0\,\kappa
=-f_3f_0f_1f_0\,\kappa$ which follows from the first line of (\ref{eq:def2}).
Moreover, as easily seen from equation (\ref{eq:hcond1}), 
the constraint $(H^1_{-1}+H^3_{-1})\kappa=0$ implies $f_0f_3f_2f_1\,f_0\,
\kappa=f_0\,f_1f_2f_3\,f_0\,\kappa$. Thus  $f_1\,f_0f_1f_3f_2f_3f_2f_1\,f_0\,\kappa=-
f_1f_2f_1f_3\,f_0f_1f_2f_3\,f_0\,\kappa$. The weight of this vector is then included
in the weight sets associated to $f_0\,f_1f_2f_3\,f_0\,\kappa$.
The six vectors $\alpha^{(-2)}_t$ 
and the sets collecting the weights attributed to the vectors obtained from them
by application of $f_1,f_2,f_3$ are given by the following table:
\bea
\begin{array}{ll}
\alpha^{(-2)}_t:\qquad&
\mbox{sets}\;\;\sigma(\bar h_1,\bar h_2,\bar h_3,\bar h_4)\;\; \mbox{attributed to}
\;\; \alpha^{(-2)}_i:\cr
\noalign{\bigskip}
\noalign{\medskip}
\alpha^{(-2)}_1=
f_0f_1\,f_0\,\hat{\kappa}\qquad\;\;&\sigma(2,-1,-2,3),\,\sigma(2,-2,-2,2),\,\sigma(1,-1,-1,
3),\,\sigma(1,-2,-1,2)\cr
\noalign{\bigskip}
\alpha^{(-2)}_2=
f_0f_1f_2f_3\,f_0\,\hat{\kappa}\qquad\;\;&\sigma(1,0,-1,2),\,\sigma(1,-1,-1,1),\,
\sigma(0,-1,0,1)\cr
\noalign{\bigskip}
\alpha^{(-2)}_3=
f_0f_1f_3f_2f_3\,f_0\,\hat{\kappa}\qquad\;\;&\sigma(1,1,-1,1),\sigma(1,0,-1,0),
\sigma(0,1,0,1),\sigma(0,0,0,0),\cr
\noalign{\bigskip}
&\sigma(0,0,0,0),\,\sigma(0,-1,0,-1),\,\sigma(-1,0,1,0),\,\sigma(-1,-1,1,-1)\cr
\noalign{\bigskip}
\alpha^{(-2)}_4=
f_0f_1f_3f_2f_3f_2f_1\,f_0\,\hat{\kappa}\qquad\;\;&\sigma(0,0,0,0)\cr
\noalign{\bigskip}
\alpha^{(-2)}_5=
f_0f_1f_3f_2\,f_0f_1f_3f_2f_3\,\hat{\kappa}\qquad\;\;&\sigma(0,1,0,-1),\,
\sigma(-1,1,1,-1),\,\sigma(-1,0,1,-2)\cr
\noalign{\bigskip}
\alpha^{(-2)}_6=
f_0f_1f_3f_2f_3f_2\,f_0f_1f_3f_2f_3\,\hat{\kappa}\qquad\;\;&\sigma(-1,2,1,-2),\,
\sigma(-1,1,1,-3),\,\sigma(-2,2,2,-2),\,\sigma(-2,1,2,-3)
\label{eq:grade2}
\end{array}
\eea
Each weight vector listed in (\ref{eq:grade2}) and in
(\ref{eq:grade3a})-(\ref{eq:grade3b}) below occurs with multiplicity one.
The weights attributed to
$\alpha^{(-2)}_{2t-1}$ and $\alpha^{(-2)}_{2t}$ for $t=1,2,3$ coincide with the
weights of the configurations $(\ldots\otimes w_{j_4}\otimes w^*_{j_3}\otimes w_{
j_2}\otimes w^*_{j_1})$ with $(\ldots,j_6,j_4,j_2)\in\bigl\{\tau\bigr\}_{t-3}$
and $(\ldots,j^*_5,j^*_3,j^*_1)\in\bigl\{\tau^*\bigr\}_{1-t}$.  At grade $-3$,
seventy-three sets $\sigma(\bar h_1,\bar h_2,\bar h_3,\bar h_4)$ and
fifteen vectors $\alpha^{(-3)}_t$ are found:

\bea
\begin{array}{ll}
\alpha^{(-3)}_t:\qquad\;\;&\mbox{sets}\;\;\sigma(\bar h_1,\bar h_2,\bar h_3,\bar h_4)
\;\;\mbox{attributed to}\;\;\alpha^{(-3)}_t\cr
\noalign{\bigskip}
\noalign{\medskip}
\alpha^{(-3)}_1=f_0f_1\,f_0f_1\,f_0\,\hat{\kappa}\qquad&\sigma(3,-2,-3,4),\,
\sigma(3,-3,-3,3),\;\sigma(2,-2,-2,4),\cr
\noalign{\bigskip}
&\sigma(2,-3,-2,3),\cr
\noalign{\bigskip}
\alpha^{(-3)}_2=f_0f_3f_2\,f_0f_1\,f_0\,\hat{\kappa}\qquad&\sigma(2,0,-2,4),\,
\sigma(2,-1,-2,3),\,\sigma(1,-1,-1,3),\cr
\noalign{\bigskip}
&\sigma(1,-2,-1,2),\,\sigma(0,-2,0,2)\cr
\noalign{\bigskip}
\alpha^{(-3)}_3=f_0f_1f_2f_3\,f_0f_1\,f_0\,\hat{\kappa}\qquad\;\;&\sigma(2,-1,-2,3),
\,\sigma(2,-2,-2,2),\,\sigma(1,-1,-1,3),\cr
\noalign{\bigskip}
&\sigma(1,-2,-1,2),\cr
\noalign{\bigskip}
\alpha^{(-3)}_4=f_0f_1f_3f_2f_3f_2\,f_0f_1\,f_0\,\hat{\kappa}\qquad&
\sigma(1,0,-1,2),\,\sigma(1,-1,-1,1),\,\sigma(0,-1,0,1)\cr
\noalign{\bigskip}
\alpha^{(-3)}_5=f_0f_1f_3f_2f_3\,f_0f_1\,f_0\,\hat{\kappa}\qquad&
\sigma(2,0,-2,2),\,\sigma(2,-1,-2,1),\,\sigma(1,0,-1,2),\cr
\noalign{\bigskip}
&\sigma(1,-1,-1,1),\,
\sigma(1,-1,-1,1),\,\sigma(1,-2,-1,0),\cr
\noalign{\bigskip}
&\sigma(0,-1,0,1),\,\sigma(0,-2,0,0)
\cr
\noalign{\bigskip}
\alpha^{(-3)}_6=f_0f_1f_2f_3\,f_0f_1f_2f_3\,f_0\,\hat{\kappa}\qquad&
\sigma(1,0,-1,2),\,\sigma(1,-1,-1,1),\,\sigma(0,0,0,2),\cr
\noalign{\bigskip}
&\sigma(0,-1,0,1)
\cr
\noalign{\bigskip}
\alpha^{(-3)}_7=f_0f_1f_3f_2f_3\,f_0f_1f_2f_3\,f_0\,\hat{\kappa}\qquad&
\sigma(1,1,-1,1),\sigma(1,0,-1,0),\,\sigma(0,1,0,1),\cr
\noalign{\bigskip}
&\sigma(0,0,0,0),
\sigma(0,0,0,0),\,\sigma(0,-1,0,-1),\cr
\noalign{\bigskip}
&\sigma(-1,0,1,0)\,\sigma(-1,-1,1,-1)\cr
\noalign{\bigskip}
\alpha^{(-3)}_8=f_0f_1f_3f_2f_1f_2\,f_0f_1f_3f_2f_3\,f_0\,\hat{\kappa}\qquad&
\sigma(0,0,0,0)
\end{array}
\label{eq:grade3a}
\eea
and
\bea
\begin{array}{ll}
\alpha^{(-3)}_t:\qquad\;\;&\mbox{sets}\;\;\sigma(\bar h_1,\bar h_2,\bar h_3,\bar h_4)\;\;
\mbox{attributed to}\;\;\alpha^{(-3)}_t\cr
\noalign{\bigskip}
\noalign{\medskip}
\alpha^{(-3)}_9=f_0f_1f_3f_2\,f_0f_1f_3f_2f_3\,f_0\,\hat{\kappa}\qquad&
\sigma(1,1,-1,1),\,\sigma(1,0,-1,0),\,\sigma(0,1,0,1),\cr
\noalign{\bigskip}
&\sigma(0,0,0,0),
\sigma(0,0,0,0),\,\sigma(0,-1,0,-1),\cr
\noalign{\bigskip}
&\sigma(-1,0,1,0),\,\sigma(-1,-1,1,-1)\cr
\noalign{\bigskip}
\alpha^{(-3)}_{10}=f_0f_1f_3f_2f_3f_2\,f_0f_1f_3f_2f_3\,f_0\,\hat{\kappa}\qquad
&\sigma(0,2,0,0),\,\sigma(0,1,0,-1),\,\sigma(-1,2,1,0),\cr
\noalign{\bigskip}
&\sigma(-1,1,1,-1),\,\sigma(-1,1,1,-1)\,\sigma(1,0,1,-2),\cr
\noalign{\bigskip}
&\sigma(-2,1,2,-1),\,\sigma(-2,0,2,-2)\cr
\noalign{\bigskip}
\alpha^{(-3)}_{11}=f_0f_1f_3f_2f_3f_2f_1\,f_0f_1f_3f_2f_3\,f_0\,\hat{\kappa}\qquad
&\sigma(0,1,0,-1),\,\sigma(0,0,0,-2),\,\sigma(-1,1,1,-1),\cr
\noalign{\bigskip}
&\sigma(-1,0,1,-2)\cr
\noalign{\bigskip}
\alpha^{(-3)}_{12}=f_0f_1f_3f_2f_3\,f_0f_1f_3f_2\,f_0f_1f_3f_2f_3\,\hat{\kappa}
\qquad&\sigma(0,2,0,-2),\,\sigma(-1,2,1,-2),\,\sigma(-1,1,1,-3),\cr
\noalign{\bigskip}
&\sigma(-2,1,2,-3),\,\sigma(-2,0,2,-4)\cr
\noalign{\bigskip}
\alpha^{(-3)}_{13}=f_0f_1f_3f_2f_3f_2\,f_0f_1f_3f_2\,f_0f_1f_3f_2f_3\,\hat{\kappa}
\qquad&\sigma(-1,2,1,-2),\sigma(-1,1,1,-3),\,\sigma(-2,2,2,-2),\cr
\noalign{\bigskip}
&\sigma(-2,1,2,-3)
\cr
\noalign{\bigskip}
\alpha^{(-3)}_{14}=f_0f_1f_3f_2\,f_0f_1f_3f_2\,f_0f_1f_3f_2f_3\,\hat{\kappa}\qquad
&\sigma(0,1,0,-1),\,\sigma(-1,1,1,-1),\,\sigma(-1,0,1,-2)\cr
\noalign{\bigskip}
\alpha^{(-3)}_{15}=f_0f_1f_3f_2f_3f_2\,f_0f_1f_3f_2f_3f_2\,f_0f_1f_3f_2f_3\,\hat{\kappa}
\qquad&\sigma(-2,3,2,-3),\,\sigma(-2,2,2,-4),\,\sigma(-3,3,3,-3),\cr
\noalign{\bigskip}
&\sigma(-3,2,3,-4)
\end{array}
\label{eq:grade3b}
\eea
The weights attributed to the vectors $\alpha^{(-3)}_{4t-3},\,\alpha^{(-3
)}_{4t-2},\,\alpha^{(-3)}_{4t-1}$ and $\alpha^{(-3)}_{4t}$ for $t=1,2$
coincide with the weights of the configurations $(\ldots\otimes w_{j_4}\otimes
w^*_{j_3}\otimes w_{j_2}\otimes w^*_{j_1})$ with $(\ldots,j_6,j_4,j_2)\in\bigl\{
\tau\bigr\}_{t-4}$ and $(\ldots,j^*_5,j^*_3,j^*_1)\in\bigl\{\tau^*\bigr\}_{1-t}$.
The weights related to $\alpha^{(-3)}_9,\,\alpha^{(-3)}_{10},\,\alpha^{(-3)}_{11}$
are the weights of the configurations with  $(\ldots,j_6,j_4,j_2)\in\bigl\{\tau
\bigr\}_{-1}$ and $(\ldots,j^*_5,j^*_3,j^*_1)\in\bigl\{\tau^*\bigr\}_{-2}$. Finally,
the weights of $\alpha^{(-3)}_{t}$ with $12\leq t\leq15$ correspond to the
weights of the configurations with $(\ldots,j_6,j_4,j_2)\in\bigl\{\tau\bigr\}_0$
and $(\ldots,j^*_5,j^*_3,j^*_1)\in\bigl\{\tau^*\bigr\}_{-3}$.
The tables (\ref{eq:grade2}),
(\ref{eq:grade3a}) and (\ref{eq:grade3b})
result from imposing both the Serre relations (\ref{eq:def2}) and the conditions
$\bigl(H^1_{-n}+H^3_{-n}\bigr)\,\hat{\kappa}=0$ with $n=1,2,$ for (\ref{eq:grade2})
and $n=1,2,3$ for (\ref{eq:grade3a}), (\ref{eq:grade3b}).

In summary, for $n=0,1,2,3$, the weights of the vectors found in $\tilde V
(\Lambda_0)$ at grade $-n$ and the weights of
all configurations $(\ldots\otimes w_{j_4}\otimes w^*_{j_3}\otimes w_{j_2}\otimes
w^*_{j_1})$ with diagonal element of $h_{CTM}$ given by $-n$
are in one-to-one correspondence. Provided that this remains true at any grade,
the following statement holds.
\vspace{0.8cm}

{\it Conjecture I} : The character of $\tilde V(\Lambda_0)$ can be written
\bea
&&ch_{\tilde V(\Lambda_0)}(\rho,p_0,p_1,p_2)\equiv tr_{\tilde V(\Lambda_0)}\rho^d
\,p_0^{{1\over2}(h_1+h_2+h_3+h_4)}\,p_1^{{1\over2}(h_1-h_2-h_3-h_4)}\,p_2^{-
{1\over2}(h_1+h_2-h_3-h_4)}=
\cr
\noalign{\bigskip}
&&=\sum_{\{\ldots,j_6,j_4,j_2\}}\rho^{-\sum_{r=1}^{\infty}ry_{j_{2r+2},j_{2r}}}
\,\prod_{j=0,1,2}p_j^{\sum_{r=1}^{\infty}\delta_{j_{2r},j}}\cdot
\sum_{\{\ldots,j^*_5,j^*_3,j^*_1\}}\rho^{-\sum_{\bar r=1}^{\infty}\bar rx_{j_{2
\bar r+1},j_{2\bar r-1}}}
\,\prod_{j'=0,1,2}p_{j'}^{-\sum_{\bar r=1}^{\infty}\delta_{j_{2\bar r-1},j'}}
\label{eq:pathchar}
\eea
with $\vert p_1\vert>1$.

As described in section \ref{sec:ctm}, the configurations
in (\ref{eq:pathchar}) are restricted by
the requirement that $j_{2r}=3\,\forall r>R$ and $j_{2r+1}^*=3^*\,\forall r>R\in
{\mathbb{N}}$.
The weights of the configurations are consistent with the
reference weight $\bar h^{A}$ of the
configuration $(\ldots,3,3^*,3,3^*)$ specified in (\ref{eq:weightch}).

Since each contribution to the sum (\ref{eq:pathchar}) is positive, the conjecture
states a character formula of quasiparticle type. Obviously, the modules
$V(\Lambda_2-\Lambda_4)$ and $\tilde V(\Lambda_0)$ do not satisfy the third requirement
for weak integrability given in Sect. \ref{sec:mod}.
Consequently,
the characters given in \cite{kacwak2}, \cite{cheng} do not include the
character of the $\widehat{gl}(2\vert2)\bigr)$-module corresponding to
$V(\Lambda_2)$.
In contrast, the irreducible module $V(\Lambda_0)$ is weakly integrable. The highest
weight $\Lambda_0$ remains the same when changing the simple root system
by odd reflections \cite{kacwak2}. Thus the
$U_q\bigl(\widehat{gl}(2\vert2)\bigr)$-module $V(\Lambda_0)$ is associated to
the weakly integrable module $F_0$ in \cite{kacwak2}.
Unfortunately,
in the configuration space the decomposition into irreducible components appears
to be rather indirect. 
An expression for the character of $V(\Lambda_0)$
will be discussed in Sect. \ref{sec:bs2}.

The R-matrix solution for the vector representation related to the standard system
of simple roots has been given in \cite{perk}. Inspection of the appropriate
$q\to0$-limits indicates that also in this case
 weight structures of nonintegrable modules are relevant to the
description of the analogous inhomogeneous vertex model. Moreover, nonintegrable
modules should be related to the analogous $U_q\bigl(\widehat{gl}(m\vert m')\bigr)$-models
with $m,m'>0$.

Alternatively, the weights may be evaluated with respect to $\bar h'^{A}$. Then the
weights of all configurations with vanishing diagonal element of $h_{CTM}$
coincide with the weights
of an infinite-dimensional, reducible $U_q\bigl(gl(2\vert2)\bigr)$-module 
$\tilde V_{-\bar{\Lambda}_3+\bar{\Lambda}_4}$. Its highest weight vector
$\nu$ is characterized by
\bea
h_j\,\nu&=&\delta_{j,4}-\delta_{j,3}\cr
\noalign{\bigskip}
e_j\,\nu&=&0\;\;\mbox{for}\;\;j=1,2,3\qquad f_3f_2f_3\,\nu\neq0\qquad f_1\,\nu=f_2\,
\nu=0
\label{eq:mu2hw}
\eea
$\tilde V_{-\bar{\Lambda}_3+\bar{\Lambda}_4}$ may be decomposed into a
four-dimensional irreducible 
$U_q\bigl(gl(2\vert2)\bigr)$-module $V_{-\bar{\Lambda}_3+\bar{\Lambda}_4}$
 with weights $(0,0,-1,1)$, $(0,1,-1,0)$,
$(-1,1,0,0)$ and $(-1,0,0,-1)$ and an infinite-dimensional irreducible module
$V_{-\bar{\Lambda}_1+2\bar{\Lambda}_2-\bar{\Lambda}_4}$
with weights $(-m,m+1,m-1,-m)$, $(-m-1,m+1,m,-m)$, $(-m,m,m-1,-m-1)$ and $(-m-1,m,
m,-m-1)$, $m=1,2,3,\ldots$. $V_{-\bar{\Lambda}_3+\bar{\Lambda}_4}$ and
$V_{-\bar{\Lambda}_1+2\bar{\Lambda}_2-\bar{\Lambda}_4}$
contain the highest weight
vectors $\lambda_{-3}$ and $\lambda_{-1,2^2}$ with $h_j\,\lambda_{-3}=(\delta_{j,4}
-\delta_{j,3})\,\lambda_{-3}$ and  $h_j\,\lambda_{-1,2^2}=(2\delta_{j,2}-\delta_{j,1}
-\delta_{j,4})\,\lambda_{-1,2^2}$.
The level-one $U_q\bigl(\widehat{sl}(2\vert2)\bigr)/{\cal H}$-module
$\tilde V(2\Lambda_0-\Lambda_{3}+\Lambda_4)$ with its grade-zero subspace given by
$\tilde V_{-\bar{\Lambda}_3+\bar{\Lambda}_4}$ can be
decomposed into the level-one irreducible modules $V(2\Lambda_0-\Lambda_{3}+\Lambda_4)$
and $V(-\Lambda_1+2\Lambda_2-\Lambda_4)$. The latter contain $V_{-\bar{\Lambda}_3+
\bar{\Lambda}_4}$ and
$V_{-\bar{\Lambda}_1+2\bar{\Lambda}_2-\bar{\Lambda}_4}$
as their grade-zero subspaces,
respectively.
An analysis based on the defining relations of $U_q\bigl(\widehat{sl}(2\vert2)
\bigr)$ reveals that shifting the weights of all vectors in $\tilde V(\Lambda_0)$
at grade $-n$ by $(0,0,-1,1)$ yields the complete set of weights found in
$\tilde V(2\Lambda_0-\Lambda_{3}+\Lambda_4)$ at grade $-n$, $n=0,1,2,3$.
Thus, for the grades $-n\geq
-3$, switching to the reference weight $\bar h'^{A}$ amounts to replacing $\tilde V(
\Lambda_0)$ by $\tilde V(2\Lambda_0-\Lambda_{3}+\Lambda_4)$.
Above the eigenvalues of the grading operator are set to zero for
the highest weight vectors of both modules $\tilde V(\Lambda_0)$
and $\tilde V(\Lambda_2-\Lambda_4)$. A different choice results in a constant
shift of all grades and it easily taken into account in each statement.
The relation between the
weight structures of $\tilde V(\Lambda_0)$ and $\tilde V(2\Lambda_0
-\Lambda_{3}+\Lambda_4)$ becomes apparent in the boson
realization of both modules constructed in Sect. \ref{sec:bos}.

From the results described above, an interpretation of
the weights associated the half-infinite configurations $\bigl(\ldots\otimes w^*_{j_3}
\otimes w_{j_4}\otimes w^*_{j_1}\otimes w_{j_2}\bigr)$ in $\Omega_B$ is readily
obtained.
As pointed out in the preceeding section, the classification of the configurations
in $\Omega_A$ and $\Omega_B$ according to the diagonal elements of the corner
transfer matrix Hamiltonians in the limit $q\to0$ is achieved by the same sets $\{\tau
\}_{-n}$ and $\{\tau^*\}_{-n}$. Choosing the reference weights $\bar h^{A}$ and
$\bar h^{B}$ specified in (\ref{eq:weizero})
and (\ref{eq:weizero2}), the weights of any two configurations 
$(\ldots,j^*_3,j_4,j^*_1,j_2)\in\Omega_B$ and
$(\ldots,j_4,j^*_3,j_2,j^*_1)\in\Omega_A$ coincide. Hence, with $\bar h^{B}$ the
description by the weights of $\tilde V(\Lambda_0)$ holds for the configurations
in $\Omega_B$, too.
The irreducible level-one
$U_q\bigl(\widehat{sl}(2\vert2)\bigr)/{\cal H}$-modules
$V(\Lambda_1+\Lambda_4)$ and $V(\Lambda_3-\Lambda_4)$ relevant for the
reference weight $\bar h'^{B}$ have highest
weight vectors $\hat{\lambda}_1$ and $\hat{\lambda}_3$ with the properties
\bea
h_j\,\hat{\lambda}_1&=&(\delta_{j,1}+\delta_{j,4})\,\hat{\lambda}_1,\;\;\;
j=0,1,2,3,4\cr
\noalign{\bigskip}
e_k\,\hat{\lambda}_1&=&0\;\;\mbox{for}\;\;j=0,1,2,3\qquad\;\;f_k\,\hat{\lambda}_1
=0\;\;\mbox{for}\;\;k=0,2,3
\label{eq:lambda1def}
\eea
and
\bea 
h_j\,\hat{\lambda}_3&=&(\delta_{j,3}-\delta_{j,4})\,\hat{\lambda}_3,\;\;\;\;j=0,1,2,3,4\cr
\noalign{\bigskip}
e_k\,\hat{\lambda}_3&=&0\;\;
\mbox{for}\;\;k=0,1,2,3
\qquad\;\;f_k\,\hat{\lambda}_3=0\;\;\mbox{for}\;\;k=0,1,2
\label{eq:lambda3def}
\eea
Applying a uniform shift by $(0,0,-1,1)$, 
the weights of all vectors present in the grade-zero subspace of $V(\Lambda_3-\Lambda_4)$
coincide with the weights collected in (\ref{eq:weizero}). 
An automorphism $\bar{\varsigma}$ of $U'_q\bigl(\widehat{gl}(2\vert2)\bigr)$ is given by
\bea
\begin{array}{llllllll}
\bar{\varsigma}(e_0)&=e_3\qquad&\bar{\varsigma}(e_1)&=e_2\qquad&\bar{\varsigma}(e_2)
&=e_1\qquad&\bar{\varsigma}(e_3)&=e_0\cr
\noalign{\bigskip}
\bar{\varsigma}(f_0)&=f_3\qquad&\bar{\varsigma}(f_1)&=f_2\qquad&\bar{\varsigma}(f_2)
&=f_1\qquad&\bar{\varsigma}(f_3)&=f_0\cr
\noalign{\bigskip}
\bar{\varsigma}(h_0)&=h_3\qquad&\bar{\varsigma}(h_1)&=h_2\qquad&\bar{\varsigma}(h_2)
&=h_1\qquad&\bar{\varsigma}(h_3)&=h_0
\end{array}
\label{eq:aut1}
\eea
$\bar{\varsigma}$ gives rise to the isomorphism $\varsigma$ of the vector spaces
$V(\Lambda_0)$ $\bigl(V(\Lambda_2-\Lambda_4)\bigr)$ and
$V(\Lambda_3-\Lambda_4)$ $\bigl(V(\Lambda_1+\Lambda_4)\bigr)$ via
\bea
\begin{array}{lllll}
\varsigma(\hat{\lambda}_0)&=\hat{\lambda}_3\qquad\;\;\;&\varsigma(au)&=\bar{\varsigma}\,
(a)\varsigma(u)\;\;&\mbox{with}\;\;u\in V(\Lambda_0)
\cr
\noalign{\bigskip}
\varsigma(\hat{\lambda}_2)&=\hat{\lambda}_1\qquad\;\;\;&\varsigma(au)&=\bar{\varsigma}\,
(a)\varsigma(u)\;\;&\mbox{with}\;\;
u\in V(\Lambda_2-\Lambda_4)
\end{array}
\label{eq:isdef}
\eea
with $a\in U'_q\bigl(\widehat{gl}(2\vert2)\bigr)$. Here $\hat{\lambda}_0$ and
$\hat{\lambda}_2$ denote the highest weight vectors of $V(\Lambda_0)$ and
$V(\Lambda_2-\Lambda_4)$.
Reducible modules may be composed from $V(\Lambda_1+\Lambda_4)$ and
$V(\Lambda_3-\Lambda_4)$ in various
ways. Obviously, the isomorphism $\varsigma$ may be applied to the module $\tilde V
(\Lambda_0)$. Another reducible module $\tilde V(\Lambda_1+\Lambda_4)$ is characterized by
the vector $\hat{\mu}$ with the properties
\bea
h_j\,\hat{\mu}&=&(\delta_{j,1}+\delta_{j,4})\,\hat{\mu},\;\;\;j=0,1,2,3,4\cr
\noalign{\bigskip}
e_j\,\hat{\mu}&=&0\;\;\mbox{for}\;\;j=1,2,3\qquad\;\;f_0\,\hat{\mu}=f_2\,\hat{\mu}=0
\qquad\;\;
E^{3,-}_1f_2f_1\,\hat{\mu}=E^{2,-}_1f_3f_1\,\hat{\mu}=qe_0\,\hat{\mu}\neq0
\label{eq:mudef}
\eea
The highest weight properties are imposed on the vector $e_0\,\hat{\mu}$:
\bea
e_je_0\,\hat{\mu}=0\qquad\;\;j=0,1,2,3
\label{eq:mudef2}
\eea
With this choice, all three reducible modules $\tilde V(\Lambda_0)$, $\tilde V(\Lambda_
1+\Lambda_4)$ and $\tilde V(2\Lambda_0-\Lambda_{3}+\Lambda_4)$
have simple realizations in terms of the same
bosonization scheme given in Sect. \ref{sec:bos}.
A second automorphism $\bar{\varrho}$ of $U_q'\bigl(\widehat{gl}(2\vert2)\bigr)$ is defined
by
\bea
\begin{array}{llllllll}
\bar{\varrho}(e_0)&=e_2\qquad&\bar{\varrho}(e_1)&=e_3\qquad&\bar{\varrho}(e_2)&=e_0
\qquad&\bar{\varrho}(e_3)&=e_1\cr
\noalign{\bigskip}
\bar{\varrho}(f_0)&=f_2\qquad&\bar{\varrho}(f_1)&=f_3\qquad&\bar{\varrho}(f_2)&=f_0
\qquad&\bar{\varrho}(f_3)&=f_1\cr
\noalign{\bigskip}
\bar{\varrho}(h_0)&=h_2\qquad&\bar{\varrho}(h_1)&=h_3\qquad&\bar{\varrho}(h_2)&=h_0
\qquad&\bar{\varrho}(h_3)&=h_1
\end{array}
\eea
$\bar{\varrho}$ yields an automorphism $\varrho$ of $\tilde V(\Lambda_0)$ via
\bea
\varrho(\hat{\mu})=f_3\,\hat{\mu}\qquad \varrho(au)=\bar{\varrho}(a)\,\varrho(u)\;\;
\mbox{with}\;\;a\in U'_q\bigl(\widehat{sl}(2\vert2)\bigr),\,u\in\tilde V(\Lambda_0)
\label{eq:auto1}
\eea
Isomorphisms of the irreducible modules $V(\Lambda_0)$, $V(\Lambda_1+\Lambda_4)$
or $V(2\Lambda_0-\Lambda_{3}+\Lambda_4)$ and $V(\Lambda_2-\Lambda_4)$,
$V(\Lambda_3-\Lambda_4)$ or $V(-\Lambda_{1}+2\Lambda_2-\Lambda_4)$
are  obtained setting
\bea
\varrho(\hat{\lambda}_0)=\hat{\lambda}_2\qquad\qquad\;\;\varrho(\hat{\lambda}_1)=
\hat{\lambda}_3\qquad\qquad\;\;\varrho(\hat{\lambda}_{-3})=\hat{\lambda}_{-1,2^2}
\label{eq:autoirr}
\eea
where $\hat{\lambda}_{-3}$ and $\hat{\lambda}_{-1,2^2}$ denote
the highest weight vectors of $V(2\Lambda_0-\Lambda_{3}+\Lambda_4)$
and $V(-\Lambda_{1}+2\Lambda_{2}-\Lambda_4)$, respectively.
Relations between the weight structures
of $\tilde V(\Lambda_0)$ and $\tilde V(\Lambda_1+\Lambda_4)$ follow from the isomorphisms
$\varsigma$ and $\varrho$ providing maps between the irreducible components of
these modules. They can also be inferred from the boson realizations of the
reducible modules.

Assuming that the above correspondences hold at any grade, the
 set of all infinite configurations
$\ldots\otimes w_{j_3}\otimes w^*_{j_2}\otimes w_{j_1}\otimes w^*_{j_0}\otimes w_{j_{-1}}
\otimes w^*_{j_{-2}}\otimes\ldots$ with $j_r=3$ for almost all $r$ may be 
interpreted as a level-zero module. Depending on the choice of the reference weight, it
is given by
\bea
\tilde V(\Lambda_0)\otimes\tilde V(\Lambda_1+\Lambda_4)^{*S}\qquad\mbox{or}\qquad\tilde V(
2\Lambda_0-\Lambda_{3}+\Lambda_4)\otimes \tilde V(\Lambda_0)^{*S}
\label{eq:qusp1}
\eea
Analogously, the set of all configurations $\ldots\otimes w^*_{j_3}\otimes w_{j_2}
\otimes w^*_{j_1}\otimes w_{j_0}\otimes w^*_{j_{-1}}\otimes w_{j_{-2}}\otimes\ldots$
with $j_r=3$ for almost all $r$ can be viewed as
\bea
\tilde V(\Lambda_0)\otimes\tilde V(2\Lambda_0-\Lambda_{3}+\Lambda_4)^{*S}
\qquad\mbox{or}\qquad\tilde V(\Lambda_1+\Lambda_4)
\otimes\tilde V(\Lambda_0)^{*S}
\label{eq:qusp2}
\eea
In the following sections, explicit
notation of the weight $\Lambda_4$ will be omitted for brevity.

\section{Bosonization}
\label{sec:bos}

This section provides a boson realization of $U_q\bigl(\widehat{gl}(2\vert2)\bigr)$ at
level one similar to the construction presented in \cite{shiretal} for $U_q\bigl(
\widehat{sl}(M+1\vert N+1)\bigr)$ with the standard choice of simple roots. 
The $U_q\bigl(\widehat{sl}(2\vert2)\bigr)/{\cal H}$-modules 
$\tilde V(\iota_{II'}\Lambda_{I'})$ with $I,I'=0,1,3$ and
$\iota_{0I}=\delta_{I,0}$, $\iota_{1,I}=\delta_{I,1}$, $\iota_{3,I}=2\delta_{I,0}
-\delta_{I,3}$ are expressed as suitable restrictions of a Fock space.

Bosonization of the currents $\Psi^{j,\pm}(z)$ at level one requires four sets
$\{\varphi^j,\varphi^j_0,\varphi^j_n,j=1,2,3,4;\,n\in{\mathbb{Z}}\}$ of deformed
bosonic oscillators with commutation relations
\bea
\bigl[\varphi^j_n,\varphi^k_m\bigr]&=&\delta_{j,k}\delta_{n+m,0}\,{[n]^2\over n}
\qquad\;\;n,m\neq0\cr
\noalign{\bigskip}
\bigl[\varphi^j,\varphi^k_0\bigr]&=&i\delta_{j,k}
\label{eq:oscsets}
\eea
In terms of the oscillators (\ref{eq:oscsets}), the currents $\Psi^{j,\pm}(z)$
are realized by
\bea
\Psi^{4,+}(z)&=&q^{\varphi^1_0-i\varphi^4_0}\,\exp\Bigl((q-q^{-1})\sum_{n>0}
\bigl(\varphi^1_n-i\varphi^4_n\bigr)\,z^{-n}\Bigr)\cr
\noalign{\bigskip}
\Psi^{4,-}(z)&=&q^{-\varphi^1_0+i\varphi^4_0}\,\exp\Bigl(-(q-q^{-1})\sum_{n>0}
\bigl(\varphi^1_{-n}-i\varphi^4_{-n}\bigr)\,z^n\Bigr)
\label{eq:bos1b}
\eea
and 
\bea
\Psi^{j,+}(z)&=&q^{-i^j(\varphi^{j+1}_0+i\varphi^j_0)}\,\exp\Bigl(
-(q-q^{-1})\sum_{n>0}i^j\bigl(\varphi^{j+1}_n+i\varphi^j_n\bigr)\,z^{-n}\Bigr)\cr
\noalign{\bigskip}
\Psi^{j,-}(z)&=&q^{i^j(\varphi^{j+1}_0+i\varphi^j_0)}\,\exp\Bigl(
(q-q^{-1})\sum_{n>0}i^j\bigl(\varphi^{j+1}_{-n}+i\varphi^j_{-n}\bigr)\,z^n\Bigr)
\label{eq:bos1}
\eea
for $j=1,2,3$.
Equations (\ref{eq:bos1b}) and  (\ref{eq:bos1}) imply
\bea
h_4=\varphi^1_0-i\varphi^4_0,\qquad\;\;
h_j=-i^j\bigl(\varphi^{j+1}_0+i\varphi^j_0\bigr)\;\;\mbox{for}\;\;j=1,2,3
\label{eq:hbos}
\eea
Combining each set of oscillators in the deformed free field
\bea
\varphi^{j,\pm}(z)=\varphi^j-i\varphi^j_0\,\ln z+i\sum_{n\neq0}{q^{\mp{1\over2}
\vert n\vert}\over[n]}\,\varphi^j_nz^{-n}\qquad\;\;j=1,2,3,4
\label{eq:freef}
\eea
the dependence of the bosonized currents $E^{k,\pm}(z)$ on the sets (\ref{eq:oscsets})
can be expressed by
\bea
E^{j,+}(z)&=&:\exp\Bigl(-i^{j+1}\bigl(\varphi^{j+1,+}(z)+i\varphi^{j,+}(z)\bigr)\Bigr):
\,\exp\bigl(i\pi\delta_{j,1}\varphi^3_0\bigr)\,X^{j,+}(z)
\cr
\noalign{\bigskip}
E^{j,-}(z)&=&:\exp\Bigl(i^{j+1}\bigl(\varphi^{j+1,-}(z)+i\varphi^{j,-}(z)\bigr)\Bigr):
\,\exp\bigl(-i\pi\delta_{j,1}\varphi^3_0\bigr)\,X^{j,-}(z)
\label{eq:bos3}
\eea
for $j=1,2,3$. The additional parts $X^{k,\pm}(z)$ are needed to impose
relations (\ref{eq:def5})-(\ref{eq:def6}). Their construction involves two further
fields
\bea
\beta^r(z)=\beta^r-i\beta^r_0\,\ln z+i\sum_{n\neq0}{1\over n}\beta^r_n\,z^{-n}
\label{eq:beta1}
\eea
with
\bea
\bigl[\beta^r_n,\beta^s_m\bigr]&=&-n\delta_{r,s}\delta_{n+m,0}\cr
\noalign{\bigskip}
\bigl[\beta^r,\beta^s_0\bigr]&=&-i\delta_{r,s}\qquad\;\;r,s=1,2
\label{eq:beta2}
\eea
A suitable realization of $X^{j,\pm}(z)$ in terms of these fields is provided by 
\bea
-X^{1,-}(z)=X^{2,+}(z)&=&{1\over z(q-q^{-1})}\Bigl(:\exp\bigl(\beta^1(q^{-1}z)\bigr):-
:\exp\bigl(\beta^1(qz)\bigr):\Bigr)
\cr
\noalign{\bigskip}
X^{1,+}(z)=X^{2,-}(z)&=&:\exp\bigl(-\beta^1(z)\bigr):
\cr
\noalign{\bigskip}
X^{3,+}(z)&=&{1\over z(q-q^{-1})}\Bigl(:\exp\bigl(\beta^2(q^{-1}z)\bigr):-:\exp\bigl(
\beta^2(qz)\bigr):\Bigr)
\cr
\noalign{\bigskip}
X^{3,-}(z)&=&:\exp\bigl(-\beta^2(z)\bigr):
\label{eq:xfields}
\eea
Finally, the expression for the grading operator $d$ is inferred from
the defining relations (\ref{eq:d2}) and (\ref{eq:bos1})-(\ref{eq:xfields}):
\bea
d=-{1\over2}\sum_{j=1}^4\bigl(\varphi^j_0\bigr)^2-\sum_{j=1}^4\sum_{n>0}{n^2
\over[n]^2}\,\varphi^j_{-n}\varphi^j_n+{1\over2}\sum_{r=1,2}\beta^r_0\bigl(
\beta^r_0-i\bigr)+\sum_{r=1,2}\sum_{n>0}\beta^r_{-n}\beta^r_n
\label{eq:dbos}
\eea
The highest weight vectors $\hat{\kappa}$ and $\hat{\nu}$ of
the reducible modules $\tilde V(\Lambda_0)$ and
$\tilde V(2\Lambda_0-\Lambda_{3})$ may be realized by
\bea
\hat{\kappa}=e^{\beta^2}\,\vert0\rangle\qquad\;\;\hat{\nu}=e^{-\varphi^4+2\beta^2}\,
\vert0\rangle
\label{eq:hwbos1}
\eea
Furthermore, the vector $\hat{\mu}\in \tilde V(\Lambda_1)$ with properties
(\ref{eq:mudef}) and (\ref{eq:mudef2}) can be expressed by
\bea
\hat{\mu}=e^{i\varphi^1+\beta^2}\,\vert0\rangle
\label{eq:hwbos2}
\eea
Here the boson Fock vacuum $\vert0\rangle$ is characterized by 
\bea
\varphi^j_0\,\vert0\rangle=\beta^r_0\,\vert0\rangle&=&0
\qquad\qquad
\varphi^j_n\,\vert0\rangle=
\beta^r_n\,\vert0\rangle=0\qquad\forall n>0,\,j=1,2,3,4,\,r=1,2
\label{eq:vacdef}
\eea
The vector $e^{\beta^2}\,\vert0
\rangle$ satisfies the highest weight properties $E^{k,+}_n\,e^{\beta^2}\,\vert0
\rangle=0\forall n\geq0$ and $E^{k,-}_n\,e^{\beta^2}\,\vert0\rangle=0\forall n>0$,
$k=1,2,3$. From (\ref{eq:bos1})-(\ref{eq:xfields}), all descendants of grade zero
are listed by
\bea
\bigl(E^{2,-}_0E^{3,-}_0\bigr)^n\,\hat{\kappa}&\sim&e^{n(\varphi^2+\varphi^4-
\beta^1-\beta^2)+\beta^2}\,\vert0\rangle\cr
\noalign{\medskip}
E^{3,-}_0\bigl(E^{2,-}_0E^{3,-}_0\bigr)^n\,\hat{\kappa}&\sim&e^{n(\varphi^2+\varphi^4-
\beta^1-\beta^2)+i\varphi^3+\varphi^4}\,\vert0\rangle\cr
\noalign{\medskip}
E^{1,-}_0\bigl(E^{2,-}_0E^{3,-}_0\bigr)^{n+1}\,\hat{\kappa}&\sim&e^{n(\varphi^2+\varphi^4-
\beta^1-\beta^2)-i\varphi^1+\varphi^4}\,\vert0\rangle\cr
\noalign{\medskip}
E^{1,-}_0E^{3,-}_0\bigl(E^{2,-}_0E^{3,-}_0\bigr)^{n+1}\,\hat{\kappa}&\sim&e^{n(\varphi^2+
\varphi^4-\beta^1-\beta^2)-i\varphi^1+i\varphi^3+2\varphi^4}\,\vert0\rangle
\label{eq:bzerodes}
\eea
with $n=0,1,2,\ldots$.
The vectors (\ref{eq:bzerodes}) coincide with the grade-zero subspace of
$\tilde V(\Lambda_0)$. Thus a suitable subspace of the Fock space ${\cal F}_0$ related
to $e^{\beta^2}
\,\vert0\rangle$ can be expected to furnish a boson realization of $\tilde V(
\Lambda_0)$. Similar considerations apply to $\tilde V(\Lambda_1)$ and $\tilde V
(2\Lambda_0-\Lambda_{3})$.
According to the boson realizations of
the currents (\ref{eq:bos1})-(\ref{eq:bos3}), the Fock spaces ${\cal F}_I$ are
\bea
{\cal F}_I={\mathbb{C}}\bigl[\tilde{\varphi}^r_{-1},\beta^s_{-1},\tilde{\varphi}^{r'}_{-2},
\beta^{s'}_{-2},
\ldots\bigr]\,\otimes\Bigl(\oplus_{s_1,s_2,s_3\in{\mathbb{Z}}}{\mathbb{C}}\,e^{s_1(i\varphi^1
+\varphi^2-\beta^1)+s_2(\varphi^2-i\varphi^3-\beta^1)+s_3(i\varphi^3+\varphi^4-
\beta^2)+\beta^2+\alpha_I}\,\vert0\rangle\Bigr)
\label{eq:fock}
\eea
with
\bea
\alpha_0=0\qquad\;\;\alpha_1=\varphi^4-\beta^2\qquad\;\;\alpha_{3}=-\varphi^4+\beta^2
\label{eq:fockadd}
\eea
and
\bea
\tilde{\varphi}^1_{-n}&\equiv&\varphi^1_{-n}-i\varphi^2_{-n}+\varphi^3_{-n}-i
\varphi^4_{-n}\cr
\noalign{\bigskip}
\tilde{\varphi}^2_{-n}&\equiv&i\varphi^2_{-n}+\varphi^3_{-n}
\label{eq:tilde}
\eea 
Only two linear combinations $\tilde{\varphi}^1_{-n}$ and $\tilde{\varphi}^2_{-n}$
are required since the Fock space ${\cal F}_I$ should contain a realization of
a $U_q\bigl(\widehat{sl}(2\vert2)\bigr)/{\cal H}$-module rather than a
$U_q\bigl(\widehat{gl}(2\vert2)\bigr)$-module.

The construction (\ref{eq:bos3}) implies the existence of two screening operators
$\eta^r(z)=\sum_{n\in{\mathbb{Z}}}\eta^r_n\,z^{-n-1}=:e^{-\beta^r(z)}:$ for $r=1,2$.

The particular form of the highest weight vectors (\ref{eq:hwbos1}) and of
(\ref{eq:hwbos2}) as well as the fact that
$\eta^1_0$ commutes with all generators of $U_q\bigl(\widehat{gl}(2\vert2)\bigr)$
indicate that the boson realizations of $\tilde V(\iota_{II'}\Lambda_{I'})$ are contained in the
restricted Fock spaces
\bea
Ker_{\eta^1_0}\,{\cal F}_I\qquad\;\;I=0,1,3
\label{eq:fres}
\eea
With the action of the grading operator $d$ on the vacuum $\vert0\rangle$ fixed
by $d\,\vert0\rangle=0$, its realization by (\ref{eq:dbos}) leads to
\bea
&&d\,\Bigl(e^{s_1(i\varphi^1
+\varphi^2-\beta^1)+s_2(\varphi^2-i\varphi^3-\beta^1)+s_3(i\varphi^3+\varphi^4-
\beta^2)+\beta^2+\alpha_I}\,\vert0\rangle\Bigr)=\cr
\noalign{\bigskip}
&&\qquad\qquad=-{1\over2}\Bigl(s_1(s_1+1)+(s_2-s_3)(s_2-s_3+1)
-\delta_{I,1}+\delta_{I,3}\Bigr)\cdot \cr
\noalign{\bigskip}
&&\qquad\qquad\qquad\qquad\qquad\qquad\qquad\qquad\cdot\; e^{s_1(i\varphi^1
+\varphi^2-\beta^1)+s_2(\varphi^2-i\varphi^3-\beta^1)+s_3(i\varphi^3+\varphi^4-
\beta^2)+\beta^2+\alpha_I}\,\vert0\rangle,\qquad I=0,1,3
\label{eq:dexp}
\eea
Equations (\ref{eq:hbos}) and (\ref{eq:dexp}) allow to collect 
the weights of all vectors in $Ker_{\eta^1_0}\,{\cal F}_I$ at a given grade.
At grades $-1,-2,-3$,
the author finds a one-to-one correspondence of the weights in $Ker_{\eta^1_0}\,
{\cal F}_0$ with the
weights of $\tilde V(\Lambda_0)$ listed in the previous section ((\ref{eq:lamlist}),
(\ref{eq:list1}), (\ref{eq:list2}), (\ref{eq:grade2})-(\ref{eq:grade3b})).
This results supports the conjecture that (\ref{eq:fres}) provides 
the boson realization of the level-one modules $\tilde V(\iota_{II'}\Lambda_{I'})$:
\bea
\tilde V(\iota_{II'}\Lambda_{I'})=Ker_{\eta^1_0}\,{\cal F}_I\qquad\;\;I=0,1,3
\label{eq:modbos}
\eea
Introduction of a further fermionic field
$\xi^2(z)=\sum_{n\in{\mathbb{Z}}}\xi^2_n\,z^{-n}=:e^{\beta^2(z)}:$ allows for a
convenient decomposition of the modules $\tilde V(\iota_{II'}\Lambda_{I'})$ into irreducible
components. Relations (\ref{eq:beta2}) lead to
$\{\xi^r_n,\eta^s_m\}=\delta_{r,s}\delta_{n+m,0}$ and $\{\xi^r_n,\xi^s_m\}=
\{\eta^r_n,\eta^s_m\}=0$. In terms of $\xi^2_0$ and $\eta^2_0$, 
direct sum decompositions are provided by $Ker_{\eta^1_0}\;{\cal F}_I=\xi^2_0\eta^2_0\,
Ker_{\eta^1_0}\;{\cal F}_I\oplus \eta^2_0\xi^2_0\;Ker_{\eta^1_0}\,{\cal F}_I$.
The second part constitutes 
 boson realizations of the irreducible $U_q\bigl(\widehat{sl}(2\vert2)\bigr)
/{\cal H}$-modules $V(\Lambda_2)$, $V(\Lambda_3)$ and $V(-\Lambda_{1}+2
\Lambda_2)$ specified in the previous section:
\bea
V(\Lambda_{2})=Ker_{\eta^2_0}\;Ker_{\eta^1_0}\,{\cal F}_0\qquad
V(\Lambda_3)=Ker_{\eta^2_0}\;Ker_{\eta^1_0}\,{\cal F}_1\qquad
V(-\Lambda_{1}+2\Lambda_2)=Ker_{\eta^2_0}\;Ker_{\eta^1_0}\,{\cal F}_{3}
\label{eq:modalt1}
\eea
Expressions for the irreducible $U_q\bigl(\widehat{sl}(2\vert2)\bigr)
/{\cal H}$-modules $V(\iota_{II'}\Lambda_{I'})$ are obtained as
\bea
V(\iota_{II'}\Lambda_{I'})=\eta^2_0\,Ker_{\eta^1_0}\,{\cal F}_I\qquad\;\;I=0,1,3
\label{eq:modalt2}
\eea
Boson realizations provide an efficient tool to obtain character formulae.
In \cite{zha}, character expressions for the modules $V(\Lambda_I)$, $V(\Lambda_2)$
and $V(-\Lambda_1+2\Lambda_2)$ are found. The
character expressions are sums of both positive and negative contributions.
Hence, unlike the character expressions in \cite{cheng} and formula (3.14) in
\cite{kacwak2}, there are not of quasiparticle type.
According to (\ref{eq:modalt2}), the highest
weight vector of $V(2\Lambda_0-\Lambda_3)$ is realized by $\beta^2_{-1}\,e^{-\varphi^4+
\beta^2}\,\vert0\rangle$. The module $V(2\Lambda_0-\Lambda_3)$ is not considered in
\cite{zha}, since there analysis is restricted to highest weight vector
given by pure exponentials in $\varphi^j$ and $\beta^1$ acting on the Fock vacuum.

\section{Border stripes and level-zero modules of $U'_q\bigl(\widehat{sl}(2\vert2)
\bigr)$}
\label{sec:bs1}

In Sect. \ref{sec:module}, the space of half-infinite configurations $(\ldots\otimes
w_{j_4}\otimes w^*_{j_3}\otimes w_{j_2}\otimes w^*_{j_1})$ has been related
to one reducible or two irreducible modules of $U_q\bigl(\widehat{sl}
(2\vert2)\bigr)/{\cal H}$ with level one. The aim of section \ref{sec:bs3} is a
description of the half-infinite configurations in terms of infinitely many
level-zero modules of $U_q\bigl(\widehat{sl}(2\vert2)\bigr)$. For this purpose,
classification by border stripes offers a useful tool.

Following \cite{macd}, 
a skew Young diagram $\lambda\backslash\mu$ may be associated to
any two partitions $\lambda$, $\mu$ with $\lambda_i\geq\mu_i$ $\forall i$. It consists of
all squares with edges $(i-1,j-1)$, $(i-1,j)$, $(i,j)$ and $(i,j-1)$ with the pair
$i,j$ satisfying the property $\lambda_j\geq i>\mu_j$. A connected skew diagram
without $2\times2$ blocks of boxes is called a border strip. Below, two sets of
numbers $\{n_i\}_{1\leq i<R}$ and $\{m_i\}_{1\leq i\leq R}$ with $R\geq1$, $n_i>0,\;
m_i>0$ for $i<R$ and $m_R\geq0$ will be used to
characterize a given border strip. If $R>1$,
the parameters $\{n_i;\,m_i\}$ and the partitions $(\lambda_1,\lambda_2,
\ldots,\lambda_{K+1})$ and $(\mu_1,\mu_2,\ldots,\mu_{K+1})$ with $K=\sum_{i=1}^{R-1}
n_i$ are related by 
\bea
\lambda_1&=&\sum_{i=1}^Rm_i\cr
\noalign{\bigskip}
\lambda_2&=&1+\sum_{i=2}^Rm_i\qquad\;\mbox{for}\;\;\;1<j\leq1+n_1\cr
\noalign{\bigskip}
\lambda_j&=&1+\sum_{i=S}^Rm_i\qquad\;\mbox{for}\;\;\;
1+\sum_{i=1}^{S-2}n_i<j\leq1+\sum_{i=1}^{S-1}n_i,\;\;3\leq S\leq R\cr
\noalign{\bigskip}
\noalign{\medskip}
\mu_{K+1}&=&0\cr
\noalign{\bigskip}
\mu_j&=&\sum_{i=2}^Rm_i\qquad\;\mbox{for}\;\;\;0<j\leq n_1\cr
\noalign{\bigskip}
\mu_j&=&\sum_{i=S}^Rm_i\qquad\;\mbox{for}\;\;\;
\sum_{i=1}^{S-2}n_i<j\leq\sum_{i=1}^{S-1}n_i,\;\;3\leq S\leq R
\label{eq:pprel}
\eea
In the case $R=1$, this reduces to $\lambda_1=M$ and $\mu_1=0$.
Thus $n_i+1$
denotes the number of boxes in the $i-th$ column containing more than one box,
counted from the right. Similarly, $m_i+1$ is the number of boxes in the $i$-th
row with more than two boxes, counted from the top to the bottom of the border strip.
$m_R+1$ and $m_1$ count the number of boxes in the lowest and in the uppermost row,
respectively. The total number of boxes is given by
$M=m_R+\sum_{i=1}^{R-1}(n_i+m_i)$. Fig. 3  illustrates an example. 

\begin{center}
\setlength{\unitlength}{1cm}
\begin{picture}(8,7)
\thicklines

\put(1,1){\line(1,0){3}}
\put(1,1.6){\line(1,0){3}}
\put(3.4,2.2){\line(1,0){2.4}}
\put(3.4,2.8){\line(1,0){2.4}}
\put(5.2,3.4){\line(1,0){0.6}}
\put(5.2,4){\line(1,0){0.6}}
\put(5.2,4.6){\line(1,0){1.2}}
\put(5.2,5.2){\line(1,0){1.2}}

\put(1,1){\line(0,1){0.6}}
\put(1.6,1){\line(0,1){0.6}}
\put(2.2,1){\line(0,1){0.6}}
\put(2.8,1){\line(0,1){0.6}}
\put(3.4,1){\line(0,1){1.8}}
\put(4,1){\line(0,1){1.8}}
\put(4.6,2.2){\line(0,1){0.6}}
\put(5.2,2.2){\line(0,1){3}}
\put(5.8,2.2){\line(0,1){3}}
\put(6.4,4.6){\line(0,1){0.6}}

\put(2,5.7){$m_3$}
\put(4.1,5.6){$m_2$}
\put(5.63,5.5){$m_1$}
\put(1.92,5.8){\vector(-1,0){0.92}}
\put(4.02,5.7){\vector(-1,0){0.62}}
\put(5.54,5.6){\vector(-1,0){0.34}}
\put(2.49,5.8){\vector(1,0){0.91}}
\put(4.59,5.7){\vector(1,0){0.61}}
\put(6.07,5.6){\vector(1,0){0.33}}

\put(6.8,1.52){$n_2$}
\put(6.7,3.32){$n_1$}
\put(6.9,1.44){\vector(0,-1){0.434}}
\put(6.8,3.24){\vector(0,-1){1.04}}
\put(6.9,1.76){\vector(0,1){0.44}}
\put(6.8,3.56){\vector(0,1){1.04}}

\end{picture}\par
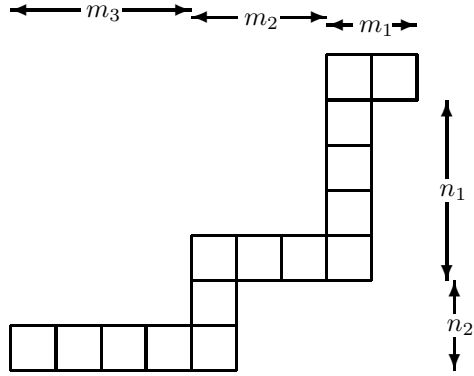
Fig. 3: Border strip with $R=3$, $m_1=2,\,m_2=3,\,m_3=4$ and $n_1=4,\,n_2=2$
\end{center}

In contrast to the case of quantum affine algebras, the length of a column
is not restricted.
From a particular border strip, a semi-standard super tableau for $U'_q\bigl(
\widehat{sl}(2\vert2)\bigr)$ is obtained by attributing a number $0,1,2$ or $3$
to each of its boxes according to three rules. Within each row (column), the
numbers are weakly decreasing from the left to the right (from the top to the
bottom). Second, any row receives each of the numbers $1$
and $3$ at most once. Third, any column obtains each of
the numbers $0$ and $2$ at most once. These three rules allow for two labelings
involving only the numbers $0$ and $1$. Each column with $n_i+1>1$ boxes obtains
the value $1$ in the upper $n_i$ boxes. Attributing the number $0$ to all
remaining boxes yields a labeling referred to as top labeling below.
A second labeling without the numbers $2$ and $3$ follows from the top
labeling by replacing the number $0$ in the lowest leftmost box of the
border strip by $1$. A semi-standard super tableau according to the
three rules specified above will be called a type $A$ labeling 
in the remainder. To the number $l$ in a type $A$ labeling,
the $U_q\bigl(gl(2\vert2)\bigr)$-weight of $w_l$ given in
(\ref{eq:wdef1}), (\ref{eq:glstr}) may be attributed. 
A $U_q\bigl(gl(2\vert2)\bigr)$-weight for a given labeling
is then introduced as the sum over all weights related to its numbers.
Denoting the multiplicity of a number $l$ in a given labeling $T_A$ of type
$A$ by $N_l(T_A)$, the (supersymmetric) skew Schur function $s_{\lambda\backslash\mu}
({\bf p})$ is defined by
\bea
s_{\lambda\backslash\mu}({\bf p})\equiv\sum_{T_A\in T_A(\lambda\backslash\mu)}
p_0^{N_0(T_A)}\,p_1^{N_1(T_A)}\,p_2^{N_2(T_A)}\,p_3^{N_3(T_A)}
\label{eq:schurdef}
\eea
where $T_A(\lambda\backslash\mu)$ denotes the set of all type $A$ labelings of
the border strip characterized by the partitions $\lambda,\mu$ given in
(\ref{eq:pprel}). The skew Schur function can be expressed as
\bea
s_{\lambda\backslash\mu}({\bf p})=det\bigl(e_{\lambda_i-\mu_j-i+j}\bigr)_{1\leq
i,j\leq K+1}
\label{eq:schur1}
\eea
where $e_0=1$, $e_{-m}=0$ and
\bea
e_m=p_0^{m-1}(p_0+p_1)+{p_2+p_3\over p_2-p_0}
\bigl(p_2^m-p_0^m+p_1p_2^{m-1}-p_1p_0^{m-1}\bigr)
\eea
for $m=1,2,3,\ldots$. Equation ({\ref{eq:schur1}) corresponds to the expression
proved for the skew Schur functions attributed to $U_q\bigl(\widehat{sl}
(n)\bigr)$ in \cite{macd}.

For a given border stripe, a subset of its semi-standard super tableaux is obtained 
by discarding any labeling of the boxes with $3$ in the lowest leftmost box.
In the following, this subset will be referred to as the restricted (type $A$) labelings.
Expressions (\ref{eq:hdiag1}) and (\ref{eq:hdiag2}) 
for the diagonal elements of the CTM Hamiltonian reveal a
one-to-one correspondence between the restricted labelings
and the sets of configurations $\{\tau\}_n$ with $n<0$. Formally,
the unique configuration in $\{\tau\}_0$ may be related to the value $R=0$. 
The boxes in a border strip of total length $M$ are counted starting from its
lower left end. Then, for a given labeling subject to the above rules,
the number in the $k-th$ box is associated with the entry $j_{2(M-k+1)}$ of a
component $\bigl(\ldots,j_6,j_4,j_2\bigr)$ in $\bigl\{\tau\bigr\}_{n_{\{n_i;\,m_i\}}}$.
The remaining entries are fixed by $j_{2r}=3\;\forall r>M$. A restricted labeling
yields $j_{2M}\neq3$. The number $n_{\{n_i;\,m_i\}}$ denoting
the contribution of this particular component
to the diagonal element of the CTM-Hamiltonian is given by
\bea
n_{\{n_i;\,m_i\}}=-M-\sum_{i=1}\Bigl\{{1\over2}m_i(m_i-1)+m_{i+1}\sum_{j=1}^i(m_j+n_j)
\Bigr\}
\label{eq:gradef1}
\eea
If $m_R>0$, the weights of the discarded labelings can be written
\bea
\bigl(M-1-t_2,-K+t_3-t_1,1+t_2,M-K-2-t_1-t_3\bigr)
\label{eq:disc1}
\eea
with suitable values of
$t_l>0$. In the case $m_R=0$, the dropped weights have the form
\bea
\bigl(M-n_{R-1}-1-t_2,-K+n_{R-1}+t_3-t_1,n_{R-1}+1+t_2,M-K-n_{R-1}-2-t_1-t_3\bigr)
\label{eq:disc2}
\eea
A restricted skew Schur function $s'_{\lambda\backslash\mu}({\bf p})$ may be defined
by replacing the sum in (\ref{eq:schurdef}) by the sum over all restricted type $A$
labelings of the border strip.
Then the determinant formula reads
\bea
s'_{\lambda\backslash\mu}({\bf p})=det\bigl(E'\bigr)_{1\leq i,j\leq K+1}
\label{eq:schur2}
\eea
where $E'_{i,j}=e_{\lambda_i-\mu_j-i+j}$ for $j\leq K$ and
$E'_{i,K+1}=\tilde e_{\lambda_i-i+K+1}$ with
\bea
\tilde e_m=p_0^{m-1}(p_0+p_1)+
{p_2\over p_2-p_0}\bigl(p_2^m-p_0^m+p_1p_2^{m-1}-p_1p_0^{m-1}\bigr)
\eea

Including the labelings with the number $3$ in the lowest leftmost box, the semi-standard
tableaux can be related to level-zero modules of $U'_q\bigl(\widehat{sl}(2\vert2)\bigr)$.
To this aim, a complex number $x_k$ is attributed to the $k-th$ box of the border
strip (see Fig. 4). With the abbreviations
\bea
t_0&=&m_R+n_{R-1}\cr
\noalign{\bigskip}
t_r&=&m_R+n_{R-r-1}+\sum_{i=1}^r(n_{R-i}+m_{R-i})\qquad\;\;1\leq r\leq R-2
\label{eq:trdef}
\eea
their relations are expressed by
\bea
x_{t_r+1}&=&q^{-2(m_{R-r-1}+1-\delta_{R-r,2})}\,x_{t_r}=q^{-2(t_r-m_R+m_{R-r-1}
-\delta_{R-r,2})}\,x_1\qquad\mbox{for}\;\;0\leq r\leq R-2
\label{eq:xspecrel1}
\eea
\bea
\begin{array}{rll}
x_{t_r+1+k}&=q^{2k}\,x_{t_r+1}\;\;&\mbox{for}\;\;1\leq k\leq t_{R-r-1}-\delta_{R-r,
2}\,,\qquad0\leq r\leq R-2\cr
\noalign{\bigskip}
x_{k+1}&=q^{2k}x_k\;\;&\mbox{for}\;\;1\leq k\leq m_R-\delta_{R,1}
\end{array}
\label{eq:xspecrel2}
\eea
and
\bea
x_{t_r+m_{R-r-1}+1+k}&=&q^{-2k}\,x_{t_r+1}=q^{-2(t_r-m_R+m_{R-r-1}+k)}\,x_1\qquad
\mbox{for}\;\;1\leq k< n_{R-r-2},\;\;\;0\leq r\leq R-3\cr
\noalign{\bigskip}
x_{m_R+1+k}&=&q^{-2k}x_1\qquad\mbox{for}\;\;1\leq k<n_{R-1}
\label{eq:xspecrel3}
\eea
Here the first (second) line of (\ref{eq:xspecrel3}) is omitted for $n_{R-r-2}=1$
($n_{R-1}=1$).
Similarly, the last line of (\ref{eq:xspecrel2}) is dropped for $m_R=0$.
The level-zero module for the border strip is contained in the tensor product
of evaluation modules 
\bea
W_{x_1}\otimes W_{x_2}\otimes\ldots\otimes W_{x_M}
\label{eq:evmopro}
\eea
$U_q\bigl(\widehat{gl}(2\vert2)\bigr)$ acts on (\ref{eq:evmopro}) via the iterated
coproduct $\Delta^{(M-1)}(a)$ introduced by $\Delta^{(m)}(a)=(\Delta\otimes{\bf 1})\,
\Delta^{(m-1)}(a)$ with $\Delta^{(1)}(a)=\Delta(a)$ and equations (\ref{eq:wdef1}),
(\ref{eq:wevmo1}).
For $x_k$ related to $x_1$ as specified by (\ref{eq:xspecrel1})-(\ref{eq:xspecrel3}),
the tensor product (\ref{eq:evmopro}) contains an element $\nu_{\{n_i;\,m_i\}}$
characterized by the properties
\bea
\Delta^{(M-1)}(
E^{l,+}_{s})\,\nu_{\{n_i;\,m_i\}}&=&0\qquad \mbox{for}\;\;s\in{\mathbb{Z}},\;l=1,2,3
\label{eq:hwlzero1}
\eea
\bea
\begin{array}{llll}
\Delta^{(M-1)}(h_1)\,\nu_{\{n_i;\,m_i\}}&=M\,\nu_{\{n_i;\,m_i\}}\qquad&
\Delta^{(M-1)}(h_2)\,\nu_{\{n_i;\,m_i\}}=-K\,\nu_{\{n_i;\,m_i\}}\cr
\noalign{\bigskip}
\Delta^{(M-1)}(h_3)\,\nu_{\{n_i;\,m_i\}}&=0\qquad&
\Delta^{(M-1)}(h_4)\,\nu_{\{n_i;\,m_i\}}=(M-K)\,\nu_{\{n_i;\,m_i\}}
\end{array}
\label{eq:hwlzero2}
\eea
and
\bea
\Delta^{(M-1)}(H^1_s)\,\nu_{\{n_i;\,m_i\}}&=&q^s{[s]\over s}\sum_{k=1}^Mx_k^s\;
\nu_{\{n_i;\,m_i\}}\cr
\noalign{\bigskip}
\Delta^{(M-1)}(H^2_s)\,\nu_{\{n_i;\,m_i\}}&=&\left\{\begin{array}{cc}
0\qquad&\mbox{if}\;\;R=1\cr
\noalign{\smallskip}
-{[s]\over s}\sum_{k\in\Upsilon_{\{n_i;\,m_i\}}}x_k^s\;
\nu_{\{n_i;\,m_i\}}\qquad&\mbox{if}\;\;R>1\end{array}\right.\cr
\noalign{\bigskip}
\Delta^{(M-1)}(H^3_s)\,\nu_{\{n_i;\,m_i\}}&=&0
\label{eq:hwlzero3}
\eea
$\forall s\neq0$.

$\Upsilon_{\{n_i;\,m_i\}}$ denotes a subset of all $\bigl\{k\in{\mathbb{N}}\,\vert\,
1\leq k\leq M
\bigr\}$ defined for all $\{n_i;\,m_i\}$ with $R>1$. It contains the value $k=1$ for any
of these sets $\{n_i;\,m_i\}$. In addition, the
value $k>1$ is included if the $k$-th box is occupied by the number $1$ in the
top labeling except for the uppermost of these boxes.  All other values are
not contained in $\Upsilon_{\{n_i;\,m_i\}}$. $\Upsilon_{\{4,2;2,3,4\}}$ for
the example illustrated in Fig. 3 is given by $\{1,6,7,11,12,13\}$.  As a second example,
for the border strip
consisting of a single column of $n_1+1$ boxes the set
$\Upsilon_{\{n_1;\,m_1=1,m_2=0\}}=(1,2,\ldots,n_1)$ results.
According to (\ref{eq:hwlzero2}), $w_0$ and $w_1$ occur $(M-K)$-times and $K$-times
in $\nu_{\{n_i;\,m_i\}}$,
respectively. This implies the last of equations (\ref{eq:hwlzero3}). Clearly,
$\nu_{\{n_i;\,m_i\}}$ corresponds to the top labeling $A$ of the border strip.
The coproduct of $(H^1_{\pm s}+H^3_{\pm s})$ takes the simple form
\bea
\Delta(H^1_{\pm s}+H^3_{\pm s})=(H^1_{\pm s}+H^3_{\pm s})\otimes q^{-{1\over2}sc(1
\mp2)}+q^{{1\over2}sc(1\pm2)}\otimes(H^1_{\pm s}+H^3_{\pm s})\qquad\;\;s> 0
\label{eq:coh1h3}
\eea
where $c$ denotes the level.
Then the first of equations (\ref{eq:hwlzero3}) follows from the third and
\bea
H^l_s\,\bigl(w_j\otimes x^{s'}\bigr)=\bigl(\delta_{l,j}+\delta_{l,j+1}\bigr)\cdot(-1)^{l-1}
q^{s(1+\delta_{l,2})}{[s]\over s}\,\bigl(w_j\otimes x^{s+s'}\bigr)
\label{eq:h1h3co}
\eea
An evaluation making use of (\ref{eq:xspecrel1})-(\ref{eq:xspecrel3}) leads to
\bea
\Delta^{(M-1)}
(H^1_s)\,\nu_{\{n_i;\,m_i\}}&=&q^{2s(m_R-\delta_{R,1}+1)-sM}\,{[sM]\over s}\,x_1^s\,
\nu_{\{n_i;\,m_i\}}\cr
\noalign{\bigskip}
\Delta^{(M-1)}(H^2_s)\,\nu_{\{n_i;\,m_i\}}&=&-{1\over s}x_1^s\left\{\sum_{i=1}^{R-1}
q^{-2s(t_{i-1}-m_R)+s(n_{R-i}+1)}\,[sn_{R-i}]\right\}\nu_{\{n_i;\,m_i\}}
\label{eq:h1eigzero}
\eea
Equation (\ref{eq:hwlzero1}) and the second of (\ref{eq:hwlzero3}) result from
a straightforward analysis employing the coproduct structure given in \cite{gade2}.
$\Delta^{(M-1)}(E^{1,-}_s)\,\nu_{\{n_i;\,m_i\}}$ and
$\Delta^{(M-1)}(E^{1,-}_{s'})\,\nu_{\{n_i;\,m_i\}}$ differ
only by a complex number for any $s$, $s'$. In contrast, there are $R-1$ linear
independent expressions $\Delta^{(M-1)}(E^{2,-}_s)\,\nu_{\{n_i;\,m_i\}}$.

The action of $U_q\bigl(\widehat{sl}(2\vert2)\bigr)$ on $\nu_{\{n_i;\,m_i\}}$
gives rise to a finite-dimensional irreducible module denoted by 
$W_{x_1,\{n_i;\,m_i\}}$ for $R>1$ and by $W_{x_1,\{M\}}$ for $R=1$.
Below, reference to a general parameter set $\{n_i;\,m_i\}$ will also include
$\{M\}$.
In analogy to the terminology used in \cite{cp},
the eigenvalues of $\Delta^{(M-1)}(h_l)$ and $\Delta^{(M-1)}(H^l_s)$ given in
(\ref{eq:hwlzero2}) and (\ref{eq:hwlzero3})
may be called the highest weight of $W_{x_1,\{n_i;\,m_i\}}$.
The $U_q\bigl(gl(2\vert2)\bigr)$-characters of $W_{x_1,\{n_i;\,m_i\}}$ 
can be inferred from the Schur functions $s_{\lambda\backslash\mu}
({\bf p})$ via
\bea
&&ch_{\{n_i;\,m_i\}}({\bf p})=\cr
\noalign{\medskip}
&&=tr_{W_{x_1,\{n_i;\,m_i\}}}p_0^{{1\over2}\Delta^{(M-1)}
(h_1+h_2+h_3+h_4)}\,p_1^{{1\over2}\Delta^{(M-1)}(h_1-h_2-h_3-h_4)}
\,p_2^{-{1\over2}\Delta^{(M-1)}(h_1+h_2-h_3-h_4)}\,
p_3^{{1\over2}\Delta^{(M-1)}(h_1+h_2+h_3-h_4)}\cr
\noalign{\medskip}
&&\qquad=s_{\lambda\backslash\mu}({\bf p})
\label{eq:chschur1}
\eea
Instead of $W_{x_1,\{n_i;\,m_i\}}$, a level-zero module contained in the tensor product
of evaluation modules
\bea
W^*_{\bar x_1}\otimes W^*_{\bar x_2}\otimes\ldots\otimes W^*_{\bar x_M}
\label{eq:evprod2}
\eea
may be associated to the border strip. To any border strip with
parameters $R$, $\{n_i;\,m_i\}$
there corresponds a reversed border strip with the same number of rows and columns and
$\{\bar n_i;\bar m_i\}$ given by $\bar n_i=n_{R-i}$ for $1\leq i<R$ and $\bar m_i=m_{R-i+1}-
\delta_{i,R}+\delta_{i,1}$ for $1\leq i\leq R$. Both the labeling and the
spectral parameters for the border strip $\{n_i;\,m_i\}$ are mapped onto (a second
type of) labeling and a set of spectral parameters for the reversed border strip
in two steps. First, the number as well as the spectral parameter in the $k$-th box
of the border strip $\{n_i;\,m_i\}$ are attributed to the $M-k$-th box of the
reversed border strip $\{\bar n_i;\,\bar m_i\}$ for $1\leq k\leq M$. In both stripes, the
counting proceeds from the lowest leftmost box to the upper rightmost box. Then
the mapping is completed by substituting each number $l$ in the reversed border
strip by $3-l$, $l=0,1,2,3$.
Thus a row obtains each of the numbers
$0$ and $2$ at most once and a column is occupied at most once by each of the
numbers $1$ and $3$. The first rule of the type $A$ labeling for the numbering is preserved.
Below, this labeling will be referred to as type $B$ labeling.
For the top labeling $B$, the  number $2$ is attributed to the $n_i$ lower boxes
of a column with $n_i+1>1$ boxes. All other boxes are
occupied by the number $3$. The spectral parameter in the $k$-th box is called
$\bar x_k$.
Fig. 4 illustrates the mapping for the
border strip with $R=3$, $m_1=m_2=m_3=2$ and $n_1=3,\,n_2=1$.

\begin{center}
\setlength{\unitlength}{1cm}
\begin{picture}(17,6)
\thicklines

\put(0.2,2){\line(1,0){1.8}}
\put(0.2,2.6){\line(1,0){3}}
\put(1.4,3.2){\line(1,0){1.8}}
\put(2.6,3.8){\line(1,0){0.6}}
\put(2.6,4.4){\line(1,0){1.2}}
\put(2.6,5){\line(1,0){1.2}}

\put(0.2,2){\line(0,1){0.6}}
\put(0.8,2){\line(0,1){0.6}}
\put(1.4,2){\line(0,1){1.2}}
\put(2,2){\line(0,1){1.2}}
\put(2.6,2.6){\line(0,1){2.4}}
\put(3.2,2.6){\line(0,1){2.4}}
\put(3.8,4.4){\line(0,1){0.6}}

\put(0.4,2.2){$0$}
\put(1,2.2){$0$}
\put(1.6,2.2){$0$}
\put(1.6,2.8){$1$}
\put(2.2,2.8){$0$}
\put(2.8,2.8){$0$}
\put(2.8,3.4){$1$}
\put(2.8,4){$1$}
\put(2.8,4.6){$1$}
\put(3.4,4.6){$0$}

\put(4.2,2){\line(1,0){1.8}}
\put(4.2,2.6){\line(1,0){3}}
\put(5.4,3.2){\line(1,0){1.8}}
\put(6.6,3.8){\line(1,0){0.6}}
\put(6.6,4.4){\line(1,0){1.2}}
\put(6.6,5){\line(1,0){1.2}}

\put(4.2,2){\line(0,1){0.6}}
\put(4.8,2){\line(0,1){0.6}}
\put(5.4,2){\line(0,1){1.2}}
\put(6,2){\line(0,1){1.2}}
\put(6.6,2.6){\line(0,1){2.4}}
\put(7.2,2.6){\line(0,1){2.4}}
\put(7.8,4.4){\line(0,1){0.6}}

\put(4.35,2.2){$x_1$}
\put(4.95,2.2){$x_2$}
\put(5.55,2.2){$x_3$}
\put(5.55,2.8){$x_4$}
\put(6.15,2.8){$x_5$}
\put(6.75,2.8){$x_6$}
\put(6.75,3.4){$x_7$}
\put(6.75,4){$x_8$}
\put(6.75,4.6){$x_9$}
\put(7.27,4.6){$x_{10}$}

\put(8.8,2){\line(1,0){1.2}}
\put(8.8,2.6){\line(1,0){1.2}}
\put(9.4,3.2){\line(1,0){0.6}}
\put(9.4,3.8){\line(1,0){1.8}}
\put(9.4,4.4){\line(1,0){3}}
\put(10.6,5){\line(1,0){1.8}}

\put(8.8,2){\line(0,1){0.6}}
\put(9.4,2){\line(0,1){2.4}}
\put(10,2){\line(0,1){2.4}}
\put(10.6,3.8){\line(0,1){1.2}}
\put(11.2,3.8){\line(0,1){1.2}}
\put(11.8,4.4){\line(0,1){0.6}}
\put(12.4,4.4){\line(0,1){0.6}}

\put(9,2.2){$3$}
\put(9.6,2.2){$2$}
\put(9.6,2.8){$2$}
\put(9.6,3.4){$2$}
\put(9.6,4){$3$}
\put(10.2,4){$3$}
\put(10.8,4){$2$}
\put(10.8,4.6){$3$}
\put(11.4,4.6){$3$}
\put(12,4.6){$3$}

\put(12.8,2){\line(1,0){1.2}}
\put(12.8,2.6){\line(1,0){1.2}}
\put(13.4,3.2){\line(1,0){0.6}}
\put(13.4,3.8){\line(1,0){1.8}}
\put(13.4,4.4){\line(1,0){3}}
\put(14.6,5){\line(1,0){1.8}}

\put(12.8,2){\line(0,1){0.6}}
\put(13.4,2){\line(0,1){2.4}}
\put(14,2){\line(0,1){2.4}}
\put(14.6,3.8){\line(0,1){1.2}}
\put(15.2,3.8){\line(0,1){1.2}}
\put(15.8,4.4){\line(0,1){0.6}}
\put(16.4,4.4){\line(0,1){0.6}}

\put(12.87,2.2){$x_{10}$}
\put(13.55,2.2){$x_9$}
\put(13.55,2.8){$x_8$}
\put(13.55,3.4){$x_7$}
\put(13.55,4){$x_6$}
\put(14.15,4){$x_5$}
\put(14.75,4){$x_4$}
\put(14.75,4.6){$x_3$}
\put(15.35,4.6){$x_2$}
\put(15.95,4.6){$x_1$}

\put(0.2,1){Fig. 4:
Top labelling $A$ and spectral parameters for $R=3$, $m_1=m_2=m_3=2$,
$n_1=3,\,n_2=1$ (left)} 

\put(1.4,0.5){and top labelling $B$ for $\bar m_1=3,\,\bar m_2=2,\,\bar m_3=1,\,
\bar n_1=3,\,\bar n_2=1$ and spectral parameters}

\put(1.4,0){$x_i=\bar x_{10-i}$ for the reversed border strip (right)}
\end{picture}
\end{center}
A border strip with parameters $R,\,M$ and $\{\bar n_i;\,\bar m_i\}$, type B
labeling 
and spectral parameters $\bar x_k$ is related to a level-zero module
contained in (\ref{eq:evprod2}).
The replacements
$x_k\rightarrow\bar x_k$ for $1\leq k\leq M$ and $q\rightarrow q^{-1}$ in
(\ref{eq:xspecrel1})-(\ref{eq:xspecrel3}) provide the dependence among the
spectral parameters in (\ref{eq:evprod2}).
With these relations, the tensor product (\ref{eq:evprod2})
has an element $\bar{\nu}_{\{\bar n_i;\,\bar m_i\}}$ satisfying
\bea
\Delta^{(M-1)}(
E^{l,+}_{s})\,\bar{\nu}_{\{\bar n_i;\,\bar m_i\}}&=&0\qquad \mbox{for}\;\;s
\in{\mathbb{Z}},\;l=1,2,3
\label{eq:hwlzero4}
\eea
\bea
\begin{array}{llll}
\Delta^{(M-1)}(h_1)\,\bar{\nu}_{\{\bar n_i;\,\bar m_i\}}&=0\qquad&
\Delta^{(M-1)}(h_2)\,\bar{\nu}_{\{\bar n_i;\,\bar m_i\}}=K\,\bar{\nu}_{\{\bar n_i;
\,\bar m_i\}}\cr
\noalign{\bigskip}
\Delta^{(M-1)}(h_3)\,\bar{\nu}_{\{\bar n_i;\,\bar m_i\}}&=-M\,\bar{\nu}_{\{\bar n_i;
\,\bar m_i\}}\qquad&
\Delta^{(M-1)}(h_4)\,\bar{\nu}_{\{\bar n_i;\,\bar m_i\}}=(M-K)\,\bar{\nu}_{\{\bar n_i;
\,\bar m_i\}}
\end{array}
\label{eq:hwlzero5}
\eea
and
\bea
\Delta^{(M-1)}(H^1_s)\,\bar{\nu}_{\{\bar n_i;\,\bar m_i\}}&=&0\cr
\noalign{\bigskip}
\Delta^{(M-1)}(H^2_s)\,\bar{\nu}_{\{\bar n_i;\,\bar m_i\}}&=&\left\{\begin{array}{cc}
0\qquad&\mbox{if}\;\;R=1\cr
\noalign{\smallskip}
{[s]\over s}\sum_{k\in\bar{\Upsilon}_{\{\bar n_i;\,\bar m_i\}}}\bar x_k^s\;
\bar{\nu}_{\{\bar n_i;\,\bar m_i\}}\qquad&\mbox{if}\;\;R>1\end{array}\right.\cr
\noalign{\bigskip}
\Delta^{(M-1)}(H^3_s)\,\bar{\nu}_{\{\bar n_i;\,\bar m_i\}}&=&-q^s{[s]\over s}
\sum_{k=1}^M\bar x_k^s\;
\bar{\nu}_{\{\bar n_i;\,\bar m_i\}}
\label{eq:hwlzero6}
\eea
$\forall s\neq0$. $\bar{\Upsilon}_{\{\bar n_i;\,\bar m_i\}}$ contains the value $k=M$ for
any $\{\bar n_i;\,\bar m_i\}$ with $R>1$
as well as each value $k<M$ provided that the $k$-th box is
occupied by the number $2$ in the top labeling $B$ except for the lowest of these.
The relations among the spectral parameters yield
\bea
\Delta^{(M-1)}(H^3_s)\,\bar{\nu}_{\{\bar n_i;\,\bar m_i\}}&=&-q^{-2s(\bar m_R-\delta_{R,1}
)+sM}\,{[sM]\over s}\,\bar x_1^s\,\bar{\nu}_{\{\bar n_i;\,\bar m_i\}}\cr
\noalign{\bigskip}
\Delta^{(M-1)}(H^2_s)\,\bar{\nu}_{\{\bar n_i;\,\bar m_i\}}&=&{[s]\over s}\bar x_1^s
\left\{\sum_{i=1}^{R-1}q^{2s(t_{R-1-i}-\bar m_{R})-s(\bar n_i-1)}\,
{[s\bar n_i]\over[s]}\right\}\bar{\nu}_{\{\bar n_i;\,\bar m_i\}}
\label{eq:h3eigzero}
\eea
There exist
$R-1$ linear independent expressions $\Delta^{(M-1)}(E^{2,-}_s)\,\bar{\nu}_{\{
\bar n_i;\,\bar m_i\}}$ 
while $\Delta^{(M-1)}(E^{3,-}_s)\,\bar{\nu}_{\{\bar n_i;\,\bar m_i\}}$ and
$\Delta^{(M-1)}(E^{3,-}_{s'})\,\bar{\nu}_{\{\bar n_i;\,
\bar m_i\}}$ are linear dependant for any $s,\,s'$. In the following, the module
generated by the action of $U'_q\bigl(\widehat{sl}(2\vert2)\bigr)$
on $\bar{\nu}_{\{\bar n_i;\,\bar m_i\}}$ will be denoted by $W^*_{\bar x_1,\{
\bar n_i;\,\bar m_i\}}$ for $\bar R>1$ and by $W^*_{\bar x_1,\{\bar M\}}$
for $\bar R=1$.

Each labeling subject to the rules given below (\ref{eq:evprod2}) can be mapped
onto a component $(\ldots,j^*_5,j^*_3,j^*_1)$ in $\bigl\{\tau^*\bigr\}_{\bar n_{\{
\bar n_i;\,\bar m_i\}}}$ with $j^*_{2r+1}=3\;\forall r\geq M$. Here the number in the $k$-th
box is associated with $j^*_{2(M-k)+1}$. The value of $\bar n_{\{\bar n_i;\,\bar m_i\}}$
indicates the contribution of this component
to the diagonal element of the CTM-Hamiltonian. In terms of the parameters
of the border stripes it is expressed by
\bea
\bar n_{\{\bar n_i;\,\bar m_i\}}=-\sum_{i=1}\bar n_i\Bigl\{-{1\over2}(\bar n_i+1)+
\sum_{j=1}^i(\bar n_j+\bar m_j)\Bigr\}
\label{eq:gradef2}
\eea
In contrast to the correspondence between border stripes and configurations in
$\bigl\{\tau\bigr\}_{n_{\{n_i;m_i\}}}$ described above, labelings of different border stripes
may correspond to the same configuration $\bigl(\ldots\otimes w^*_{j_5}\otimes
w^*_{j_3}\otimes w^*_{j_1}\bigr)$. For example, all border stripes given by
a single row with $\bar m_1$ boxes admit the  top labeling $B$ with the number $3$ in
each box. Thus they are mapped on the configuration $\bigl(\ldots\otimes w^*_3
\otimes w^*_3\otimes w^*_3\bigr)$ for arbitrary $\bar m_1$.

\section{Border stripes and the level-one
 module $V(\Lambda_0)$}
\label{sec:bs2}

To pursue further the relations between border stripes, level-one modules
and the space of states
investigated in Sect. \ref{sec:ctm} and \ref{sec:module}, tensor products of
$W_{x_1,\{n_i;\,m_i\}}$ and $W^*_{\bar x_1,\{\bar n_i;\,\bar m_i\}}$ with independent sets
of $\{n_i;\,m_i\}$ and $\{\bar n_i;\,\bar m_i\}$
need to be considered.
Any two parameterizations $\{n_i;\,m_i\}$ and $\{\bar n_i;\,\bar m_i\}$ of border stripes
with $n_i>0,\,m_i>0$ for $1\leq i<R$ and $\bar n_i>0,\,\bar m_i>0$ for $1\leq i
<\bar R$ may be selected with the properties
\bea
M=\sum_{i=1}^R(n_i+m_i)=\sum_{i=1}^{\bar R}(\bar n_i+\bar m_i)
\label{eq:zerocond1}
\eea
and
\bea
K=\sum_{i=1}^{R-1}n_i\geq\bar K=\sum_{i=1}^{\bar R-1}\bar n_i
\label{eq:zerocond2}
\eea
A generator $a\in U'_q\bigl(\widehat{sl}(2\vert2)\bigr)$ acts on 
$W_{x_1,\{n_i;\,m_i\}}\otimes W^*_{\bar x_1,\{\bar n_i;\,\bar m_i\}}$ via the
coproduct $\Delta^{(2M-1)}(a)$ and equations (\ref{eq:wdef1}), (\ref{eq:wevmo1}).
For a more compact notation, explicit reference to the spectral parameters
in (\ref{eq:wevmo1}) is suppressed in the remainder.
The collection of all vectors in
$W_{x_1,\{n_i;\,m_i\}}\otimes W^*_{\bar x_1,\{\bar n_i;\,\bar m_i\}}$ 
obtained by the action of $U'_q\bigl(\widehat{sl}(2\vert2)\bigr)$
on $\nu_{\{n_i;\,m_i\}}\otimes\bar{\nu}_{\{\bar n_i;\,\bar m_i\}}$ may be denoted
by $W_{x_1,\{n_i;\,m_i\},\,\bar x_1,\{\bar n_i;\,\bar m_i\}}$.
An element $\sigma\in
W_{x_1,\{n_i;\,m_i\}}\otimes W^*_{\bar x_1,\{\bar n_i;\,\bar m_i\}}$
is called a highest weight vector if it is an
eigenvector of $\{h_j,\,H^j_s\}$ and satisfies
$E^{j,+}_s\,\sigma=0\;\forall s$ with $j=1,2,3$.

Due to the first equation and (\ref{eq:hwlzero2}), (\ref{eq:hwlzero5}), the
eigenvalue of
$\Delta^{(2M-1)}(h_1+h_3)$  vanishes on the tensor product
\bea
W_{x_1,\{n_i;\,m_i\}}\otimes W^*_{\bar x_1,\{\bar n_i;\,\bar m_i\}}
\label{eq:evprod3}
\eea
In particular, if
\bea
\bar x_1=q^{-2(M-1)+2(m_R+\bar m_{\bar R}-\delta_{R,1}-\delta_{\bar R,1})}\,x_1
\label{eq:zeroxrel}
\eea
the linear combination $\Delta^{(2M-1)}(H^1_s+H^3_s)$ annihilates
all elements of (\ref{eq:evprod3}) for any $s\neq0$
as a consequence of (\ref{eq:h1h3co}), (\ref{eq:hwlzero3}) and (\ref{eq:hwlzero6}).
Moreover, with (\ref{eq:zeroxrel}) the spectral parameters $x_k$, $\bar x_k$ take
values in the set
\bea
q^{2(m_R-\delta_{R,1})}\,x_1,\;q^{2(m_R-\delta_{R,1}-1)}\,x_1,\ldots,
q^{2(m_R-\delta_{R,1}-M+1)}\,x_1
\label{eq:zeroset}
\eea
Since $x_k\neq x_{k'}$ and $\bar x_k\neq \bar x_{k'}$ if $k\neq k'$, there exists a
$\bar k$ for any $k$ such that $x_k=\bar x_{\bar k}$. 
The following list collects all
combinations of $\{n_i;\,m_i\}$ and $\{\bar n_i;\,\bar m_i\}$ with
$n_{\{n_i;\,m_i\}}+\bar n_{\{\bar n_i;\,\bar m_i\}}>-4$ and the properties
(\ref{eq:zerocond1}), (\ref{eq:zerocond2}):

\bigskip
 
\begin{tabular}{cc|l|l|cc|l}
$R$&$\bar R$&$\;\{n_i;\,m_i\}$\qquad&$\;\{\bar n_i;\,\bar m_i\}$&$n_{\{n_i;\,m_i\}}$&$
\bar n_{\{n_i;\,m_i\}}$&$\;(\bar h_1,\bar h_2,\bar h_3,\bar h_4)_{h.w.}$\\
\hline
&&&&&&\\
$1$&$1$&$\;m_1=1$&$\;\bar m_1=1$&$-1$&$0$&$\;(1,0,-1,2)$\\
&&&&&&\\
$2$&$1$&$\;m_1=1,\,m_2=0,\,n_1=1$\qquad&$\;\bar m_1=2$&$-2$&$0$&$\;(2,-1,-2,3)$\\
&&&&&&\\
$2$&$1$&$\;m_1=1,\,m_2=0,\,n_1=2$&$\;\bar m_1=3$&$-3$&$0$&$\;(3,-2,-3,4)$\\
&&&&&&\\
$1$&$1$&$\;m_1=2$&$\;\bar m_1=2$&$-3$&$0$&$\;(2,0,-2,4)$\\
&&&&&&\\
$2$&$2$&$\;m_1=1,\,m_2=0,\,n_1=1$&$\;\bar m_1=1,\,\bar m_2=0,\,\bar n_1=1$&$-2$&$-1$&
$\;(2,0,-2,2)$
\end{tabular}

\bigskip

Here the rightmost column specifies the $U_q\bigl(gl(2\vert2)\bigr)$-weight
of $\nu_{\{n_i;\,m_i\}}\otimes \bar{\nu}_{\{\bar n_i;\,\bar m_i\}}$. According to
(\ref{eq:hwlzero2}), (\ref{eq:hwlzero5}) this is given by $(M,\bar K-K,-M,
2M-K-\bar K)$. The remaining part of the highest weight
follows from
\bea
\Delta^{(2M-1)}\bigl(H^l_s\bigr)\,(\nu_{\{n_i;\,m_i\}}\otimes\bar{\nu}_{\{
\bar n_i;\,\bar m_i\}})=\Delta^{(M-1)}\bigl(H^l_s\bigr)\nu_{\{n_i;\,m_i\}}\otimes
\bar{\nu}_{\{\bar n_i;\,\bar m_i\}}+\nu_{\{n_i;\,m_i\}}\otimes \Delta^{(M-1)}
\bigl(H^l_s\bigr)\bar{\nu}_{\{\bar n_i;\,\bar m_i\}}
\label{eq:hwwwstar}
\eea
for $l=1,2,3$
and (\ref{eq:hwlzero3}), (\ref{eq:hwlzero6}). In general, 
the module $W_{x_1,\{n_i;\,m_i\}}\otimes W^*_{\bar x_1,\{\bar n_i;\,\bar m_i\}}$
with $x_1$ and $\bar x_1$ related by (\ref{eq:zeroxrel}) is reducible. This is
easily demonstrated for the simplest case $W_{x_1}\otimes W^*_{\bar x_1}$ with
$x_1=\bar x_1$. The highest weight vector conditions are satisfied by the
two vectors
\bea
w_0\otimes w^*_3\qquad\mbox{and}\qquad
\sigma_0=\sum_{i=1}^3w_i\otimes w^*_i=q^{-1}\Delta(f_3f_2f_1-f_1f_2f_3)\;w_0\otimes
w^*_3
\label{eq:hwvec1}
\eea
At $x_1=\bar x_1$, acting with the coproduct of the $U'_q\bigl(\widehat{sl}
(2\vert2)\bigr)$-generators on
$w_0\otimes w^*_3$ does not produce the vector $w_0\otimes w^*_0-w_1\otimes w^*_1
+w_2\otimes w^*_2-w_3\otimes w^*_3$. In turn, action
of $U'_q\bigl(\widehat{sl}(2\vert2)\bigr)$
on the latter gives rise to all $w_l\otimes w^*_{l'}$ with $0\leq l,l'\leq3$. Taking
the quotient of all vectors emerging under the action of $U'_q\bigl(\widehat{sl}(2
\vert2)\bigr)$ on $w_0\otimes w^*_3$ by the vector $\sigma_0$ provides an
irreducible level-zero module $W_{x_1,\{1\},\,x_1,\{1\}}$. 
The vector $\sigma_0$ may be seen as the first example of a
particular type of highest weight vector occurring in each tensor product
$W_{x_1,\{n_i;\,m_i\}}\otimes W^*_{\bar x_1,\{\bar n_i;\,\bar m_i\}}$
with $R=2$, $\bar R=1$, $m_1=1,\;m_2=0$. These values correspond to border stripes
given by the single column of $M=K+1=n_1+1$ boxes for the $W$-part and the single
row of $M$ boxes for the $W^*$-part. Their spectral parameters are related by
$x_k=\bar x_k=q^{-2(k-1)}x_1$ for $1\leq k\leq K+1$. The highest weight vector
properties are satisfied by
\bea
\nu_{\{K;\,1,0\}}\otimes\bar{\nu}_{\{K+1\}}=\left\{\sum_{k=1}^K(-1)^kq^{-k}
w_1^{\otimes(K-k)}\otimes w_0\otimes w_1^{\otimes k}\right\}\;\otimes {w^*_3}^{
\otimes(K+1)}
\label{eq:hwvec2}
\eea
with $U_q\bigl(gl(2\vert2)\bigr)$-weight $(K+1,-K,-K-1,K+2)$ and by 
\bea
\sigma_K&=&\Delta^{(2M-1)}\bigl(f_3f_2f_1-f_1f_2f_3\bigr)\cdot\cr
\noalign{\bigskip}
&&\cdot\Delta^{(2M-1)}
\left\{(e_1e_0)^K+\sum_{k=1}^KA_k(f_3f_2)^k(e_1e_0)^{K-k}+
\sum_{k=1}^KB_k(f_2f_3)^k(e_1e_0)^{K-k}\right\}\bigl(\nu_{\{K;\,1,0\}}\otimes
\bar{\nu}_{\{K+1\}}\bigr),\cr
\noalign{\bigskip}
&&\qquad A_k=(-1)^kx_1^kq^{-k(K-1)}\,{[K][K-1]\ldots[K-k+1]\over[k][k-1]\ldots[1]}
\qquad\qquad B_k={k+1\over[k+1]}\,A_k
\label{eq:hwvec3}
\eea
with $U_q\bigl(gl(2\vert2)\bigr)$-weight $(0,0,0,0)$. The expression in the
large parenthesis in (\ref{eq:hwvec3}) is annihilated by $\Delta^{(2M-1)}(e_j)$
with $j=1,2,3$
but not by $\Delta^{(2M-1)}\bigl(E^{j,+}_s\bigr)$ with general $s$.
To include the value $K=0$, it should be replaced by the unit.
$\sigma_K$ is a one-dimensional $U'_q\bigl(\widehat{sl}(2\vert2)\bigr)$-module. 
With the spectral parameters specified above,
no further elements with highest weight vector properties are found in
$W_{x_1,\{K;\,1,0\},\,x_1,\{K+1\}}$.
For any $K$, the quotient $W_{x_1,\{K;\,1,0\},\,x_1,\{K+1\}}/\sigma_K$
forms an irreducible level-zero module of $U'_q\bigl(\widehat{sl}(2\vert2)
\bigr)$. Removing $\sigma_K$ corresponds to elimination of a vector of weight
$(0,0,0,0)$ in the level-one module $V(\Lambda_0)$ at grade $-K-1$. This
is done by imposing the
condition $\bigl(H^1_{-K-1}+H^3_{-K-1}\bigr)\,\hat{\lambda}_0=0$. Here $\hat{\lambda}_0$
denotes the highest weight vector of $V(\Lambda_0)$ with $h_j\,\hat{\lambda}_0
=\delta_{j,0}\hat{\lambda}_0$, $0\leq j\leq4$.

In the remaining two cases listed in the above table, application of the $U'_q\bigl(
\widehat{sl}(2\vert2)\bigr)$-generators on $\nu_{\{n_i,m_i\}}\otimes \bar{\nu}_{\{
\bar n_i,\bar m_i\}}$ produces only a part of 
$W_{x_1,\{n_i;\,m_i\}}\otimes W^*_{\bar x_1,\{\bar n_i;\,\bar m_i\}}$.
If $R=\bar R=1$, equation (\ref{eq:zeroxrel}) leads to $x_k=q^{2(k-1)}x_1$, $\bar x_k
=q^{2(M-k)}x_1$ with $1\leq k\leq M$. With these relations,
\bea
\Delta^{(2M-1)}\left(x_1^{-1}e_0-f_3f_2f_1+f_1f_2f_3+qf_2f_1f_3\right)(\nu_{\{M\}}
\otimes\bar{\nu}_{\{M\}})=0\qquad\;\;M=1,2,3,\ldots
\eea
A highest weight vector with $U_q\bigl(gl(2\vert2)
\bigr)$-weight $(M-1,0,-M+1,2M-2)$ 
is found in $W_{x_1,\{M\}}\otimes W^*_{q^{2(M-1)}x_1,\{M\}}$
but not present in $W_{x_1,\{M\},\,q^{2(M-1)}x_1,\{M\}}$.
For $M=2$ it is given by
\bea
\sum_{l=0}^3(-1)^{\vert l\vert}w_0\otimes w_l\otimes w^*_l\otimes w^*_3-q^{-1}
\sum_{l=0}^3q^{-\delta_{l,0}-\delta_{l,3}}\,w_l\otimes w_0\otimes w^*_3\otimes w^*_l
\label{eq:hwvec4}
\eea
Furthermore, the tensor product 
$W_{x_1,\{2\}}\otimes W^*_{q^2x_1,\{2\}}$ contains the
vector $\tilde{\sigma}=\bigl(w_3\otimes w_1-qw_1\otimes w_3\bigr)\otimes
\bigl(w^*_2\otimes w^*_0-qw^*_0\otimes w^*_2\bigr)$ with $U_q\bigl(gl(2\vert2)
\bigr)$-weight $(0,0,0,-2)$ which is not reached by the action of $U_q\bigl(
\widehat{sl}(2\vert2)\bigr)$ on $\nu_{\{2\}}\otimes\bar{\nu}_{\{2\}}$ or on
(\ref{eq:hwvec4}). Application of $U'_q\bigl(\widehat{sl}(2\vert2)\bigr)$-generators
on $\tilde{\sigma}$ leads to (\ref{eq:hwvec4}) but not to $\nu_{\{2\}}\otimes
\bar{\nu}_{\{2\}}$. As a result, sixteen elements of
$W_{x_1,\{2\}}\otimes W^*_{q^2x_1,\{2\}}$ are not
found in $W_{x_1,\{2\},\,q^{2}x_1,\{2\}}$ which contains 48 
different vectors.

For $R=\bar R=2$ and $m_1-1=m_2=\bar m_1-1=\bar m_2=0$, the spectral parameters
are related by
$x_k=q^{-2(k-1)}x_1$ and $\bar x_k=q^{-2(M-k)}x_1$ with $1\leq k\leq M$. Then
\bea
\Delta^{(2M-1)}\bigl(x_1^{-1}e_3e_1e_0-qf_2\bigr)(\nu_{\{K;\,0,0\}}\otimes\bar{\nu}_{
\{K;\,0,0\}})=0\qquad\;\;K=1,2,3,\ldots
\eea
and $W_{x_1,\{K;\,1,0\}}\otimes W^*_{q^{-2K}x_1,\{K;\,1,0\}}$
has a vector with $U_q\bigl(gl(2\vert2)\bigr)$-weight $(K,0,-K,2)$ not
found in $W_{x_1,\{K;\,1,0\},\,q^{-2K}x_1,\{K;\,1,0\}}$. Though it
is annihilated by $\Delta^{(2K+1)}(e_j)$ with $j=1,2,3$, it does not satisfy the highest weight
vector properties. At $K=1$ it is
given by
\bea
\bigl(w_0\otimes w_1-qw_1\otimes w_0\bigr)\otimes\bigl(w^*_1\otimes w^*_3-qw^*_3
\otimes w^*_1\bigr)-\bigl(w_0\otimes w_2-qw_2\otimes w_0\bigr)\otimes\bigl(w^*_2
\otimes w^*_3+qw^*_3\otimes w^*_2\bigr)
\label{eq:hwvec5}
\eea
In addition, $W_{x_1,\{1;\,1,0\}}\otimes W^*_{q^{-2}x_1,\{1;\,1,0\}}$ contains the
vector $\hat{\sigma}=\bigl(w_0\otimes w_2-qw_2\otimes w_0\bigr)\otimes\bigl(w^*_1
\otimes w^*_3-qw^*_3\otimes w^*_1\bigr)$ which does not result from the action of
$U'_q\bigl(\widehat{sl}(2\vert2)\bigr)$ on (\ref{eq:hwvec5}) or $\nu_{\{1;\,1,0\}}
\otimes\bar{\nu}_{\{1;\,1,0\}}$. Action of the $U'_q\bigl(\widehat{sl}(2\vert2)
\bigr)$-generators on $\hat{\sigma}$ creates the whole tensor product 
$W_{x_1,\{1;\,1,0\}}\otimes W^*_{q^{-2}x_1,\{1;\,1,0\}}$. Sixteen vectors are
present in $W_{x_1,\{1;\,1,0\}}\otimes W^*_{q^{-2}x_1,\{1;\,1,0\}}$ but not in
$W_{x_1,\{1;\,1,0\},\,q^{-2}x_1,\{1;\,1,0\}}$ which has 48 different
vectors. No highest weight vectors  besides $\nu_{\{2\}}\otimes\bar{\nu}_{\{2\}}$
and $\nu_{\{2;\,1,0\}}\otimes\bar{\nu}_{\{2;\,1,0\}}$ are found in
$W_{x_1,\{2\},\,q^2x_1,\{2\}}$ and $W_{x_1,\{1;\,1,0\},\,q^{-2}x_1,\{1;\,1,0\}}$,
respectively.

The quotient of $W_{x_1,\{n_i;\,m_i\},\,\bar x_1,\{\bar n_i;\,\bar m_i\}}$ by all
highest weight
vectors $\rho_{x_1,\{n_i;\,m_i\},\,\bar x_1,\{\bar n_i;\,\bar m_i\}}\in
W_{x_1,\{n_i;\,m_i\},\,\bar x_1,\{\bar n_i;\,\bar m_i\}}$ different from
$\nu_{\{n_i;\,m_i\}}\otimes\bar{\nu}_{\{\bar n_i;\,\bar m_i\}}$ and the vectors obtained
by the action of $U'_q\bigl(\widehat{sl}(2\vert2)\bigr)$ on 
$\rho_{x_1,\{n_i;\,m_i\},\,\bar x_1,\{\bar n_i;\,\bar m_i\}}$ may be called
$\hat W_{x_1,\{n_i;\,m_i\},\,\bar x_1,\{\bar n_i;\,\bar m_i\}}$. It is easily verified
that for $n_{\{n_i;\,m_i\}}+\bar n_{\{\bar n_i;\,\bar m_i\}}\geq-3$ the $U_q\bigl(gl
(2\vert2)\bigr)$-weights of the vectors present in 
$\hat W_{x_1,\{n_i;\,m_i\},\,\bar x_1,\{\bar n_i;\,\bar m_i\}}$ are in one-to-one
correspondence with the $U_q\bigl(gl(2\vert2)\bigr)$-weights of the vectors found in
the level-one
$U_q\bigl(\widehat{sl}(2\vert2)\bigr)/{\cal H}$-module $V(\Lambda_0)$ at grade
$n_{\{n_i;\,m_i\}}+\bar n_{\{\bar n_i;\,\bar m_i\}}$. This result may be assumed to
hold true at any grade.
\vspace{0,5cm}

{\it Conjecture II}: The character of the $U_q\bigl(\widehat{sl}(2\vert2)\bigr)/
{\cal H}$-module $V(\Lambda_0)$ at level one can be written
\bea
ch_{V(\Lambda_0)}(\rho,{\bf p})=1+\sideset{}{'}
\sum_{\{n_i;\,m_i\},\{\bar n_i;\,\bar m_i\}}
\rho^{n_{\{n_i;\,m_i\}}+\bar n_{\{\bar n_i;\,\bar m_i\}}}\;ch_{\{n_i;\,m_i\},
\{\bar n_i;\,\bar m_i\}}({\bf p})
\label{eq:conj2}
\eea
Here $\sum'$ denotes the sum over all $\{n_i;\,m_i\}$ and $\{\bar n_i;\,\bar m_i\}$
with $\sum_i(n_i+m_i)=\sum_i(\bar n_i+\bar m_i)$ and $\sum_in_i\geq\sum_i\bar n_i$.
The character on the rhs of (\ref{eq:conj2}) is defined by
\bea
ch_{\{n_i;\,m_i\},\{\bar n_i;\,\bar m_i\}}&\equiv&
tr_{\hat W_{x_1,\{n_i;\,m_i\},\bar x_1,\{\bar n_i;\,\bar m_i\}}}p_0^{{1\over2}
\Delta^{(2M-1)}(h_1+h_2+h_3+h_4)}
p_1^{{1\over2}\Delta^{(2M-1)}(h_1-h_2-h_3-h_4)}\cdot\cr
\noalign{\bigskip}
&&\qquad\qquad\qquad\qquad\qquad\qquad\qquad \cdot\, p_2^{-{1\over2}\Delta^{(2M-1)}
(h_1+h_2-h_3-h_4)}p_3^{{1\over2}\Delta^{(2M-1)}(h_1+h_2+h_3-h_4)}
\label{eq:co2ch}
\eea
with $x_1$ and $\bar x_1$ related by (\ref{eq:zeroxrel}).
Its general evaluation requires further investigation. The submodules
$\hat W_{x_1,\{n_i;\,m_i\},\bar x_1,\{\bar n_i;\,\bar m_i\}}$ are  special cases of
the modules introduced in the last section of \cite{dipe}.
Apparently, their structure has not been
studied so fare. In this context, it may
be helpful to observe that the $U_q\bigl(\widehat{gl}(2\vert2)\bigr)$-characters
in \cite{cheng} are written as sums of products of supersymmetric (Hook) Schur functions. 
Multiplying (\ref{eq:conj2})
by the factor $\prod_{t=1}^{\infty}(1-q^t)^{-2}$ gives an expression for
the $U_q\bigl(\widehat{gl}(2\vert2)\bigr)$-character of $V(\Lambda_0)$.
It is easily checked that at grades $\geq-3$ the expression reproduces the character formula
3.14 in \cite{kacwak2} with $s=0$.
Combinatorial identities can be obtained from comparison with the results in 
\cite{cheng}.

\section{Border stripes and the  module $\tilde V(\Lambda_0)$}
\label{sec:bs3}
\subsection{Infinite border stripes}
\label{sec:bos3a}

A one-to-one mapping between the complete space of configurations
$(\ldots\otimes w_{j_4}\otimes w^*_{j_3}\otimes w_{j_2}\otimes w^*_{j_1})$
and the semi-standard super tableaux of border stripes can be achieved by
identifying the type $B$ labelings of infinite border stripes with the components
$(\ldots\otimes w^*_{j_5}\otimes w^*_{j_3}\otimes w^*_{j_1})$. 
To this aim, the infinite border strip parameterized
by $\bar R,\;\{\bar n_i,\,\bar m_i\}$ with $\bar m_{\bar R}=0$
 is introduced as the limit $\bar m_{\bar R}\to
\infty$ of the border stripes characterized by fixed values of $\bar R>1$ 
and (finite) $\bar n_i$, $\bar m_i$, $1\leq i\leq \bar R-1$.
For $\bar R=1$, the infinite border strip parameterized by $\{1\}$ is a half-infinite row.
The values of $\bar n_{\{\bar n_i;\,\bar m_i
\}}$ in (\ref{eq:gradef2}) to be equated with the diagonal elements of the CTM  
 don't depend on the value of $\bar m_{\bar R}$.
In contrast to the infinite
border stripes considered for $U_q\bigl(\widehat{sl}(N)\bigr)$-models in
\cite{naka1}, each border strip $\bar R,\;\{\bar n_i;\,\bar m_i\}$
 accommodates an infinite number of semi-standard super
tableaux. Due to the boundary condition imposed on the configurations,
almost all boxes are occupied by the number $3$. For $l=0,1,2$ the multiplicity
of the number $l$ in a given type $B$ labeling $T_B$ may be denoted by $N_l(T_B)$.
The values of $N_0(T_B)$ and $N_2(T_B)$ are 
bound from above for a given set of parameters
$\bar R$, $\bar n_i$, $\bar m_i$, $1\leq i\leq \bar R-1$.
Hence a skew Schur function $\tilde s_{\lambda\backslash\mu}(\tilde{\bf p})$
can be introduced as
\bea
\tilde s_{\lambda\backslash\mu}(\tilde{\bf p})\equiv\sum_{T_B\in T
_B(\lambda\backslash\mu)}\tilde p_0^{N_0(T_B)}\,\tilde p_1^{N_1(T_B)}\,
\tilde p_2^{N_2(T_B)}
\label{eq:schurdef2}
\eea
with $\vert \tilde p_1\vert<1$. Here the partitions $\lambda=(\lambda_1,\lambda_2,\ldots,
\lambda_{\bar K+1})$ and $\mu=(\mu_1,\mu_2,\dots,\mu_{\bar K+1})$ satisfy
$\lambda_{\bar K+1}=1$ and $\mu_{\bar K+1}=0$. The remaining values $(\lambda_1,\lambda_2,
\ldots,\lambda_{\bar K})$, $(\mu_1,\mu_2,\ldots,\mu_{\bar K})$ are
related to the sets $\{\bar n_i\}_{1\leq i\leq \bar R-1}$ and $\{\bar m_i
\}_{1\leq i\leq \bar R-1}$ by (\ref{eq:pprel}) with $R$, $K$, $m_i$, $n_i$
replaced by $\bar R$, $\bar K$, $\bar m_i$, $\bar n_i$ for $1\leq i\leq R-1$ and
$m_R$ set to zero. If $\bar R=1$, the partitions are $\lambda=(1)$ and $\mu=(0)$.
$T_B(\lambda\backslash\mu)$
denotes the set of all type $B$ labelings of the infinite border strip
characterized by $\bar R,\, \{\bar n_i,\,\bar m_i\}$ with $\bar m_{\bar R}=\delta_{\bar R,1}$. 
A determinant formula for
$\tilde s_{\lambda\backslash\mu}(\tilde{\bf p})$ is written as
\bea
\tilde s_{\lambda\backslash\mu}(\tilde{\bf p})=det\bigl(\tilde E\bigr)_{1
\leq i,j\leq \bar K+1}
\label{eq:schur3}
\eea
where $\tilde E_{i,j}=\tilde e_{\lambda_i-\mu_j-i+j}$ for $i,j\leq \bar K$ with
$\tilde e_0=1$, $\tilde e_{-m}=0$ and
\bea
\tilde e_m=\tilde p_1^m+\tilde p_0\tilde p_1^{m-1}+{1+\tilde p_2\over 1-\tilde p_1}
\bigl(1-\tilde p_1^m+\tilde p_0-\tilde p_0\tilde p_1^{m-1}\bigr)
\eea
for $m=1,2,3,\ldots$ and
 $\tilde E_{\bar K+1,i}=0$ for $1\leq i<\bar K$, $\tilde E_{\bar K+1,\bar K}=1$,
$\tilde E_{i,\bar K+1}=(1-\tilde p_1)^{-1}(1+\tilde p_0)(1+\tilde p_2)$.
For any type $B$ labeling $T_B$ of an infinite border strip, a
$U_q\bigl(gl(2\vert2)\bigr)$-weight may be defined by
\bea
\bigl(-N_0(T_B)-N_1(T_B),\,N_1(T_B)+N_2(T_B),\,N_0(T_B)+N_1(T_B),\,-2N_0(T_B)-N_1(T_B)-
N_2(T_B)\bigr)
\label{eq:typebw}
\eea 
The boxes in the infinite border strip may be counted starting from the upper
rightmost box. Then the number in the $k$-th box is associated with the entry
$j^*_{2k-1}$ of the component $(\ldots,j^*_5,j^*_3,j^*_1)$.
The restricted type $A$ labelings of finite border stripes are
associated with the components $(\ldots,j_6,j_4,j_2)\neq(\ldots,3,3,3)$ as specified
in Sect. \ref{sec:bs1}. 
Thus pairs of border stripes parameterized by $R,\,\{n_i;\,m_i\}$
and $\bar R,\,\{\bar n_i;\,\bar m_i\}$ with $\bar m_{\bar R}=\delta_{\bar R,1}$
allow to characterize the space of half-infinite configurations contained
in $\{\tau\}_n\otimes\{\tau^*\}_{\bar n}$ with $n\neq0$. A labeling
$T_{A,B}$ of such a pair consists of a restricted type $A$ labeling $T'_A$ of the
finite border strip $R,\,\{n_i;\,m_i\}$ and of a type $B$ labeling $T_B$ of the
infinite border strip $\bar R,\,\{\bar n_i;\,\bar m_i\}$.
The value $R=0$ related to the configuration $(\ldots,3,3,3)$
may be included in the set of finite border stripes. In this case, the labelings
$T_{A,B}$ are given by the type $B$ labelings $T_B$ of the infinite border strip.
Clearly, the labelings $T_{A,B}$ are in one-to-one correspondence with
the half-infinite configurations $(\ldots,j_4,j^*_3,j_2,j^*_1)$ with $j_r=3$ for
almost all $r$. The number 
$n_{\{n_i;\,m_i\}}+\bar n_{\{\bar n_i;\,\bar m_i\}}$ for $R>0$ (or
$\bar n_{\{\bar n_i;\,\bar m_i\}}$ for $R=0$) equals the diagonal element of
the corner transfer matrix Hamiltonian acting on the corresponding configuration.
The multiplicities of the
numbers $l=0,1,2$ in the restricted type $A$ labeling $T'_A$ may be denoted by
$N_l(T'_A)$. Then a $U_q\bigl(gl(2\vert2)\bigr)$-weight of the labeling $T_{A,B}$
is introduced as the sum of (\ref{eq:typebw}) and
\bea
\bigl(N_0(T'_A)+N_1(T'_A),\,-N_1(T'_A)-N_2(T'_A),\,-N_0(T'_A)-N_1(T'_A),\,2N_0(T'_A)+
N_1(T'_A)+N_2(T'_A)\bigr)
\label{eq:typeaw}
\eea
If $R=0$, (\ref{eq:typeaw}) is replaced by $(0,0,0,0)$.

This weight coincides with the $U_q\bigl(gl(2\vert2)\bigr)$-weight of the
half-infinite configuration corresponding to the labeling $T_{A,B}$
provided that the former is evaluated with respect to the reference weight
$\bar h^A$.
The set of the $U_q\bigl(gl(2\vert2)\bigr)$-weights attributed to all labelings
$T_{A,B}$ of a given pair of border stripes $R,\,\{n_i;\,m_i\}$ and $\bar R,\,\{
\bar n_i;\,\bar m_i\}$ with $\bar m_{\bar R}=\delta_{\bar R,1}$ will be denoted by
$\Xi_{\{n_i;\,m_i\},\{\bar n_i;\,\bar m_i\}}$. For $R=0$, the set of the weights
(\ref{eq:typebw}) attributed to all labelings $T_B$ is denoted by
$\Xi_{\{\bar n_i;\,\bar m_i\}}$. The simplest set $\Xi_{\{1\}}$ coincides with
$\sigma(0,0,0,0)$ introduced in (\ref{eq:setdef}).

Conjecture {\it I} and the correspondence between labelings $T_{A,B}$
and the half-infinite configurations imply a decomposition
of the affine character of $\tilde V(\Lambda_0)$ in terms of skew Schur functions.
\vspace{0.5cm}
 
{\it Conjecture III}\,:
A spectral decomposition for the character (\ref{eq:pathchar})
 of the $U_q\bigl(\widehat{sl}(2\vert2)\bigr)/{\cal H}$-module
$\tilde V(\Lambda_0)$ at level one is given by
\bea
&&ch_{\tilde V(\Lambda_0)}(\rho,p_0,p_1,p_2)
=\biggl(1+\sum_{\lambda\backslash\mu}\rho^{n_{\lambda\backslash\mu}}\,
s'_{\lambda\backslash\mu}(p_0,p_1,p_2,1)\biggr)
\cdot\sum_{\bar{\lambda}\backslash\bar{\mu}}\rho^{\bar n_{\bar{\lambda}
\backslash\bar{\mu}}}\,
\tilde s_{\bar{\lambda}\backslash\bar{\mu}}
\bigl(p_0^{-1},p_1^{-1},p_2^{-1}\bigr)
\label{eq:charf1}
\eea
where $\vert p_1\vert>1$. In (\ref{eq:charf1}), $\sum_{\lambda\backslash\mu}$ means
the sum over all partitions related to
finite border stripes $R,\,\{n_i;\,m_i\}$ with $R>0$.
$\sum_{\bar{\lambda}\backslash\bar{\mu}}$ is
the sum over all partitions related to infinite border stripes 
$\bar R,\;\{\bar n_i;\,\bar m_i\}$ with $\bar m_{\bar R}=\delta_{\bar R,1}$.
The values of $n_{\lambda\backslash\mu}$ and $\bar n_{\bar{\lambda}\backslash
\bar{\mu}}$ are given by equations (\ref{eq:gradef1}) and (\ref{eq:gradef2})
with the parameters $\{n_i;\,m_i\}$ and $\{\bar n_i;\,\bar m_i\}$ associated to
$\lambda\backslash\mu$ and $\bar{\lambda}\backslash\bar{\mu}$, respectively.

Infinite Young skew diagrams of the type described above and the associated
characters or skew Schur functions have not been introduced so fare. In the
following section, another classification of the half-infinite
configurations is proposed.

\subsection{Infinite-dimensional $U_q\bigl(gl(2\vert2)\bigr)$-modules}
\label{sec:bos3b}

The aim of this section is a description of the complete space $\Omega_A$
of configurations $(\ldots\otimes w_{j_4}\otimes w^*_{j_3}\otimes w_{j_2}
\otimes w^*_{j_1})$ in terms of level-zero modules of $U_q\bigl(\widehat{sl}(2
\vert2)\bigr)$. These modules can be constructed from finite tensor products of
both finite- and infinite-dimensional evaluation modules.
A suitable infinite-dimensional evaluation module is obtained from the
$U'_q\bigl(\widehat{sl}(2\vert2)\bigr)$-module $V$ with basis
$\bigl\{v_{j,t}\bigr\}_{0\leq j\leq3,\,t\in{\mathbb{N}}_0}$ and
\bea
\begin{array}{llll}
h_0\,v_{j,t}&=-(t+1-\delta_{j,0})\,v_{j,t}\qquad\qquad&
h_1\,v_{j,t}&=-(t+1-\delta_{j,2})\,v_{j,t}\cr
\noalign{\bigskip}
h_2\,v_{j,t}&=(t+1-\delta_{j,0})\,v_{j,t}\qquad\qquad&
h_3\,v_{j,t}&=(t+1-\delta_{j,2})\,v_{j,t}\cr
\noalign{\bigskip}
\noalign{\bigskip}
f_3\,v_{3,t}&=v_{2,t+1}\qquad\qquad&e_3\,v_{2,t+1}&=[t+1]\,v_{3,t}\cr
\noalign{\bigskip}
f_3\,v_{0,t}&=-v_{1,t}\qquad\qquad&e_3\,v_{1,t}&=-[t+1]\,v_{0,t}\cr
\noalign{\bigskip}
f_2\,v_{2,t}&=v_{3,t}\qquad\qquad&e_2\,v_{3,t}&=[t+1]\,v_{2,t}\cr
\noalign{\bigskip}
f_2\,v_{1,t}&=-{[t+1]\over[t+2]}\,v_{0,t+1}\qquad\qquad&e_2\,v_{0,t+1}&=-[t+2]\,v_{1,t}\cr
\noalign{\bigskip}
f_1\,v_{3,t}&=v_{0,t}\qquad\qquad&e_1\,v_{0,t}&=-[t+1]\,v_{3,t}\cr
\noalign{\bigskip}
f_1\,v_{2,t}&=v_{1,t-1}\qquad\qquad&e_1\,v_{1,t}&=-[t+1]\,v_{2,t+1}\cr
\noalign{\bigskip}
f_0\,v_{1,t}&=q^{-2}[t+1]^2\,v_{2,t}\qquad\qquad&e_0\,v_{2,t}&=-q^2[t+1]^{-1}\,v_{1,t}\cr
\noalign{\bigskip}
f_0\,v_{0,t+1}&=q^{-2}[t+1][t+2]\,v_{3,t}\qquad\qquad&e_0\,v_{3,t}&=-q^2[t+2]^{-1}
\,v_{0,t+1}
\end{array}
\label{eq:vdef1}
\eea
A $U_q\bigl(gl(2\vert2)\bigr)$-weight matching the first choice in (\ref{eq:weightch})
for the weights of the configurations is given by
\bea
h_4\,v_{j,t}=-(t+1+\delta_{j,0}+2\delta_{j,1})\,v_{j,t}
\label{eq:vdef2}
\eea
On $V_z=V\otimes {\mathbb{C}}[z,z^{-1}]$ a $U_q\bigl(\widehat{sl}(2\vert2)\bigr)$-structure
is defined by (\ref{eq:wevmo1}) with $v_j$ replaced by $v_{j,t}$. Adding $(0,0,0,0)$
to the $U_q\bigl(gl(2\vert2)\bigr)$-weights of $V_z$ yields the set $\Xi_{\{1\}}$.
In terms of $V_z$, level-zero modules with the $U_q\bigl(gl(2\vert2)\bigr)$-weights
given by the weight sets $\Xi_{\{\bar n_i;\,\bar m_i\}}$ with $\bar R>1$ can be specified.
There are two type $B$ labelings of the infinite border strip $\bar R,\,\{\bar n_i;\,
\bar m_i\}$ attributing only the numbers $2$
and $3$ to the boxes. One of them
coincides with the top labeling $B$ for the finite border strip $\bar R,\,\{\bar n_i;\,
\bar m_i\}$
in the first $\bar M=\sum_i(\bar n_i+\bar m_i)$ boxes (counted from the right end).
All remaining boxes receive the number $3$. The $U_q\bigl(
gl(2\vert2)\bigr)$-weight (\ref{eq:typebw})  attributed to this labeling provides
the highest weight for a level-zero module related to $\Xi_{\{\bar n_i;\,\bar m_i\}}$. 
This weight is given by $(0,\bar K,0,-\bar K)$ with $\bar K=\sum_i\bar n_i$.
Replacing the number $3$ in the upper rightmost box
by the number $2$ yields the second labeling.
Its $U_q\bigl(gl(2\vert2)\bigr)$-weight $(0,\bar K+1,0,-\bar K-1)$
provides a highest weight for another level-zero module.

A simple example is provided by the infinite border strip $\bar R=2,\,\{\bar K;1,0\}$.
The weights of the set $\Xi_{\{\bar K;1,0\}}$ coincide with
the weights of the two $U_q\bigl(gl(2\vert2)\bigr)$-modules with highest weights
$(0,\bar K,0,-\bar K)$ and $(0,\bar K+1,0,-\bar K-1)$. In terms of the evaluation
modules $V_z$, the related level-zero $U_q\bigl(\widehat{sl}(2\vert2)
\bigr)$-modules are expressed by
\bea
V_{y_1}\otimes V_{q^2y_1}\otimes V_{q^4y_1}\otimes\ldots\otimes V_{q^{2(\bar K-1)}
y_1}\qquad\mbox{and}\qquad
V_{y_1}\otimes V_{q^2y_1}\otimes V_{q^4y_1}\otimes\ldots\otimes V_{q^{2\bar K}y_1}
\label{eq:vmod1}
\eea
A second example is the infinite border strip $\bar R=2,\,\{1;\bar m_1>1,0\}$. 
The weights of the set $\Xi_{\{1;\bar m_1,0\}}$
exactly match the weights of the $U_q\bigl(gl(2\vert2)\bigr)$-modules with
highest weights $(0,1,0,-1)$ and $(-k,2,k,-2)$ with $0\leq k\leq \bar m_1
-1$. The related level-zero $U_q\bigl(\widehat{sl}(2\vert2)\bigr)$-modules read
\bea
V_{y_1}\qquad\mbox{and}\qquad V_{y_1}\otimes V_{q^{2\bar m_1}y_1}
\label{eq:vmod2}
\eea
Generally, an arbitrary infinite border strip with $\bar m_{\bar R}=\delta_{\bar R,1}$
corresponds to two
irreducible level-zero $U_q\bigl(\widehat{sl}(2\vert2)\bigr)$-modules
with highest weights
\bea
(0,\bar K,0,-\bar K)
\label{eq:vmod3}
\eea
and
\bea
\qquad(0,\bar K+1,0,-\bar K-1)
\label{eq:vmod4}
\eea
The weight (\ref{eq:vmod3}) refers to a tensor product of $\bar K$ evaluation modules
$V_z$. Their spectral parameters are given by the spectral parameters $\bar x_k$
with $k$ contained in the set $\bar{\Upsilon}_{\{\bar n_i;\,\bar m_i\}}$
in equation
(\ref{eq:hwlzero6}) multiplied by a factor $q^{-2}$. According to Sect. \ref{sec:bs1},
these spectral parameters can be expressed as $q^{2r_j}\bar x_1$ for $\bar K$
suitable values of $r_j$ where $1\leq r_j\leq\bar M$ and
$r_j>r_{j'}$ if $j>j'$ and $k,k'\in\bar{\Upsilon}_{\{\bar n_i;\bar m_i\}}$
for $q^{2r_j}\bar x_1=\bar x_k$, $q^{2r_{j'}}\bar x_1=\bar x_{k'}$.
With the notation $y_j=q^{2(r_j-1)}\bar x_1$, the level-zero module can be written
\bea
V_{y_1}\otimes V_{y_2}\otimes V_{y_3}\otimes\ldots\otimes V_{y_{\bar K}}
\label{eq:vmod5}
\eea
Similarly, 
the second level-zero module related to (\ref{eq:vmod4}) is expressed by
\bea
V_{y_1}\otimes V_{y_2}\otimes V_{y_3}\otimes\ldots\otimes V_{y_{\bar K}}\otimes
V_{q^{2(\bar M-1)}y_1}
\label{eq:vmod6}
\eea
where $y_{\bar K}=q^{2(\bar M-\bar m_1-1)}y_1$ according to the previous section.
For $\bar M=\bar K+1=1$, the tensor product $V_{y_1}\otimes V_{y_2}\otimes\ldots
\otimes V_{y_{\bar K}}$ in (\ref{eq:vmod5}) and (\ref{eq:vmod6}) is replaced
by the one-dimensional module with $U_q\bigl(gl(2\vert2)\bigr)$-weight
$(0,0,0,0)$.
A tensor product $V_{z_1}\otimes V_{z_2}\otimes\ldots\otimes V_{z_L}$ with
arbitrary spectral parameters $z_i$ and $L\geq1$ may be denoted by $V_{\{z_i\}}$.
Its highest weight vector $\tilde{\nu}_{\{z_i\}}$ satisfies
\bea
H^1_s\,\tilde{\nu}_{\{z_i\}}&=&0\cr
\noalign{\bigskip}
H^2_s\,\tilde{\nu}_{\{z_i\}}&=&q^{2s}{[s]\over s}\sum_{i=1}^Lz_i^s\;\tilde{\nu}_{
\{z_i\}}\cr
\noalign{\bigskip}
H^3_s\,\tilde{\nu}_{\{z_i\}}&=&0
\label{eq:veig1}
\eea
Explicit analysis of the weight structure reveals that the $U_q\bigl(gl(2
\vert2)\bigr)$-weights
of (\ref{eq:vmod5}) and (\ref{eq:vmod6}) exactly coincide with the weights of
the set $\Xi_{\{\bar n_i;\,\bar m_i\}}$.
Level-zero modules with the weight structure described by the sets $\Xi_{\{n_i;\,
m_i\},\{\bar n_i;\,\bar m_i\}}$ are obtained in two steps. First, appropriate
tensor products
of $M$ evaluation modules $W^*_{z}$ with (\ref{eq:vmod5}), (\ref{eq:vmod6})
are considered. They contain level-zero modules with the $U_q\bigl(gl(2\vert2)
\bigr)$-weights given by the sets
\bea
&&\Xi^{(M)}_{\{\bar n_i;\,\bar m_i\}}=\cr
\noalign{\bigskip}
&&\;\;\Bigl\{
\bigl(-N_0(T_B)-N_1(T_B),\,N_1(T_B)+N_2(T_B),\,-M+N_0(T_B)+N_1(T_B),\,M-2N_0(T_B)-N_1(T_B)-
N_2(T_B)\bigr)
\Bigr\}
\label{eq:shiftset}
\eea
Here $M=\sum_i(n_i+m_i)$ and $N_l(T_B)$ is defined as
in equation (\ref{eq:typebw}).
Stated briefly, a uniform shift by $(0,0,-M,M)$
applied on $\Xi_{\{\bar n_i;\,\bar m_i\}}$
yields the set $\Xi^{(M)}_{\{\bar n_i;\,\bar m_i\}}$.
A priori, several choices for the tensor products are possible.
In the construction outlined below, all finite-dimensional modules
$W^*_{\bar x_i,\,\{\bar n_i;\,\bar m_i\}}$ with $\bar m_{\bar R}=\delta_{\bar R,
1}$ involved in the decomposition of $V(\Lambda_0)$ in Sect.\ref{sec:bs2}
 appear as submodules of suitable quotients.
For $M<\bar m_1$, the level-zero module contains the whole tensor product
\bea
W^*_{q^{2\bar m_1}y_{\bar K}}\otimes W^*_{q^{2(\bar m_1-1)}y_{\bar K}}\otimes
W^*_{q^{2(\bar m_1-2)}y_{\bar K}}\otimes\ldots\otimes W^*_{q^{2(\bar m_1-M+1)}
y_{\bar K}}\otimes V_{y_1}\otimes V_{y_2}\otimes\ldots\otimes V_{y_{\bar K}}
\label{eq:prod2}
\eea
If $M\geq\bar m_1$ and $\bar M>1$, the tensor product (\ref{eq:prod2}) 
has a vector $\breve{\nu}_{\{M;\,\bar m_1;\,y_i\}}$ with $U_q\bigl(gl(2\vert2)
\bigr)$-weight $(-\bar m_1,\bar K+\bar m_1,-M+\bar m_1,M-\bar K-\bar m_1)$
and the highest weight properties
\bea
\Delta^{(M+\bar K-1)}(E^{l,+}_s)\,\breve{\nu}_{\{M;\,\bar m_1;\,y_i\}}=0\qquad\;\;
\mbox{for}\;\;s\in{\mathbb{Z}},\;l=1,2,3
\label{eq:hwpropa}
\eea
\bea
\Delta^{(M+\bar K-1)}(H^1_s)\,\breve{\nu}_{\{M;\,\bar m_1;\,y_i\}}&=&
-q^{s(2\bar M-\bar m_1)}{[s\bar m_1]\over s}\,y_1^s\;
\breve{\nu}_{\{M;\,\bar m_1;\,y_i\}}\cr
\noalign{\bigskip}
\Delta^{(M+\bar K-1)}(H^2_s)\,\breve{\nu}_{\{M;\,\bar m_1;\,y_i\}}&=&
q^{2s}\,{[s]\over s}\Biggl\{q^{s\bar m_1}{\bigl[s(\bar m_1+1)\bigr]\over[s]}
y_1^s
+\sum_{i=1}^{\bar K-1}y_i^s\Biggr\}\breve{\nu}_{\{M;\,\bar m_1;\,y_i\}}\cr
\noalign{\bigskip}
\Delta^{(M+\bar K-1)}(H^3_s)\,\breve{\nu}_{\{M;\,\bar m_1;\,y_i\}}&=&
-q^{s(2\bar M-M-\bar m_1)}{\bigl[s(M-\bar m_1)\bigr]\over s}\,y_1^s\;
\breve{\nu}_{\{M;\,\bar m_1;\,y_i\}}
\label{eq:hwpropb}
\eea
The level-zero module corresponding to the set $\Xi^{(M)}_{\bar n_i;\,\bar m_i\}}$
with $M\geq \bar m_1$, $\bar M>1$ is the quotient
$V_{\{M;\,\bar M;\,\bar m_1;\,y_i\}}$ of
the tensor product (\ref{eq:prod2}) by the module generated from $\breve{\nu}_{\{
M;\,\bar m_1;\,y_i\}}$ by the level-zero action of $U_q\bigl(\widehat{sl}(2
\vert2)\bigr)$. 
In the following, $V_{\{M;\,\bar M;\,\bar m_1;\,y_i\}}$ with $M<\bar m_1$
refers to the tensor product (\ref{eq:prod2}).
$V_{\{M;\,\bar M;\,\bar m_1;\,y_i\}}$ need not be irreducible.
In particular, 
for $\bar M=M$ a highest weight vector is found
at the $U_q\bigl(gl(2\vert2)\bigr)$-weight $(-M+\bar K,M,-\bar K,0)$. Then
the quotient of $V_{\{M;\,M;\,\bar m_1;\,y_i\}}$ by
all vectors resulting from the action of $U_q\bigl(\widehat{sl}(2
\vert2)\bigr)$ on this vector includes a $U_q\bigl(\widehat{sl}(2\vert2)
\bigr)$-submodule isomorphic to the finite-dimensional
module $W^*_{q^{2(M-1)}y_{1},\{\bar n_i;\,\bar m_i\}}$. This part arises from the
action on the vector ${w^*_3}^{\otimes M}\otimes v_{2,0}^{\otimes\bar K}$.

In addition to $V_{\{M;\,\bar M;\,\bar m_1;\,y_i\}}$, a part of the tensor product
\bea
W^*_{q^{2\bar m_1}y_{\bar K}}\otimes W^*_{q^{2(\bar m_1-1)}y_{\bar K}}\otimes
W^*_{q^{2(\bar m_1-2)}y_{\bar K}}\otimes\ldots\otimes W^*_{q^{2(\bar m_1-M+1)}
y_{\bar K}}\otimes V_{y_1}\otimes V_{y_2}\otimes\ldots\otimes V_{y_{\bar K}}
\otimes V_{q^{2(\bar M-1)}y_1}
\label{eq:prod3}
\eea
is required in the case $M<\bar m_1$.
The level-zero action of $U_q\bigl(\widehat{sl}(2\vert2)\bigr)$ on the highest
weight vector ${w^*_3}^{\otimes M}\otimes v_{2,0}^{\otimes(\bar K+1)}$
in (\ref{eq:prod3}) does not exhaust the whole tensor product. The quotient
$\hat V_{\{M;\,\bar M;\,\bar m_1;\,y_i\}}$ of the tensor product (\ref{eq:prod3})
by all vectors emerging from the action of $U_q\bigl(\widehat{sl}(2\vert2)\bigr)$
on  ${w^*_3}^{\otimes M}\otimes v_{2,0}^{\otimes(\bar K+1)}$ contains a vector
$\hat{\nu}_{\{M;\,\bar M;\,\bar m_1;\,y_i\}}$ with $U_q\bigl(gl(2\vert2)
\bigr)$-weight $(-M,\bar K+M+1,0,-\bar K-1)$ and the properties
\bea
\Delta^{(M+\bar K-1)}(E^{l,+}_s)\,\hat{\nu}_{\{M;\,\bar M;\,\bar m_1;\,y_i\}}=0\qquad\;\;
\mbox{for}\;\;s\in{\mathbb{Z}},\;l=1,2,3
\label{eq:hwpropc}
\eea
\bea
\Delta^{(M+\bar K-1)}(H^1_s)\,\hat{\nu}_{\{M;\,\bar M;\,\bar m_1;\,y_i\}}&=&
-q^{s(2\bar M-M)}\,{[Ms]\over s}\,y_{1}^s\;
\hat{\nu}_{\{M;\,\bar M;\,\bar m_1;\,y_i\}}\cr
\noalign{\bigskip}
\Delta^{(M+\bar K-1)}(H^2_s)\,\hat{\nu}_{\{M;\,\bar M;\,\bar m_1;\,y_i\}}&=&
q^{2s}\,{[s]\over s}\Biggl\{
q^{s(2\bar M-M-2)}{\bigl[s(M+1)\bigr]\over[s]}y_1^s+\sum_{i=1}^{\bar K}y_i^s\Biggr\}\;
\hat{\nu}_{\{M;\,\bar M;\,\bar m_1;\,y_i\}}\cr
\noalign{\bigskip}
\Delta^{(M+\bar K-1)}(H^3_s)\,\hat{\nu}_{\{M;\,\bar M;\,\bar m_1;\,y_i\}}&=&0
\label{eq:hwpropd}
\eea
For $M<\bar m_1$, the $U_q\bigl(gl(2\vert2)\bigr)$-weights of both modules
$V_{\{M;\bar M;\bar m_1;y_i\}}$ and $\hat V_{\{M;\bar M;\bar m_1;y_i\}}$ are
given by the set $\Xi^{(M)}_{\{\bar n_i;\,\bar m_i\}}$.

The level-zero modules related to $X^{(M)}_{\{1\}}$
consist of the finite-dimensional
module $W^*_{y_1,\{M\}}$ and the infinite-dimensional module
$\hat V_{\{M;\,1;\,1;\,y_1\}}$. The latter is defined by (\ref{eq:prod3})
with the factor $V_{y_1}\otimes V_{y_2}\otimes\ldots\otimes V_{y_{\bar K}}$ omitted
and by (\ref{eq:hwpropc}), (\ref{eq:hwpropd}) without the sum on the rhs of the second line
in (\ref{eq:hwpropd}).

In a second step, level-zero modules corresponding to infinite sets of configurations
$(\ldots\otimes w_{j_4}\otimes w^*_{j_3}\otimes w_{j_2}\otimes w^*_{j_1})$ or
$(\ldots\otimes w^*_{j_4}\otimes w_{j_3}\otimes w^*_{j_2}\otimes w_{j_1})$
are constructed from tensor products 
$W_{x_1,\{n_i;\,m_i\}}\otimes V_{\{M;\,\bar M;\,\bar m_1;y_i\}}$ and
$W_{x_1,\{n_i;\,m_i\}}\otimes \hat V_{\{M;\,\bar M;\,\bar m_1;y_i\}}$.
First, for two border stripes $R,\,M,\,K,\,\{n_i;\,m_i\}$
and $\bar R,\,\bar M,\,\bar K,\,\{\bar n_i;\,\bar m_i\}$ with $\bar m_{\bar R}=
\delta_{\bar R,1}$ the product
\bea
W_{x_1,\{n_i;\,m_i\}}\otimes V_{\{M;\,\bar M;\,\bar m_1;y_i\}}
\label{eq:prod5}
\eea
will be considered. With
\bea
x_1=q^{2(\bar M-m_R+\delta_{R,1}-1)}\,y_1
\label{eq:xyrel}
\eea
the entire tensor product (\ref{eq:prod5}) is annihilated by $\Delta^{(2M+\bar K-1)}\bigl(
H^1_s+H^3_s\bigr)$ for any $s\neq 0$. Its highest weight vector is given by
$\nu_{\{n_i;\,m_i\},\{\bar n_i;\,\bar m_i\}}\equiv
\nu_{\{n_i;\,m_i\}}\otimes\tilde{\nu}_{\{M;\bar m_1;y_i\}}
=\nu_{\{n_i;\,m_i\}}\otimes {w^*_3}^{\otimes M}\otimes v_{2,0}^{\otimes\bar K}$.
The eigenvalues of the coproducts of $H^j_s$
acting on the highest weight vector 
follow from the highest weight properties of $\nu_{\{n_i;\,m_i\}}$ and  
$\tilde{\nu}_{\{M;\bar m_1;y_i\}}$:
\bea
\Delta^{(2M+\bar K-1)}\bigl(H^1_s\bigr)\,v_{\{n_i;\,m_i\},\{\bar n_i;\,\bar m_i\}}
&=&q^{2s(m_R-\delta_{R,1}+1)-sM}{[sM]\over s}x_1^s\,
v_{\{n_i;\,m_i\},\{\bar n_i;\,\bar m_i\}}\cr
\noalign{\bigskip}
\Delta^{(2M+\bar K-1)}\bigl(H^2_s\bigr)\,v_{\{n_i;\,m_i\},\{\bar n_i;\,\bar m_i\}}
&=&
{1\over s}x_1^s\Biggl\{q^{-2s(\bar M-m_R+
\delta_{R,1}-2)}\sum_{i=1}^{\bar R-1}q^{2s\bar t_{i-1}-s(\bar n_{\bar R-i}+1)}
[s\bar n_{\bar R-i}]\cr
\noalign{\bigskip}
&&\qquad-\sum_{i=1}^{R-1}q^{-2s(t_{i-1}-m_R)+s(n_{R-i}+1)}[sn_{R-i}]
\Biggr\}v_{\{n_i;\,m_i\},\{\bar n_i;\,\bar m_i\}}
\label{eq:firsthweig}
\eea
Here (\ref{eq:xyrel}) is taken into account.
For $\bar M=1$, $m_R=\delta_{R,1}$ and $R,\,\bar R
\leq2$, a highest weight vector with $U_q\bigl(gl(2\vert2)\bigr)$-weight
$(M-n_{R-1}-1,-K+n_{R-1},-M+n_{R-1}+1,2M-K-n_{R-1}-2)$ is present in (\ref{eq:prod5})
with $x_1$ and $y_1$ related by (\ref{eq:xyrel}). The quotient of $W_{x_1,\{n_i;\,
m_i\}}\otimes W^*_{x_1,\{M\}}$ by the $U_q\bigl(\widehat{sl}(2\vert2)
\bigr)$-module generated on this highest weight vector may be called
$W_{x_1,\{n_i;m_i\},M}$. 
If $m_R\neq\delta_{R,1}$, a vector
with $U_q\bigl(gl(2\vert2)\bigr)$-weight $(M-1,-K,-M+1,2M-K-2)$ is found in
(\ref{eq:prod5}), (\ref{eq:xyrel}) but not among the vectors obtained by applying
the $U_q\bigl(\widehat{sl}(2\vert2)\bigr)$-generators on the highest weight
vector $\nu_{\{n_i;\,m_i\}}\otimes {w^*_3}^{\otimes M}$.
In this case, $W_{x_1,\{n_i;m_i\},M}$ denotes the module generated on
$\nu_{\{n_i;\,m_i\}}\otimes {w^*_3}^{\otimes M}$ by the level-zero action of
$U_q\bigl(\widehat{sl}(2\vert2)\bigr)$.

Generally, the $U_q\bigl(\widehat{sl}(2\vert2)\bigr)$-module created on the
highest weight vector $\nu_{\{n_i;\,m_i\}}\otimes\tilde{\nu}_{\{M;\bar m_1;y_i
\}}\in W_{x_1,\{n_i;\,m_i\}}\otimes V_{\{M;\bar M;\bar m_1;y_i\}}$ with
(\ref{eq:xyrel}) may contain only a part of the $U_q\bigl(\widehat{sl}(2\vert2)
\bigr)$-weights of the corresponding part in $\{\tau\}_{n_{\{n_i;m_i\}}}
\otimes\{\tau^*\}_{\bar n_{\{\bar n_i;\bar m_i\}}}$. A simple example is
provided by the case $R=\bar R=M=\bar M=2$. The level-zero action of $U_q\bigl(
\widehat{sl}(2\vert2)\bigr)$ on $\nu_{\{1;1,0\}}\otimes\tilde{\nu}_{\{2;1;q^{-2}
x_1\}}\in W_{x_1,\{1;1,0\}}\otimes V_{\{2;2;1;q^{-2}x_1\}}
$ does not yield a vector 
$a\,f_2\nu_{\{1;1,0\}}\otimes\tilde{\nu}_{\{2;1,q^{-2}x_1\}}+b
\nu_{\{1;1,0\}}\otimes f_2\tilde{\nu}_{\{2;1,q^{-2}x_1\}}$ with $a\neq q^{-1}b$.
The $U_q\bigl(gl(2\vert2)\bigr)$-weight structure of the
$U_q\bigl(\widehat{sl}(2\vert2)\bigr)$-module $V_{x_1,\{1;1,0\},\{1;1,0\}}$
generated from such a vector
coincides with the weight structure of the part of $\{\tau\}_{-2}\otimes\{\tau^*\}_{-1}$
associated to two border stripes $\{1;1,0\}$.
For general $R,\,\bar R>1$, there are $\left({L\atop L'}\right)$
vectors with $U_q\bigl(gl(2\vert2)\bigr)$-weight
$(M-L',\bar K-K,-M+L',2M-K-\bar K)$ in $W_{x_1,\{n_i;\,m_i\}}\otimes V_{\{M;\bar M;
\bar m_1;y_i\}}$ missing among the vectors generated from all vectors with weight
$(M-L'+1,\bar K-K,-M+L'-1,2M-K-\bar K)$ if
\bea
q^{-2(t_{i_l}-m_R)}\,x_1=q^{2\bar t_{j_l}}\,y_1
\label{eq:xyrel2}
\eea
with $1\leq l\leq L<min(R,\bar R)$, $1\leq L'\leq L$,
$0\leq i_l\leq R-2,\;\;0\leq j_l\leq \bar R-2$ and $i_l\neq i_{l'},\;
j_l\neq j_{l'}\;\mbox{for}\;l\neq l'$.
In (\ref{eq:xyrel2}),  $t_i$ is defined by (\ref{eq:trdef}), $\bar t_0=0$ and
$\bar t_j=\sum_{j'=1}^j(\bar n_{\bar R-j'}+\bar m_{\bar R-j'}
)$, $1\leq j\leq \bar R-2$.
These may be regarded as vectors in the quotient of $W_{x_1,\{n_i;\,m_i\}}\otimes V_{\{M;
\bar M;\bar m_1;y_i\}}$ by the $U_q\bigl(\widehat{sl}(2\vert2)\bigr)$-modules generated
on the vectors with $U_q\bigl(gl(2\vert2)\bigr)$-weight $(M-L'+1,\bar K-K,-M+L'-1,
2-K-\bar K)$ for a given $L'$. This allows to write them as linear combinations of
eigenvectors $\nu^{(L';r_1,r_2,\ldots,r_{L'})}_{\{n_i;\,m_i\},\{\bar n_i;\,\bar m_i
\}}$ of $\Delta^{(2M+K-1)}\bigl(H^j_s\bigr)$, $j=1,2,3$:
\bea
\Delta^{(2M+\bar K-1)}\bigl(H^1_s\bigr)\,v^{(L';r_1,r_2,\ldots,r_{L'})}_{
\{n_i;\,m_i\},\{\bar n_i;\,\bar m_i\}}
&=&{1\over s}x_1^sq^{2s(m_R+1)}\Bigl(q^{-sM}[Ms]-\sum_{l=1}^{L'}
q^{-s(2t_{r_l}+1)}[s]\Bigr)\,
v^{(L';r_1,r_2,\ldots,r_{L'})}_{\{n_i;\,m_i\},\{\bar n_i;\,\bar m_i\}}\cr
\noalign{\bigskip}
&=&-
\Delta^{(2M+\bar K-1)}\bigl(H^3_s\bigr)\,v^{(L';r_1,r_2,\ldots,r_{L'}
)}_{\{n_i;\,m_i\},\{\bar n_i;\,\bar m_i\}}\cr
\noalign{\bigskip}
\Delta^{(2M+\bar K-1)}\bigl(H^2_s\bigr)\,v^{(L',r_1,r_2,\ldots,r_{L'})}_{
\{n_i;\,m_i\},\{\bar n_i;\,\bar m_i\}}&=&
{1\over s}x_1^s\Biggl\{q^{-2s(\bar M-m_R+
\delta_{R,1}-2)}\sum_{i=1}^{\bar R-1}q^{2s\bar t_{i-1}-s(\bar n_{\bar R-i}+1)}
[s\bar n_{\bar R-i}]\cr
\noalign{\bigskip}
&&\qquad-\sum_{i=1}^{R-1}q^{-2s(t_{i-1}-m_R)+s(n_{R-i}+1)}[sn_{R-i}]
\Biggr\}v^{(L';r_1,r_2,\ldots,r_{L'})}_{\{n_i;\,m_i\},\{\bar n_i;\,\bar m_i\}}
\label{eq:eigsechw}
\eea
where the parameters $r_l$ are $L'$ different numbers chosen in $\{i_{l'}\}_{
1\leq l'\leq L}$. The $U_q\bigl(\widehat{sl}(2\vert2)\bigr)$-module
$V^{(L)}_{x_1,\{n_i;\,m_i\},\{\bar n_i;\,\bar m_i\}}$ generated
on $v^{(L;l_1,l_2,\ldots, l_L)}_{\{n_i;\,m_i\},\{\bar n_i;\,\bar m_i\}}$
will be denoted by $\tilde V_{x_1,\{n_i;\,m_i\},\{\bar n_i;\,\bar m_i\}}$
It contains
all vectors in $W_{x_1,\{n_i;\,m_i\}}\otimes V_{\{M;\bar M;\bar m_1;y_i\}}$
with $U_q\bigl(gl(2\vert2)\bigr)$-weights $(M-\hat L,\bar K-K,-M+\hat L,2M-K
-\bar K)$, $\hat L=0,1,2,\ldots$. 

For a pair of border stripes not satisfying any condition of the form (\ref{eq:xyrel2}),
$\tilde V_{x_1,\{n_i;\,m_i\},\{\bar m_i;\,\bar n_i\}}$ denotes 
the $U_q\bigl(\widehat{sl}(2\vert2)\bigr)$-module obtained on $\nu_{\{n_i;\,m_i
\}}\otimes {w^*_3}^{\otimes M}\otimes v_{2,0}^{\otimes\bar K}\in W_{x_1,\{n_i;\,m_i\}}\otimes
V_{\{M;\bar M;\bar m_1;y_i\}}$ with relation (\ref{eq:xyrel}).
In particular, the equations (\ref{eq:xyrel2}) are incompatible
with the requirement (\ref{eq:xyrel}) if  $M<\bar m_1$.
Explicit examination reveals that in the cases $1\leq R,\bar R\leq2$ with $\bar M>1$
and $K=0,1$ for $m_R\neq\delta_{R,1}$
the  $U_q\bigl(gl(2\vert2)\bigr)$-weights of
$\tilde V_{x_1,\{n_i;\,m_i\},\{\bar n_i;\,\bar m_i\}}$ 
exactly correspond 
to the weights of the set $\Xi_{\{n_i;\,m_i\},\{\bar n_i;\,\bar m_i\}}$ if
$M\geq \bar m_1$.

If  $\bar M=1$ or $M<\bar m_1$, the tensor product
\bea
W_{x_1,\{n_i;\,m_i\}}\otimes\hat V_{\{M;\bar M;\bar m_1;y_i\}}
\label{eq:prod6}
\eea
remains to be considered. 
Provided that $x_1$ and $y_1$ are related by (\ref{eq:xyrel}), some general statements
on the structure of the product (\ref{eq:prod6}) can be formulated.
It contains a vector
$\grave{\nu}_{\{n_i;\,m_i\},\{\bar n_i;\,\bar m_i\}}$ with
$U_q\bigl(gl(2\vert2)\bigr)$-weight $(0,\bar K-K+M,0,M-K-\bar K-2)$
which does not result from the level-zero action of $U_q\bigl(\widehat{sl}(2\vert
2)\bigr)$ on $\nu_{\{n_i;\,m_i\}}\otimes\hat{\nu}_{\{M;\,\bar M;\,\bar m_1;\,
y_i\}}$. The vector is characterized by
\bea
&&\Delta^{(2M+\bar K-1)}\bigl(H^1_s\bigr)\,\grave{\nu}_{\{n_i;\,m_i\},\{\bar n_i;\,
\bar m_i\}}=
\Delta^{(2M+\bar K-1)}\bigl(H^3_s\bigr)\,\grave{\nu}_{\{n_i;\,m_i\},\{\bar n_i;\,
\bar m_i\}}=0\cr
\noalign{\bigskip}
&&\Delta^{(2M+\bar K-1)}\bigl(H^2_s\bigr)\,\grave{\nu}_{\{n_i;\,m_i\},\{\bar n_i;\,
\bar m_i\}}=\cr
\noalign{\bigskip}
&&\qquad\qquad
={[s]\over s}\Biggl\{q^{s(2\bar M-1)}y_1^s\biggl(q^{-sm_R}[sm_R]+\sum_{i=1}^{
R-2}q^{-s(2t_i+m_{R-i-1})}[sm_{R-i-1}]\biggr)
+q^{2s}\sum_{i=1}^{\bar K}y_i^s\Biggr\}\;
\grave{\nu}_{\{n_i;\,m_i\},\{\bar n_i;\,\bar m_i\}}
\label{eq:eigabs}
\eea
The level-zero action of $U_q\bigl(\widehat{sl}(2\vert2)\bigr)$ on
$\grave{\nu}_{\{n_i;\,m_i\},\{\bar n_i;\,\bar m_i\}}$ provides the whole
tensor product (\ref{eq:prod6}). Due to (\ref{eq:xyrel}),
both $\nu_{\{n_i;\,m_i\}}\otimes\hat{\nu}_{\{M;\,\bar M;\,\bar m_1;\,y_i\}}$
and $\Delta^{(2M+\bar K-1)}(f_1)\,\bigl(
\nu_{\{n_i;\,m_i\}}\otimes\hat{\nu}_{\{M;\,\bar M;\,\bar m_1;\,y_i\}}\bigr)$
satisfy the highest weight properties. For the latter they read
\bea
\Delta^{(2M+\bar K-1)}\bigl(E^{l,+}_s\cdot f_1\bigr)\,
(\nu_{\{n_i;\,m_i\}}\otimes\hat{\nu}_{\{M;\,\bar M;\,\bar m_1;\,y_i\}})=0
\qquad\;\;\forall s,\;l=1,2,3
\eea
\bea
&&\Delta^{(2M+\bar K-1)}\bigl(H^1_s\cdot f_1\bigr)\,
(\nu_{\{n_i;\,m_i\}}\otimes\hat{\nu}_{\{M;\,\bar M;\,\bar m_1;\,y_i\}})
=\Delta^{(2M+\bar K-1)}\bigl(H^3_s\cdot f_1\bigr)\,
(\nu_{\{n_i;\,m_i\}}\otimes\hat{\nu}_{\{M;\,\bar M;\,\bar m_1;\,y_i\}})
=0\cr
\noalign{\bigskip}
&&\Delta^{(2M+\bar K-1)}\bigl(H^2_s\cdot f_1\bigr)\,
(\nu_{\{n_i;\,m_i\}}\otimes\hat{\nu}_{\{M;\,\bar M;\,\bar m_1;\,y_i\}})
=\cr
\noalign{\bigskip}
&&\qquad\qquad
={[s]\over s}\Biggl\{q^{s(2\bar M-1)}y_1^s\biggl(q^{-sm_R}[sm_R]+\sum_{i=1}^{
R-2}q^{-s(2t_i+m_{R-i-1}-\delta_{i,R-2})}\bigl[s(m_{R-i-1}-\delta_{i,R-2})
\bigr]\biggr)+\cr
\noalign{\bigskip}
&&\qquad\qquad\qquad
+q^{2s\bar M}y_1^s+q^{2s}\sum_{i=1}^{\bar K}y_i^s\Biggr\}\;
\Delta^{(2M+\bar K-1)}(f_1)\,
(\nu_{\{n_i;\,m_i\}}\otimes\hat{\nu}_{\{M;\,\bar M;\,\bar m_1;\,y_i\}})
\label{eq:hwproper1}
\eea
with $s\neq0$.
The vector $\Delta^{(M-1)}(f_1)\nu_{\{n_i;\,m_i\}}
\otimes\Delta^{(M+\bar K-1)}(f_1)\hat{\nu}_{\{M;\,\bar M;\,\bar m_1;\,y_i\}}$
is annihilated by $\Delta^{(2M+\bar K-1)}(H^1_s)$ and
$\Delta^{(2M+\bar K-1)}(H^3_s)$. Moreover, it satisfies
\bea
&&\Delta^{(2M+\bar K-1)}(E^{l,+}_s)\,\Bigl(\Delta^{(M-1)}(f_1)\nu_{\{n_i;\,m_i\}}
\otimes\Delta^{(M+\bar K-1)}(f_1)\hat{\nu}_{\{M;\,\bar M;\,\bar m_1;\,y_i\}}\Bigr)
=\cr
\noalign{\bigskip}
&&\qquad\qquad\qquad\qquad\qquad\qquad\qquad\qquad
=\delta_{l,1}\,q^{2s\bar M}y_1^s[M]\cdot\Delta^{(2M+\bar K-1)}(f_1)
\bigl(\nu_{\{n_i;\,m_i\}}\otimes\hat{\nu}_{\{M;\,\bar M;\,\bar m_1;\,y_i\}}\bigr)
\eea
for $l=1,2,3$ and
\bea
&&\Delta^{(2M+\bar K-1)}(H^2_s)\,\Bigl(\Delta^{(M-1)}(f_1)\nu_{\{n_i;\,m_i\}}
\otimes\Delta^{(M+\bar K-1)}(f_1)\hat{\nu}_{\{M;\,\bar M;\,\bar m_1;\,y_i\}}\Bigr)
=\cr
\noalign{\bigskip}
&&\qquad\qquad
={[s]\over s}\Biggl\{q^{s(2\bar M-1)}y_1^s\biggl(q^{-sm_R}[sm_R]+\sum_{i=1}^{
R-2}q^{-s(2t_i+m_{R-i-1}-\delta_{i,R-2})}\bigl[s(m_{R-i-1}-\delta_{i,R-2})
\bigr]\biggr)+q^{2s}\sum_{i=1}^{\bar K}y_i^s\Biggr\}\cdot\cr
\noalign{\bigskip}
&&\qquad\qquad\qquad\qquad\cdot
\Bigl(\Delta^{(M-1)}(f_1)
\nu_{\{n_i;\,m_i\}}\otimes\Delta^{(M+\bar K-1)}(f_1)
\hat{\nu}_{\{M;\,\bar M;\,\bar m_1;\,y_i\}}\Bigr)
\label{eq:hwproper2}
\eea
Thus $\Delta^{(M-1)}(f_1)\nu_{\{n_i;\,m_i\}}
\otimes\Delta^{(M+\bar K-1)}(f_1)\hat{\nu}_{\{M;\,\bar M;\,\bar m_1;\,y_i\}}$ is
a highest weight vector in the quotient of
$W_{x_1,\{n_i;\,m_i\}}\otimes\hat V_{\{M;\bar M;\bar m_1;y_i\}}$ by the
$U_q\bigl(\widehat{sl}(2\vert2)\bigr)$-module generated on $\Delta^{(2M+\bar K
-1)}(f_1)\bigl(\nu_{\{n_i;\,m_i\}}\otimes\hat{\nu}_{\{M;\,\bar M;\,\bar m_1;\,y_i\}}
\bigr)$. At $\bar M=1$, the last sum on the rhs of equations (\ref{eq:eigabs}),
(\ref{eq:hwproper1}) and (\ref{eq:hwproper2}) is dropped.

These observations allow to specify particular level-zero modules contained in
(\ref{eq:prod6}) with (\ref{eq:xyrel}).
The cases $M<\bar m_1$ and $\bar M=1$ need to be distinguished.

If $\bar M=1$, ${\hat V}_{x_1,\{n_i;\,m_i\},M}$ is defined as the level-zero
$U_q\bigl(\widehat{sl}(2\vert2)\bigr)$-module generated on
$v_{\{n_i;\,m_i\}}\otimes\hat{\nu}_{\{M;\,1;\,1;\,y_i\}}$. 
For $1\leq R\leq2$, the $U_q\bigl(gl(2\vert2)\bigr)$-weights found in
$W_{x_1,\{n_i;\,m_i\},M}$ and ${\hat V}_{x_1,\{n_i;\,m_i\},M}$ are given by
the weights of the set $\Xi_{\{n_i;\,m_i\},\{1\}}$.

In the case $M<\bar m_1$,
the quotient of $W_{x_1,\{n_i;\,m_i\}}\otimes\hat V_{\{M;\bar M;
\bar m_1;y_i\}}$ with (\ref{eq:xyrel})
by the $U_q\bigl(\widehat{sl}(2\vert2)\bigr)$-module generated
on $\Delta^{(M-1)}(f_1)\nu_{\{n_i;\,m_i\}}\otimes\Delta^{(M+\bar K-1)}(f_1)
\hat{\nu}_{\{M;\,\bar M;\,\bar m_1;\,y_i\}}$ will be denoted by $\tilde{\hat V}_{x_1,
\{n_i;\,m_i\},\{\bar n_i;\,\bar m_i\}}$.
For $M<\bar m_1$ and $1\leq R,\bar R\leq2$ it is readily verified that
$\tilde V_{x_1,\{n_i;\,m_i\},\{\bar n_i;\,\bar m_i\}}$ and 
$\tilde{\hat V}_{x_1,\{n_i;\,m_i\},\{\bar n_i;\,\bar m_i\}}$ together
contain exactly the same $U_q\bigl(gl(2\vert2)\bigr)$-weights
as the set $\Xi_{\{n_i;\,m_i\},\{\bar n_i;\,\bar m_i\}}$.
Formally, the statement can be extended to the parameters $R=0$, $\{\bar n_i;\,
\bar m_i\}$. 
Then the related modules (\ref{eq:vmod5}) and (\ref{eq:vmod6})
are denoted by $\tilde V_{x_1,\{\bar n_i;\,\bar m_i\}}$ and $\tilde{\hat V}_{x_1,
\{\bar n_i;\,\bar m_i\}}$, respectively. Here the spectral parameter $x_1$ 
is fixed by $x_1=q^{2(\bar M-1)}\,y_1$. $\tilde V_{x_1,\{1\}}$ is the one-dimensional
module with $U_q\bigl(gl(2\vert2)\bigr)$-weight $(0,0,0,0)$.

It seems that the modules constructed from the tensor products (\ref{eq:prod5})
and (\ref{eq:prod6}) have not been considered before.
In general, they are reducible as $U_q\bigl(sl(2\vert2)\bigr)/(h_1+h_3)$-modules.
The $sl(2\vert2)$-modules associated to the irreducible components
are found in \cite{cheng2} and some of the references therein.

The above statements on the weight structures of the level-zero modules
may be assumed to hold true for all $R$ and $\bar R$.
\vspace{0.5cm}

{\it Conjecture IV\,:}
The $U_q\bigl(gl(2\vert2)\bigr)$-weights of the level-zero modules attributed to
a pair of border stripes 
exactly coincide with the weights in the sets $\Xi_{\{\bar n_i;\,\bar m_i\}}$
(for $R=0$) or $\Xi_{\{n_i;\,m_i\},\{\bar n_i;\,\bar m_i\}}$ (for $R>0$).

Hence the weight-preserving one-to-one correspondence between the labelings $T_{A,B}$
and the space $\Omega_A$ of half-infinite
configurations implies a decomposition of the $U_q\bigl(gl(2\vert2)\bigr)$-weight
structure of $\Omega_A$ into the weight structures of the level-zero modules. 
\vspace{0.5cm}

The pairs are formed by a finite border strip parameterized either by $R=0$ or
by $R>0,\,\{n_i;\,m_i\}$ and an infinite border strip with parameters
$\bar R>0,\,\{\bar n_i;\,\bar m_i\},\,\bar m_{\bar R}=\delta_{\bar R,1}$.
A complete list of pairs of border stripes and the related level-zero modules
is provided by the following table:

\bigskip
 
\begin{tabular}{c|c}
border stripes&level-zero modules of $U_q\bigl(\widehat{sl}(2\vert2)\bigr)$\\
\hline
&\\
$\qquad R=0;\;\bar R,\,\{\bar n_i;\,\bar m_i\},\;\bar m_{\bar R}=\delta_{\bar R,1}\qquad
$&$\qquad\tilde V_{x_1,\{\bar n_i;\,\bar m_i\}}\oplus
\tilde{\hat V}_{x_1,\{\bar n_i;\,\bar m_i\}}\qquad$\\
&\\
\hline
&\\
$\qquad R,\,\{n_i;\,m_i\};\,\bar R,\,\{\bar n_i;\,\bar m_i\},\;m<\bar m_1,\,\bar M>1,\,
\bar m_{\bar R}=0\qquad
$&$\qquad\tilde V_{x_1,\{n_i;\,m_i\},\{\bar n_i;\,\bar m_i\}}\oplus
\tilde{\hat V}_{x_1,\{n_i;\,m_i\},\{\bar n_i;\,\bar m_i\}}\qquad$\\
&\\
\hline
&\\
$\qquad R,\,\{n_i;\,m_i\};\,\bar R=\bar M=1\qquad$&$W_{x_1,\{n_i;\,m_i\},M}\oplus
\hat V_{x_1,\{n_i;\,m_i\},M}$\\
&\\
\hline
&\\
$R,\,\{n_i;\,m_i\};\,\bar R,\,\{\bar n_i;\,\bar m_i\},\;M\geq \bar m_1,\,\bar M>1,\,\bar m_{
\bar R}=0$&$\tilde V_{x_1,\{n_i;\,m_i\},\{\bar n_i;\,\bar m_i\}}$\\
&\\
\hline
&\\
\end{tabular}

\bigskip

All modules except $\tilde V_{x_1,\{1\}}$ and $W_{x_1,\{n_i;\,m_i\},M}$ are
infinite-dimensional. The labelings $T_{A,B}$ consist of a restricted type $A$
labeling of the finite border strip and a type $B$ labeling of the infinite
border strip. $\Omega_A$ is the space of all half-infinite configurations
$(\ldots\otimes w_{j_4}\otimes w^*_{j_3}\otimes w_{j_2}\otimes w^*_{j_1})$
with $j_r=3$ for almost all $r$. Here the  $U_q\bigl(gl(2\vert2)\bigr)$-weights of
the configurations are determined with respect to the reference weight $\bar h^A$
specified in (\ref{eq:weightch}). The statement also applies to the space
$\Omega_B$ of half-infinite configurations with the reference weight $\bar h^B$
given in (\ref{eq:weightch2}).

The modules in this table have been introduced with respect to the evaluation action.
Another level-zero action of $U_q\bigl(\widehat{sl}(2\vert2)\bigr)$
is furnished by a suitable generalization of the $q$-deformation of the Yangian
action given in \cite{pet1,tug1} for $U_q\bigl(\widehat{sl}(N)\bigr)$-models.
With respect to this action, the module $\tilde V(\Lambda_0)$ may decompose
into the level-zero modules collected in the above table. A similar decomposition
has been constructed in \cite{tak} in the $U_q\bigl(\widehat{sl}(N)\bigr)$-case.
Details for the $U_q\bigl(\widehat{sl}(2\vert2)\bigr)$-model studied here will
be published separately.

\vskip 1cm
{\bf Acknowledgment}. This research project is supported by Michael Krautg\"artner.

\end{document}